\documentclass[11pt]{article}
\usepackage{tikz}
\usetikzlibrary{decorations.pathmorphing,cd,decorations.markings}
\usepackage{tikz,tkz-euclide,tikz-cd}
\usepackage{circuitikz}

\usepackage{latexsym,amsmath,amsfonts,amssymb}
\usepackage[latin1]{inputenc}
\usepackage[american]{babel}
\usepackage{graphicx}
\usepackage{bbm}
\usepackage{cite}
\usepackage{tcolorbox}
\usetikzlibrary{calc}
\usetikzlibrary{decorations.pathmorphing, decorations.markings}
\usetikzlibrary{patterns}
\usepackage{soul}
\usepackage{comment}
\usepackage{array}
\usepackage{tikz-3dplot}
\usepackage{pgfplots}
\usepackage[normalem]{ulem} %for sout, strikethrough text

\usepackage[colorlinks,linkcolor=black,citecolor=blue,urlcolor=blue,linktocpage,pagebackref]{hyperref}

\makeatletter
\def\section{\@startsection {section}{1}{\z@}{-3.5ex plus -1ex minus
 -.2ex}{2.3ex plus .2ex}{\large\bf}}
\def\subsection{\@startsection{subsection}{2}{\z@}{-3.25ex plus -1ex
minus -.2ex}{1.5ex plus .2ex}{\normalsize\bf}}
\makeatother
\makeatletter

\@addtoreset{equation}{section}

\makeatother

\newcommand\eea{\end{eqnarray}}
\newcommand\bea{\begin{eqnarray}}

\def\beq{\begin{equation}}
\def\eeq{\end{equation}}

\newcommand{\be}{\begin{equation}}
\newcommand{\ee}{\end{equation}}
\newcommand{\ba}{\begin{align}}
\newcommand{\ea}{\end{align}}
\newcommand{\bg}{\begin{gather}}
\newcommand{\eg}{\end{gather}}
\newcommand{\bseq}{\begin{subequations}}
\newcommand{\eseq}{\end{subequations}}

\newcommand{\mat}[1]{\begin{pmatrix} #1 \end{pmatrix}}

\newcommand{\ie}{\textit{i.e.}}

\newcommand{\cA}{\mathcal{A}}
\newcommand{\cB}{\mathcal{B}}
\newcommand{\cC}{\mathcal{C}}
\newcommand{\cD}{\mathcal{D}}
\newcommand{\cE}{\mathcal{E}}

\newcommand{\cL}{\mathcal{L}}

\newcommand{\cT}{\mathcal{T}}

\newcommand{\cV}{\mathcal{V}}

\newcommand{\bA}{\mathbb{A}}

\newcommand{\bC}{\mathbb{C}}

\newcommand{\bQ}{\mathbb{Q}}
\newcommand{\bR}{\mathbb{R}}

\newcommand{\bZ}{\mathbb{Z}}

\newcommand{\sD}{\mathsf{D}}

\newcommand{\sO}{\mathsf{O}}

\begin{document}
\thispagestyle{empty}

\begin{center}

	\vspace*{-.6cm}

	\begin{center}

		\vspace*{1.1cm}

		{\centering \Large\textbf{Exploring duality symmetries, multicriticality and RG flows at $c=2$}}

	\end{center}

	\vspace{0.8cm}
	{\bf Jeremias Aguilera Damia$^a$, Giovanni Galati$^a$,\\ Ondrej Hulik$^b$, and Salvo Mancani$^a$}

	\bigskip
	{\it
		$^a$ Physique Th\'eorique et Math\'ematique and International Solvay Institutes\\
Universit\'e Libre de Bruxelles, C.P. 231, 1050 Brussels, Belgium \\[.0em]
		$^b$ Theoretische Natuurkunde, Vrije Universiteit Brussel\\ Pleinlaan 2, B-1050 Brussels, Belgium  \\[.6em]
	}

	\vspace{.3cm}

%{jeremias.aguilera.damai@ulb.be , giovanni.galati@ulb.be, mancanisalvo@gmail.com}
\end{center}

\begin{abstract}
In this work, we study the realization of non-invertible duality symmetries along the toroidal branch of the $c=2$ conformal manifold. A systematic procedure to construct symmetry defects is implemented to show that all Rational Conformal Field Theories along this branch enjoy duality symmetries. Furthermore, we delve into an in-depth analysis of two representative cases of multicritical theories, where the toroidal branch meets various orbifold branches. For these particular examples, the categorical data and the defect Hilbert spaces associated with the duality symmetries are obtained by resorting to modular covariance. Finally, we study the interplay between these novel symmetries and the various exactly marginal and relevant deformations, including some representative examples of Renormalization Group flows where the infrared is constrained by the non-invertible symmetries and their anomalies.
\end{abstract}

\newpage

\tableofcontents

\newpage

\section{Introduction}
Quantum Field Theories in two spacetime dimensions often provide a suitable arena to explore interesting dynamical aspects while still retaining some analytical control. This becomes more explicit when the field theory enjoys conformal invariance, {\it i.e.} it is a Conformal Field Theory (CFT). In such a case, the theory dynamics get highly constrained by the infinite dimensional Virasoro algebra \cite{DiFrancesco:1997nk, Ginsparg:1988ui}. Moreover, when the spacetime is taken to be the Eucliedean torus\footnote{Of course, this notion extends to any Riemann surface of arbitrary genus, with the  large diffeomorphisms comprised in the Mapping Class Group (see for instance \cite{Dijkgraaf:1987vp}). However, considering the two-dimensional torus will be enough for the purposes of this work.}, covariance under large diffeomorphisms, encoded in the modular group, places additional restrictions on the space of states of the theory. For some purposes, this proves enough to determine the structure of the Hilbert space or the behavior of some observables analytically over certain parametric regimes \cite{Hellerman:2009bu}, remarkably the high temperature density of states as pioneered by Cardy \cite{Cardy:1986ie} (see also \cite{Benjamin:2023qsc} for a novel generalization to higher dimensions). More generally, non-trivial solutions of these constraints can be found either analytically or numerically via more refined conformal bootstrap techniques \cite{Ferrara:1972kab,Polyakov:1974gs,Belavin:1984vu,Rattazzi:2008pe,Poland:2018epd}.  
A special set of conformal field theories in two dimensions corresponds to the Rational Conformal Field Theories (RCFT) \cite{Belavin:1984vu}. A key property of these theories is that the Hilbert space decomposes into a finite set of representations of some chiral algebra. For $c<1$, all unitary conformal field theories are of this kind and the partition function is accounted by a finite set of representations of the Virasoro algebra. These are the minimal models. For higher values of the central charge, {\it i.e.} $c\geq1$, certain points on the conformal manifold meet some extra structure due to the occurrence of additional, generically higher spin, conserved currents assembling an enhanced chiral algebra. Notably, the dynamics of these remarkable theories is described in terms of a finite set of conformal primaries with respect to the enhanced chiral algebra, hence corresponding to RCFTs. We will focus on examples of this kind throughout this work. 

Concomitantly, rational CFTs played a prominent role in the discovery and characterization of topological defects, leading to novel forms of global symmetry, and the development of alternative descriptions in terms of higher dimensional topological field theories (TQFT) (see e.g. \cite{Verlinde:1988sn,Fuchs:2002cm,Fuchs:2007tx,Freed:2022qnc,Benini:2022hzx,Bhardwaj:2023ayw,Chang:2018iay,Thorngren:2019iar,Thorngren:2021yso,Cordova:2023qei,Baume:2023kkf,Bhardwaj:2023bbf, Yu:2023nyn}). In this work, we will focus on the former aspect, namely the description of interesting global symmetry structures and the exploration of their consequences. A comprehensive treatment of the general formalism and some applications for $c\leq1$ CFTs can be found in e.g. \cite{Chang:2018iay,Thorngren:2019iar,Thorngren:2021yso,Bhardwaj:2023idu}, whereas we will be mainly concerned about examples at higher values of the central charge, specifically bosonic RCFTs \footnote{Recently, non-invertible duality symmetries were constructed for a particular subset of irrational field theories at $c=2$ \cite{Nagoya:2023zky}. We will not consider this type of theories.} at $c=2$. A prominent role is played by {\it duality symmetries}. This type of global symmetry arise in theories featuring self-duality upon gauging a particular subset of its global symmetry. In particular, the conformal manifold of $c\geq 1$ CFTs accommodates the action of non-trivial duality groups (see e.g. \cite{Giveon:1994fu} for a review), making the occurrence of duality symmetries quite natural\footnote{ Let us mention that duality symmetries are not exclusive of lower dimensional theories, see for instance \cite{Choi:2021kmx,Choi:2022zal,Antinucci:2022vyk,Bashmakov:2022uek,Damia:2023ses} for some constructions and applications in four dimensions}. 

Due to the fact that the action associated to duality symmetries usually involves an orbifold by some (invertible)\footnote{See the recent work \cite{Choi:2023vgk,Diatlyk:2023fwf, Perez-Lona:2023djo} for some alternative constructions involving gauging non-invertible symmetries.} global symmetry, the fusion realized by its associated topological operators does not follow a group law. The underlying structure is generically accounted for by Category Theory.  
A distinctive feature of conformal field theories in two dimensions is that the construction of topological defects is highly conditioned by the modular bootstrap\footnote{A priori, these conditions find a natural generalization in higher dimensions, achieved by putting the theory on appropriate spacetime manifolds. However, we are not aware of a precise treatment along these lines for higher dimensional theories.}, as a natural consequence of modular covariance \cite{Verlinde:1988sn,Petkova:2000ip,Fuchs:2002cm,Chang:2018iay,Thorngren:2019iar}. This program not only enables the determination of consistent topological defects, but also allows to obtain the additional topological data involved in the underlying categorical structure \cite{Thorngren:2021yso}. We will apply these tools extensively for the examples studied in this paper. Remarkably, solutions to the various constraints imposed by associativity of the OPE and modular covariance belong to a finite discrete set, a fact referred to as Ocneanu rigidity \cite{ostrik2002fusion}. This naturally endows categorical symmetries with a notion of robustness, rendering  well defined Renormalization Group (RG) invariants, such as ordinary symmetries.    
Concomitantly, duality symmetries may participate in anomalies \cite{Thorngren:2021yso,Apte:2022xtu,Zhang:2023wlu,Choi:2023xjw,Antinucci:2023ezl,Cordova:2023bja}, hence leading to interesting constraints on RG flows triggered by duality symmetric operators. In this work we explicitly show how anomalous duality symmetries constraint symmetry preserving RG flows.

A special loci pertaining to the conformal manifold of CFTs with central charge $c\geq 1$ corresponds to the set of multicritical points. These (generically rational) theories may connect various branches of the conformal manifold. A prototypical example is the Kosterlitz-Thouless (KT) theory sitting at the merging point between the circle and the orbifold branches of the $c=1$ bosonic conformal manifold. In particular, it has been found that the KT point preserves a series of non-invertible duality symmetries some of which are further preserved by the exactly marginal deformation spanning the orbifold branch \cite{Thorngren:2021yso}. Inspired by this kind of analysis, we present a characterization of the duality symmetry structure featured at certain multicritical points along the toroidal branch of the conformal manifold of $c=2$ bosonic CFTs. Contrary to what happens at $c=1$, the richer duality group  at $c=2$ allows for the occurrence of categorical symmetries graded by a non-commutative structure, as we will illustrate in the examples below (see \cite{Bashmakov:2022uek,Antinucci:2022cdi} for recent examples of non-commutative duality symmetries constructed in higher dimensions). 

Topological defects in CFT are a subset of the conformal defects. On general grounds, the characterization of conformal defects plays a significant role in many fields, ranging from the description of finite size effects in critical systems \cite{Cardy:1989ir,Ishibashi:1988kg,Cardy:2004hm} to the classification of branes in String Theory \cite{Dai:1989ua,Polchinski:1995mt}. Furthermore, recent studies established a connection between self-duality symmetries in bosonic CFT's and interesting topological transitions between fermionic theories by means of a generalized Jordan-Wigner map \cite{Hsieh:2020uwb,Ji:2019ugf}. By similar considerations, the structures uncovered in this work may shed light onto more intricate transitions taking place along the fermionic conformal manifold at higher values of the central charge. This type of systems have engendered some interest due to their potential application in designing quantum computing devices \cite{PhysRevB.84.144509}. We hope to explore these aspects in more detail in the near future.

This paper is organized as follows. In section \ref{sec:c=1} we briefly review some properties of the $c=1$ theory, emphasizing its symmetry structure and the modular properties of its (twisted) torus partition function that will be important for the discussion of the $c=2$ theory. In section \ref{sec: c=2} we review properties of the $c=2$ toroidal branch of the conformal manifold and we explicitly show that any RCFT point within this branch enjoys non-invertible duality symmetries arising form the self-duality of these theories under gaugings of subgroups of their symmetries. To achieve that, we introduce the generalized metric $\mathcal{E}$ as a useful parametrization of the conformal manifold. We rephrase the generic condition for the presence of a duality symmetry defect $\cD$ as the existence of an associated matrix $\sD \in O(2,2,\bQ)$, parametrizing a given topological manipulation, such that
\be
(\sD^{-1})^{\intercal} \mathcal{E} \sD^{-1} =\mathcal{E}\,.
\ee

In section \ref{sec: multicritical} we delve into the detailed analysis of some multicritical points of the toroidal branch which we dub quadri-critical and bi-critical point, where the theory possess extra exactly marginal parameters generating orbifold branches. Being RCFTs, these points have non-invertible duality symmetries which we analyze in detail, describing the underlying categorical structure and their consistency conditions coming from the modular bootstrap. We then describe duality preserving marginal and relevant deformations. In the quadri-critical point the theory factorize in the product of two $c=1$ RCFT. Therefore we can explicitly check some of the constraints coming from the above-mentioned duality symmetry, for instance verifying the ones coming from the non-invertible 't Hooft anomalies. On the other hand, in the bi-critical point the theory is genuinely $c=2$, in the sense that it is not a product of $c=1$ theories, and some considerations coming from the duality symmetry are actually non-trivial predictions.

In section \ref{sec: WZW} we briefly comment on the enhanced symmetry point where the theory is equivalent to $SU(3)_1$ WZW model. As opposed to the $c=1$ case, this point still enjoys some non-invertible duality symmetries which are not of the Verlinde type and that we briefly analyze. We end with some appendices containing some technical details.

\section{Review of duality symmetries for the $c=1$ bosonic CFT}\label{sec:c=1}

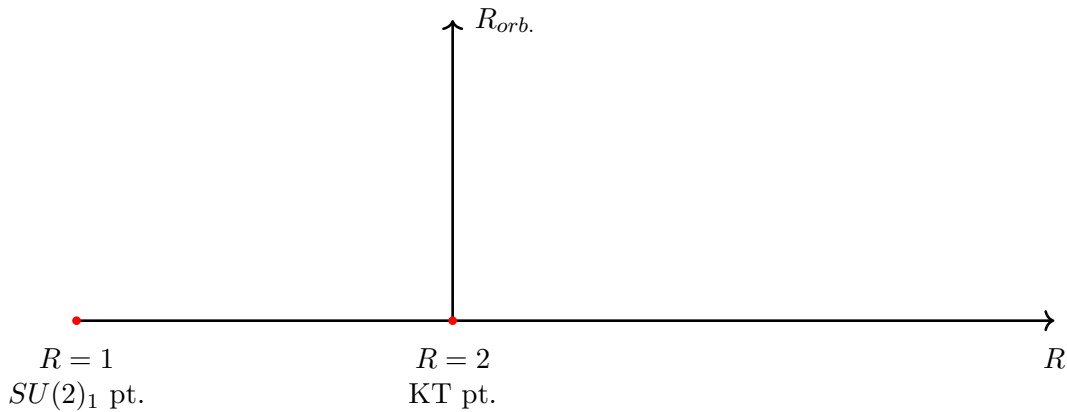
\begin{figure}
    \centering
  \begin{tikzpicture}
		\draw[line width=1,black,->] (-3,0)--(10,0);
		\draw[line width=1,black,->] (2,0)--(2,4);
		\filldraw [red] (2, 0) circle [radius = 0.05];
		\filldraw [red] (-3, 0) circle [radius = 0.05];
		\node[] at (10,-0.5){$R$};
		\node[] at (2.7,4){$R_{orb.}$};
		\node[] at (2,-0.5){$R=2$};
		\node[] at (2,-1.){KT pt.};
		\node[] at (-3,-0.5){$R=1$};
		\node[] at (-3,-1.){$SU(2)_1$ pt.};
   \end{tikzpicture}
    \caption{Pictorial representation of the $c=1$ conformal manifold.}
    \label{fig: c=1 cm}
\end{figure}

We begin by summarizing the main aspects concerning the dynamics of non-invertible duality symmetries in bosonic $c=1$ CFT's, for a comprehensive analysis see \cite{Thorngren:2021yso,Chang:2018iay}. These theories are realized in terms of a compact scalar $\phi\sim\phi+2\pi$ with action
\be
S=\frac{R^2}{4\pi}\int {\rm d}^2 x \, d\phi\wedge\star d\phi
\ee
where $R$ is the radius of the target space circle. We chose the conventions in which the self-dual radius is $R=1$ and the circle branch in the conformal manifold is parametrized by $R\geq 1$. The connected piece\footnote{In addition, there are three disconnected exceptional orbifold points pertaining to the $c=1$ conformal manifold. See \cite{Thorngren:2021yso} (section 5) for a description of the categorical structures featured by these theories.} of the conformal manifold comprises an additional branch, namely the orbifold branch, parametrized by an exactly marginal parameter $R_{orb}\geq 1$. These two branches merge at $R=2$ ($R_{orb}=1$), the Kosterlitz-Thouless (KT) point \cite{Ginsparg:1987eb}, which is then a multicritical point, see Figure \ref{fig: c=1 cm}.
Theories in this conformal manifold admit a $T$-dual description in terms of a dual variable $\widetilde\phi$ with radius $\widetilde R=R^{-1}$. In terms of the chiral left (right) moving fields $X$ ($\overline{X}$), the scalar field and its dual have the following decomposition
\be
\phi=\frac{1}{\sqrt{2}R}(X+\overline{X}) \quad , \quad \tilde\phi=\frac{R}{\sqrt{2}}(X-\overline{X})
\ee
hence $T$-duality acts on the right movers as\footnote{There are actually multiple consistent definitions of $T$-duality. In this article we chose the one that acts on the right movers and moreover, defines a (non-anomalous) ${\mathbb Z}_2$ action at the self-dual point. For a detailed discussion about alternative choices see \cite{Harvey:2017rko}.} $\overline{X}\to -\overline{X}$.

A generic point along the circle branch features a $U_n(1)\times U_w(1)$ symmetry , associated to shifts of $\phi$ and $\widetilde \phi$ respectively\footnote{Equivalently, there are two chiral symmetries $U(1)\times\overline{U(1)}$ acting respectively on $X$ and $\overline{X}$.}. The spectrum of (genuine) conformal primaries along the circle branch contains vertex operators of the form\footnote{Throughout this paper, normal ordering is implicit in the definition of vertex operators.} 
\be
V_{n,w}=e^{in\phi}e^{iw\widetilde\phi}=e^{ip X}e^{i\overline{p} \overline X} \quad , \quad p=\frac{1}{\sqrt{2}}\left(\frac{n}{R}+wR\right)
\quad , \quad \overline p=\frac{1}{\sqrt{2}}\left(\frac{n}{R}-wR\right)
\ee
with $n,w\in \mathbb{Z}$. The conformal dimension and spin reads
\be
h=\frac12 p^2 \quad , \quad \overline h=\frac12\overline p^2 \quad , \quad \Delta= h+\overline h=\frac12\left(\frac{n^2}{R^2}+w^2R^2\right) \quad , \quad s=h-\overline h=nw\,.
\ee
In this presentation, $n$ and $w$ span the charges of $V_{n,w}$ under the $U(1)_n$ and $U(1)_w$ respectively and by Dirac quatization condition we have $nw \in \mathbb{Z}$. 
Upon gauging a ${\mathbb Z}_N\subset U_n(1)$, the theory gets mapped to one at $R'=\frac{R}{N}$. This is so due to the fact that the effect of the gauging is twofold. On the one hand, it projects the spectrum to invariant states, namely the ones with $n=N n'$ ($n'\in {\mathbb Z}$). In addition, it also incorporates twisted sectors, namely Virasoro multiplets corresponding to vertex operators with fractional winding number $w=\frac{w'}{N}$ ($w'\in {\mathbb Z}$ mod $N$). Remarkably, the incorporation of the twisted sectors is required by modular invariance in two dimensions, see for instance \cite{DiFrancesco:1997nk}. One can retrieve the charge lattice to its original form by rescaling $n\to n'=\frac{n}{N}$ and $w\to w'=N w$, hence leading to the aforementioned rescaling of the radius. Analogously, gauging ${\mathbb Z}_N\times {\mathbb Z}_M\subset U_n(1)\times U_w(1)$ leads to $R'\to \frac{M}{N}R$. Notice that the latter combined gauging is only consistent for ${\rm gcd}(N,M)=1$ due to the mixed anomaly involving the two $U(1)$ symmetries\footnote{\label{ftnote} By the same token, operators of the form $V_{n,0}$ and $V_{0,w}$ are not mutually local except for $nw\in 2 {\mathbb Z}$. See Appendix \ref{App:somedetails}.}. In the following, we will use the symbol $\sigma_{N,M}$ to denote this kind of operations.

At the special values $R=\sqrt{N/M}$ with $N, M\in{\mathbb Z}$, the chiral algebra is enhanced to $U(1)_{2K}$ ($K=NM$) by additional holomorphic conserved currents $J_{\pm}=e^{\pm i\sqrt{2K}X}$ of spin $K$. In particular, for $N=M=1$, this corresponds to the self-dual point, where the additional currents have spin one and combine into an $SU(2)_1$ chiral algebra (see Appendix \ref{App:somedetails}). The theory at any of these special points becomes rational and the Hilbert space decompose into a finite number of representations of the chiral algebra $U(1)_{2K}$. At $R=\sqrt{N}$, the partition function is given by a diagonal modular invariant in terms of the characters of such representations. More generally, the partition function on the Euclidean torus with complex structure $\tau$ takes the form
\be\label{eq:c=1 RCFT part function}
Z=\sum_{a=0}^{2M-1}\sum_{b=0}^{2N-1}\chi_{Na+Mb}\overline{\chi}_{Na-Mb} 
\quad , 
\ee
with the chiral characters defined as
\be
\chi_k(\tau)\sim \chi_{-k}(\tau)\sim \chi_{k+2K}(\tau)=\frac{1}{\eta(\tau)}\sum_{r\in{\mathbb Z}} q^{\frac{1}{4K}(k+2K r)^2} \quad , \quad q=e^{2\pi i \tau} \, .
\ee

Each block in the sum \eqref{eq:c=1 RCFT part function} accounts for the contribution of primary states with $U(1)_n\times U(1)_w$ charges satisfying $Mn=Na+MN(r+r')$ and $Nw=Mb+MN(r-r')$, for $r,r'\in{\mathbb Z}$ \footnote{More generally, by equating the conformal weights of the operators involved in the chiral characters of $U(1)_{2MN}$ with their expressions in terms of the global charges one finds
$$
\chi_k \overline{\chi}_{k'} \,\, \Rightarrow \,\, Mn= \frac{k+k'}{2}+MN(r+r') \quad , \quad Nw= \frac{k-k'}{2}+MN(r-r')  
$$
where the integers $r,r'$ correspond to each term in the sums defining the characters.
}.

Interestingly, the circle branch theories at $R=\sqrt{N/M}$ feature non-trivial duality symmetries \cite{Thorngren:2021yso,Chang:2018iay, Benini:2022hzx}. These are  found by combining $T$-duality with gauging ${\mathbb Z}_N\times{\mathbb Z}_M\subset U_m(1)\times U_w(1)$, {\it i.e.} $T\circ \sigma_{N,M}$. In fact, it is straightforward to check that these two combined actions leave the radius invariant. Away from the self-dual point, the defects implementing these global symmetries are non-invertible and, together with the generator $\eta$($\tilde{\eta}$) of the invertible ${\mathbb Z}_N$(${\mathbb Z}_M$), describe a Tambara-Yamagami (TY) category with fusion 
\be
{\cal D}^2=\sum_{n=0}^{N-1}\sum_{m=0}^{M-1}\eta^n\tilde{\eta}^m \quad , \quad \eta^{N}=\tilde\eta^{M}=1 \quad , \quad {\cal D}\eta=\eta{\cal D}={\cal D}\tilde\eta=\tilde\eta{\cal D}={\cal D}
\ee
where ${\cal D}$ denotes the defect implementing the $T\circ \sigma_{N,M}$ duality symmetry. The TY category described by the above objects is also characterized by additional topological data, namely a bicharacter $\chi_{ab}$ ($a,b\in{\mathbb Z}_N\times {\mathbb Z}_M$) and Frobenius-Schur (FS) indicator $\epsilon\in\{-1,1\}$. We refer the reader to appendix \ref{App:TY} for a more detailed description of these categories. When acting on vertex operators, the action reads\footnote{More generally, there is a continuous set of consistent duality defects constructed by stacking ${\cal D}$ with invertible operators implementing chiral rotations and/or charge conjugation. See \cite{Thorngren:2021yso} for a detailed discussion on this point. }
\be\label{eq:c=1 duality action}
{\cal D}\, : \,\, V_{n,w} \, \to \,
\sqrt{K}(-1)^{nw} V_{\tfrac{Nw}{M},\tfrac{Mn}{N}} \quad  , \quad  n\in N{\mathbb Z} \, , \, w\in M{\mathbb Z}  
\ee
and mapping it to a non-genuine operator otherwise.
Note the occurrence of the $\sqrt{K}$ prefactor associated to the quantum dimension of ${\cal D}$. The additional phase $(-1)^{nm}$ arises due to the fact that momentum and winding modes, which are exchanged by $T$-duality, are not mutually local (see appendix \ref{App:somedetails}). 

A topological defect implementing a symmetry in quantum field theory must obey certain consistency conditions. In two dimensions, these are imposed by modular invariance or, more generally, modular covariance, finding their natural implementation in terms of the {\it modular bootstrap}. More precisely, torus partition functions twisted by topological defects along non-trivial cycles are mapped to each other by the modular group $SL(2,{\mathbb Z})$
\be\label{eq: modular convariance}
Z_{({\cal D},{\widetilde{ \cal D}})}(\gamma\cdot\tau) =e^{i\phi(\gamma)}Z_{({\cal \widetilde D}^{-b}{\cal D}^d,{\cal \widetilde{D}}^a{\cal D}^{-c})}(\tau) \quad , \quad \gamma\cdot\tau=\frac{a\tau+b}{c\tau+d}
\ee
where we allowed for the occurrence of an anomalous phase $\phi(\gamma)$. This phase is not going to play any role in the following, hence being omitted from now on. Whenever negative exponents occur in the right hand side of \eqref{eq: modular convariance}, those should be interpreted as the insertion of the orientation reversed defect, as the inverse might not exist.
An important implication of the above is depicted in the following diagram:
\begin{equation}
Z_{(1,\cD)}\;=\;\raisebox{-3.5 em}{ \begin{tikzpicture}
		\draw[line width=1] (0,0)--(0,3)--(3,3)--(3,0)--(0,0);
		\draw[line width=1] (0,1.5)--(3,1.5)node[right]{$\cD$};
   \end{tikzpicture}}
   \qquad \xrightarrow{\tau \rightarrow -\frac{1}{\tau}}\qquad
   \raisebox{-4.8 em}{  \begin{tikzpicture}
		\draw[line width=1] (0,0)--(0,3)--(3,3)--(3,0)--(0,0);
		\draw[line width=1] (1.5,0)node[below]{$\cD$}--(1.5,3);
   \end{tikzpicture}}\;\equiv\; Z_{(\cD,1)} 
    \label{eq: S transf}
\end{equation}
where, in the last equality, we emphasized the fact that a consistent topological defect should lead to a well defined twisted Hilbert space. More precisely, being topological, it commutes with the stress tensor, hence the defect Hilbert space must have a natural decomposition in terms of Virasoro characters
\be\label{eq: trace defect H space}
Z_{(\cD,1)}=\sum_{h,\overline h} n_{h,\overline h}\cV_h \overline{\cV}_{\overline h}
\ee
where, by the above considerations, the coefficients $n_{h,\overline h}$ must be positive integers. This becomes a key test under which any proposed symmetry defect should be contrasted, as we will show in many examples in this paper. Moreover, similar applications of \eqref{eq: modular convariance} enable extracting more refined topological data associated to a given symmetry structure \cite{Chang:2018iay,Thorngren:2019iar,Thorngren:2021yso,Lin:2023uvm,Lin:2019kpn,Lin:2019hks,Lin:2021udi}. 

As an instructive example, we apply the above analysis to the defect implementing the symmetry action \eqref{eq:c=1 duality action}. Similar computations can be found in \cite{Thorngren:2021yso}, though we are not aware of an explicit application to the non-diagonal case.  
By the action of $T$-duality, namely $\overline X\to-\overline X$, the insertion of the duality defect ${\cal D}$ along the spatial cycle projects into the $\overline p=0$ subspace.
This condition sets $Na-Mb=0$, hence the twisted partition function gets contributions only from the identity characters, spanned by the states $|p=\sqrt{2MN}r, \overline p=0 , {\bf N} ,\overline {\bf N}\rangle$, $r\in {\mathbb Z}$. Here ${\bf N}$ ($\overline{\bf N}$) denote, as usual, the number of (anti-)holomorphic oscillator modes. In addition, $T$-duality induces a $(-1)$ phase for states with an odd number of right-moving oscillator modes \cite{Harvey:2017rko}.  
Finally, there is the phase factor in \eqref{eq:c=1 duality action} due to the non-locality between winding and momentum modes. Putting all together one obtains
\be
Z_{(1,{\cal D})}(\tau)= \frac{\sqrt{NM}}{|\eta(\tau)|^2}\left(\sum_{r\in{\mathbb Z}}(-1)^{MN r}q^{MN r^2}\right) \vartheta_4(2\overline\tau) \, ,
\ee
where we already performed the sum over the oscillator modes. See appendix \ref{App:modular functions} for a list of the relevant modular functions and their modular properties. 

For the case of even $MN$ we get 
\be
Z_{(1,{\cal D})}= \frac{\sqrt{NM}}{|\eta|^2}
\vartheta_3(2NM\tau)\vartheta_4(2\overline\tau) \, \xrightarrow{\tau\to-\frac{1}{\tau}} \, Z_{({\cal D},1)}(\tau)=\frac{1}{2|\eta|^2}\vartheta_3\left(\tfrac{\tau}{2NM}\right)\vartheta_2\left(\tfrac{\overline\tau}{2}\right) \, ,
\ee
whereas for odd $MN$ one gets instead
\be
Z_{(1,{\cal D})}= \frac{\sqrt{NM}}{|\eta|^2}\vartheta_4(2NM\tau)\vartheta_4(2\overline\tau)
\, \xrightarrow{\tau\to-\frac{1}{\tau}} \, Z_{({\cal D},1)}(\tau)=\frac{1}{2|\eta|^2}\vartheta_2\left(\tfrac{\tau}{2NM}\right)\vartheta_2\left(\tfrac{\overline\tau}{2}\right)\,.
\ee
It can be easily checked that both expressions comply with the condition shown in \eqref{eq: S transf}, hence leading to a well defined trace over a defect Hilbert space.

\section{$c=2$ Toroidal CFT and Duality symmetries}\label{sec: c=2}
The conformal manifold of $c=2$ bosonic CFT's is significantly richer than the one for $c=1$ described in the previous section. In particular, the connected non-exceptional component comprises 28 non-equivalent branches, meeting at various multicritical regions. We refer the reader to \cite{Dulat:2000xj} for a comprehensive overview. A distinguished branch is the {\it toroidal} branch, where the target space corresponds to a torus with complex structure $\tau$ and Kahler modulus  $\rho$, hence spanning a four-dimensional variety. This branch also accommodates a collection of special points with enhanced global symmetry, as we will comment momentarily.

Before delving into particular examples, let us review some aspects pertaining to generic theories along the $c=2$ toroidal branch (for more details we refer to \cite{Giveon:1994fu, Dulat:2000xj}). The field content of these theories consists on two compact scalar fields $\phi^i\sim \phi^i+2\pi$ ($i=1,2$) with action
\begin{equation}
\label{lag}
S= \frac{1}{4\pi}\int \delta^{\mu \nu} G_{ij} \partial_{\mu} \phi^i  \partial_{\nu} \phi^j +\frac{i}{4\pi} \int \epsilon^{\mu \nu} B_{ij} \partial_{\mu} \phi^i  \partial_{\nu} \phi^j %\qquad i=1,2 
\,,
\end{equation}
where $G,B$ are respectively the metric and Kalb-Ramond fields parametrizing the target space two-torus
\begin{equation}
\label{param}
G = R^2
    \begin{pmatrix}
        1 & \tau_1 
        \\
        \tau_1  & \tau_1^2 + \tau_2^2
    \end{pmatrix}
    \qquad
    B = 
    \begin{pmatrix}
        0 & b 
        \\
        -b  &0 
    \end{pmatrix}\,.
\end{equation}
It is customary to repackage the geometric data in terms of the complex structure and Kahler modulus  
\begin{equation}
\tau = \tau_1 + i \tau_2 \quad, \quad
\rho=\rho_1+i\rho_2=b+i\sqrt{G}\quad , \quad \left(\sqrt{G}=R^2\tau_2\right)\,.
\end{equation}
These two complex variables parametrize the $4$-dimensional toroidal branch of the $c=2$ conformal manifold. Motion along this branch is generated by the four exactly marginal operators of the form $\partial\phi^i\partial\phi^j$. Note that when $\tau_1=\rho_1=0$ the theory factorizes as a tensor product of two $c=1$ compact bosons at radii $R_1 = R$ and $R_2 = R\tau_2$.  

These theories admit equivalent descriptions in terms of dual variables $\widetilde\phi^i\sim \widetilde\phi^i+2\pi$. In addition, the real fields $\phi^i$, $\widetilde\phi^i$ admit the following decomposition into left and right movers
\begin{equation}
    \begin{split}
        &\phi^1=\frac{1}{\sqrt{2\tau_2\rho_2}}\big[\tau_2(X^1+\overline X^1)-\tau_1(X^2+\overline X^2)\big] 
\\
&\phi^2=\frac{1}{\sqrt{2\tau_2\rho_2}}(X^2+\overline X^2) 
\\
&\widetilde\phi^1=\frac{1}{\sqrt{2\tau_2\rho_2}}\big[\rho_2(X^1-\overline X^1)-\rho_1(X^2+\overline X^2)\big]
\\
&\widetilde\phi^2=\frac{1}{\sqrt{2\tau_2\rho_2}}\big[(\rho_1\tau_2+\rho_2\tau_1)X^1+(\rho_1\tau_2-\rho_2\tau_1)\overline X^1+(\rho_2\tau_2-\rho_1\tau_1)X^2-(\rho_1\tau_1+\rho_2\tau_2)\overline X^2\big] \,.
    \end{split}
\end{equation}
At a generic point along this branch, the theory possess a continuous global symmetry of the form
\be
U(1)_{\bf n}\times U(1)_{\bf w}\equiv U(1)_{n_1}\times U(1)_{n_2}\times U(1)_{w_1}\times U(1)_{w_2}
\ee
with $U_{n_i}(1): \,\phi^i\to\phi^i+\alpha^i$ and $U_{w_i}(1): \,\widetilde\phi^i\to\widetilde\phi^i+\widetilde\alpha^i$, ($\alpha^i,\widetilde\alpha^i\in [0,2\pi)$). 
Note that the exactly marginal operators spanning the toroidal branch are naturally neutral under these symmetries. 
The spectrum of charged conformal primaries is therefore determined by four integer charges ${\bf n}=(n_1,n_2)^T$ and ${\bf w}=(w_1,w_2)^T$
\be
V_{{\bf n},{\bf w}}=e^{i{\bf n}^T{\boldsymbol \phi}}e^{i{\bf w}^T\widetilde{\boldsymbol \phi}}\,,
\ee
where $\boldsymbol{\phi}= \left( \phi^1 , \, \phi^2 \right)$, $\widetilde{\boldsymbol{\phi}}=( \widetilde{\phi}^1, \, \widetilde{\phi}^2)$.
Equivalently, these primaries are in one to one correspondence with sites in the even, self-dual, integer charge lattice with signature $(2,2)$ spanned by the following left- and right-moving momenta
\begin{align}\label{eq: momenta in terms of charges}
{\bf p}&=\frac{1}{\sqrt{2\tau_2\rho_2}}
\left(\begin{array}{c}
n_1\tau_2+\rho_1\tau_2 w_2+ \rho_2(w_1+w_2\tau_1)\\
n_2-\tau_1 n_1-\rho_1(w_1+w_2\tau_1) + \rho_2\tau_2 w_2
\end{array}\right)\\
\overline{\bf p}&=\frac{1}{\sqrt{2\tau_2\rho_2}}\left(\begin{array}{c}
n_1\tau_2+\rho_1\tau_2 w_2- \rho_2(w_1+w_2\tau_1)\\
n_2-\tau_1 n_1-\rho_1(w_1+w_2\tau_1)-\rho_2\tau_2 w_2
\end{array}\right) \,.
\nonumber
\end{align}
In terms of these momenta, the conformal dimensions read
\begin{equation}
    \begin{split}
        &h=\frac12(p_1^2+p_2^2) = \frac{1}{4}\left[n + (G+B) w\right] ^{\intercal} G^{-1}\left[n + (G+B)w\right]\\
        &\overline h=\frac12(\overline p_1^2+\overline p_2^2) = \frac{1}{4}\left[n - (G-B) w\right] ^{\intercal} G^{-1}\left[n - (G-B)w\right] \,
    \end{split}
\end{equation}
and the spins $s=h-\overline{h}=n_iw_i\in \bZ$ are integrally quantized since the theory is bosonic.

There are several criteria for defining rationality of a generic CFT. A natural one states that a theory is rational if its Hilbert space decompose into a {\it finite} number of non-equivalent irreducible representations of a certain (generally enhanced) chiral algebra \cite{Nahm:1996zn,Gannon:1996hp}\footnote{More precisely, in such a case we say that the theory is rational with respect to such enhanced chiral algebra, even if it might not be rational with respect to Virasoro. These are the prototypical examples that occur at $c\geq1$.}.

Analogously to the $c=1$ theories at $R^2\in\bQ$, there are distinguished points in the $c>1$ conformal manifold where the theory becomes rational. At $c=2$, the criterion for rationality of a toroidal CFT described by \eqref{lag} can be rephrased in terms of geometric properties of its charge lattice. Alternatively, it can be proven that a $c=2$ toroidal CFT is rational if $G\in GL(2,\bQ)$ and $B\in Skew(2,\bQ)$. The several criteria just mentioned can be proven to be equivalent. Geometrically, it has been shown in  \cite{Gukov:2002nw} that rationality is also achieved if the target space torus admits Complex Multiplication\footnote{Given a torus $\cT^2=\bC/\Lambda$ determined by some lattice $\Lambda$, $\cT^2$ admits Complex Multiplication if there exists a complex number $z\in \bC-\bR$ such that $z\Lambda\subset\Lambda$.}. The latter condition implies that the moduli $\tau, \rho$ satisfy
\be 
\tau\, , \, \rho \, \in {\mathbb Q}\left(\sqrt{D}\right)\;,
\ee
where ${\mathbb Q}\left(\sqrt{D}\right)$ denotes the imaginary quadratic number field, for some {\it negative} integer $D$. The latter is obtained from the field of the rational numbers by introducing $\sqrt{D}$, {\it i.e.} $x\in {\mathbb Q}\left(\sqrt{D}\right)$ then $x=x_1+x_2\sqrt{D}$ with $x_1,x_2\in \bQ$. The field structure of $\bQ\left(\sqrt{D}\right)$ then naturally descends from the one of $\bQ$. In particular, if $\tau$ belongs to ${\mathbb Q}\left(\sqrt{D}\right)$, then there exist some integers $a$, $b$ and $c$ such that 
\be\label{eq:quadratic tau}
a \tau^2+b\tau+c=0 \quad , \quad a,b,c\in {\mathbb Z} \quad , \quad {\rm gcd}(a,b,c)=1
\ee 
hence 
\be
\tau=-\frac{b}{2a}+\frac{i}{2a}\sqrt{-D} \quad , \quad D=b^2-4ac
\ee
and similarly for $\rho$. Even if the equations for both complex variables may have different integer coefficients, they must have the same discriminant $D$ for the theory to be an RCFT.
The condition $D<0$ is implied in order for the solutions to correspond to a physical theory, {\it i.e.} $\tau_2,\rho_2>0$. \footnote{As an illustrative example, consider the case of a product of two $c=1$ RCFTs at $R_1=\sqrt{N_1/M_1}$ and $R_2=\sqrt{N_2/M_2}$, for which $\tau=iR_1/R_2$, $\rho=iR_1 R_2$. This situation trivially fits into the class described above since 
$$
N_2 M_1\tau^2+N_1M_2=0 \quad , \quad M_1M_2\rho^2+N_1N_2=0 \quad , \quad D=-N_1N_2M_1M_2\; .
$$
}  It can be proven that the set of RCFTs is dense within the toroidal branch, similarly to the rational theories along the $c=1$ circle branch.

An RCFT in the toroidal branch features an extended chiral algebra of the form
\be\label{eq:K chiral alg mt}
u(1)^2_{K_L}\times u(1)^2_{K_R}
\ee
where $K_{L,R}$ are some matrices constructed out of $G$ and $B$ at each particular rational point. We refer to appendix \ref{app:PartFunctionc2} for some details on the determination of these matrices.   
The partition function of a rational theory is accounted for by a finite modular invariant combination of characters of \eqref{eq:K chiral alg mt}. Furthermore, the (anti)holomorphic representations of the chiral algebra are in one to one correspondence with the sites in the following lattices
\be
{\cL}_{L,R}\equiv \left\lbrace {\boldsymbol{\lambda}_{L,R}}\in {\mathbb Z}^2 \, , \, {\boldsymbol{\lambda}_{L,R}}\sim {\boldsymbol{\lambda}_{L,R}}+K_{L,R}{\bf v} \, , \, {\bf v} \in {\mathbb Z}^2
 \right\rbrace \,.
\ee
The partition function then reads
\be
Z=\sum_{\vec{\lambda}_{L}\in{\cal D}_{L}}\chi_{{\boldsymbol\lambda}_L}\overline{\chi}_{\widehat\omega{\boldsymbol\lambda}_L}
\ee
where $\hat\omega:\cL_L\to\cL_R$ is a particular group isomorphism. We refer to appendix \ref{app:PartFunctionc2} for a detailed account on the construction of the above partition function.
The characters corresponding to the representations labelled ${\boldsymbol{\lambda}_{L,R}}$ read\footnote{We denote $|v|^2_M=v^iM_{ij}v^j$.}
\be
\chi_{{\boldsymbol\lambda}_L}(\tau)=\frac{1}{\eta(\tau)^2}\sum_{{\bf l}\in{\mathbb Z}^2} q^{\frac12|{\boldsymbol\lambda}_L+K_L{\bf l}|^2_{K_L^{-1}}}
\quad , \quad 
\bar{\chi}_{{\boldsymbol\lambda}_R}(\bar\tau)=\frac{1}{\eta(\bar\tau)^2}\sum_{{\bf r}\in{\mathbb Z}^2} \bar{q}^{\frac12|{\boldsymbol\lambda}_R+K_R{\bf r}|^2_{K_R^{-1}}}\,.
\ee
Furthermore, the partition function of the theory is given by a diagonal modular invariant ({\it i.e.} $\widehat\omega=1$) whenever there exists a duality frame such that $\tau=f a \rho$ with $a$ the same as in \eqref{eq:quadratic tau} and $f$ some positive integer \cite{Gukov:2002nw}. 

Within the toroidal branch, there are special (generically rational) points where the global symmetry gets enhanced to higher rank chiral algebra, hence corresponding to enhanced symmetry points. More precisely, the algebras realized at these special loci at $c=2$ are of the A-type of maximal rank 2, \ie{} $su(2)$, $su(2)^2$ or $su(3)$. Theories featuring these algebras have been classified (see for instance \cite{Dulat:2000xj})
\begin{align}\label{eq:EnhancedPtsc2}
    (\tau ,\rho) &= (i,i) \; : \qquad SU(2)^2 \times \overline{SU(2)}^2 \; , \nonumber \\
    (\tau ,\rho) &= (e^{i 2 \pi/3},e^{i 2 \pi/3}) \; : \qquad SU(3) \times \overline{SU(3)} \; , \nonumber \\
    (\tau , \rho) &= (\tau , \tau) \neq \{ i, e^{i 2 \pi/3}\} \; , \, \tau_1 = \{ 0, \frac{1}{2} \} \; : \qquad SU(2) \times U(1) \times \overline{SU(2)} \times \overline{U(1)} \; , \nonumber \\
    (\tau , \rho) &= (\tau , \tau) \; , \tau_1 \neq \{ 0, \frac{1}{2} \} \; : \qquad U(1) \times U(1) \times \overline{SU(2)} \times \overline{U(1)} \; , \nonumber \\
    (\tau , \rho) &= (\tau , - \overline{\tau}) \; , \tau_1 \neq \{ 0, \frac{1}{2} \} \; : \qquad SU(2) \times U(1) \times \overline{U(1)} \times \overline{U(1)} \; . 
\end{align}
Note that the last two lines above correspond to continuous sets of theories, hence comprising both rational and irrational theories. 

\begin{figure}
    \centering
    \begin{tikzpicture}
\draw[draw] (0,0) to[out=-50,in=150] (6,-2);
\draw[draw] (12,1) to[out=150,in=-50] (5.5,3.2);
\draw[draw] (6,-2) -- (12,1);
\draw[draw] (5.5,3.2) -- (0,0);

\filldraw[white] (5,0)--(4,4)--(6,4)--(5,0);
\draw (5,0)--(4,4);
\draw (5,0)--(6,4);
\draw[] (5,4) ellipse [x radius = 1,y radius = 0.2];

\filldraw[white] (9,1)--(8,4)--(10,4)--(9,1);
\draw (9,1)--(8,4);
\draw (9,1)--(10,4);
\draw[] (9,4) ellipse [x radius = 1,y radius = 0.2];
\begin {scope}[shift={(2.75,-5.55)},rotate=40]
	\filldraw[white] (9,1)--(8.5,4)--(9.5,4)--(9,1);
	\draw (9,1)--(8.5,4);
	\draw (9,1)--(9.5,4);
	\draw[] (9,4) ellipse [x radius = 0.5,y radius = 0.2];
\end{scope}
\begin {scope}[shift={(1.45,6.02)},rotate=-40]
	\filldraw[white] (9,1)--(8.5,4)--(9.5,4)--(9,1);
	\draw (9,1)--(8.5,4);
	\draw (9,1)--(9.5,4);
	\draw[] (9,4) ellipse [x radius = 0.5,y radius = 0.2];
\end{scope}

\filldraw [black] (5,0) circle [radius = 0.05] node[below, black] {$(\omega,\alpha)$} ;	
\filldraw [black] (2,0) circle [radius = 0.05] node[below, black] {$(\omega,\omega)$} ;
\draw[line width=0.8, color=red,->] (2,0) arc (120: 60: 2.8);
\node[] at (3.5,0.7){$\mathbb{Z}_3^s$};
\node[] at (2,0.5){$SU(3)_1$};

\filldraw [black] (6.7,0.3) circle [radius = 0.05] node[below, black] {$(i,i)$} ;
\draw[line width=0.8, color=blue,->] (6.7,0.3)--(8.8,1);
\node[] at (7.9,0.3){$\mathbb{Z}_2^s$};
\node[] at (6.5,0.7){$SU(2)^2_1$};
\filldraw [black] (9,1) circle [radius = 0.05] node[below, black] {$(i,2i)$} ;	

\node[] at (6,-2.3){$(\tau,\rho)$};

\end{tikzpicture}
    \caption{Pictorial representation of a slice of the toroidal branch containing two multicritical points. Here $\omega = e^{\frac{2\pi i}{3}}$ while $\alpha = -\frac{1}{2} + 3 \frac{\sqrt{3}}{2}$ as in the main text. The red and blue lines represent respectively the gauging of the $\bZ_3^s$ and $\bZ_2^s$ shift symmetries defined in \eqref{eq:Z3 autom} and \eqref{eq:Z2 autom}. At the $4$-critical point $(i,2i)$ three orbifold branches (one 4d and two 2d) join while at the bi-critical point $(\omega,\alpha)$ we get just one 2d orbifold branch. }
    \label{fig: c=2 manifold}
\end{figure}
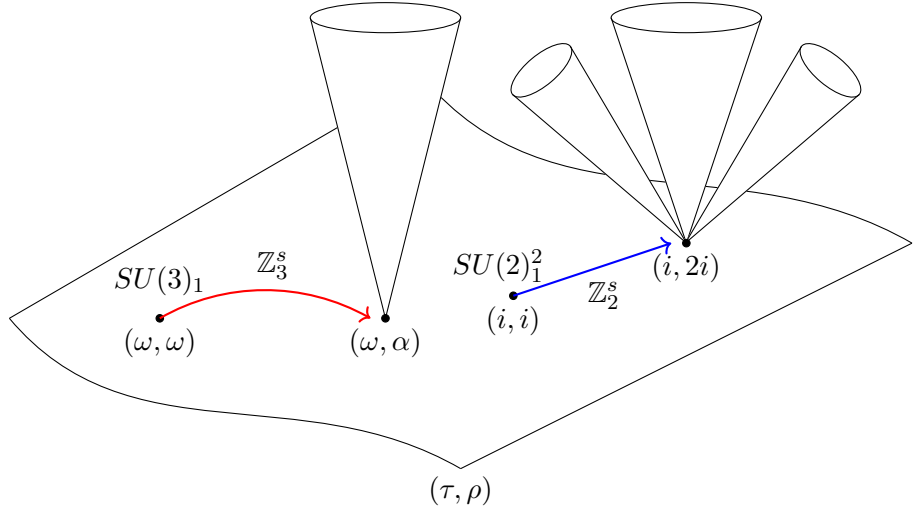

In spite of being interesting by themselves, the enhanced symmetry points play a crucial role in the characterization of the conformal manifold. In particular, it has been shown that all multicritical theories, i.e. where different branches of the conformal manifold intersect, can be obtained from non-trivial quotients of theories with enhanced symmetry \cite{Rostand:1990bv,Rostand:1991fn}. Along these lines, the additional exactly marginal deformations featured by multicritical theories are easily constructed in terms of conserved currents at enhanced symmetry points (see appendix \ref{App:somedetails} for an example of this construction at the $c=1$ KT point). In this work, we will mainly focus on two particular examples of multicritical theories, namely the quadri-critical and the bi-critical points, which are obtained by taking certain $\bZ_2$ and $\bZ_3$ quotients of the $SU(2)^2$ and the $SU(3)$ theories respectively (see figure \ref{fig: c=2 manifold} for a pictorial representation of the points we are interested in). 

The duality groups along the $c>1$ toroidal branch are significantly larger than the $T$-duality at $c=1$, hence making room for richer structures of duality symmetries featured at special points. In particular, for $c=2$ the duality group is of the form
\begin{equation}
\label{eq:duality c=2}
{\rm O}(2,2,{\mathbb Z}) \simeq {\rm P} \left( {\rm SL}(2,{\mathbb Z})_\tau\times  {\rm SL} (2,{\mathbb Z})_\rho\right)\rtimes \left({\mathbb Z}_2^M\times {\mathbb Z}_2^I\right) \, ,
\end{equation}
with the first two factors $SL(2,{\mathbb Z})_{\tau , \rho}$ acting separately on $\tau$ and  $\rho$ via
\begin{align}
\tau\to \frac{a\tau+b}{c\tau+d}   \quad &, \quad \begin{pmatrix}
    a & b \\ c & d
\end{pmatrix}\in SL(2,{\mathbb Z})_{\tau} \nonumber\\
\rho \to \frac{a'\rho +b'}{c'\rho +d'}  \quad &, \quad \begin{pmatrix}
    a' & b' \\ c' & d'
\end{pmatrix}\in SL(2,{\mathbb Z})_{\rho} 
\,.
\end{align}
These two actions are quotiented by the central element $(\tau,\rho)\sim (-\tau,-\rho)$. 
The remaining two ${\mathbb Z}_2$ actions correspond to mirror symmetry and target space inversion respectively
\begin{equation}
{\mathbb Z}_2^M \, : \, (\tau,\rho) \, \to \, (\rho,\tau) \quad , \quad {\mathbb Z}_2^I \, : \, (\tau,\rho) \, \to \, (-\overline\tau,-\overline\rho)\,.
\end{equation}
The various orbifold branches in the $c=2$ conformal manifold generically preserve a subgroup of the duality group \eqref{eq:duality c=2} of the toroidal branch (see for instance \cite{Dulat:2000xj}). 

The four dimensional representation of the duality group \eqref{eq:duality c=2} has a natural action by conjugation over the {\it generalized metric} $\cE$ (see e.g. \cite{Hohm:2010pp,Berman:2020tqn}). The latter corresponds to an alternative description of these theories in terms of a $4\times 4$ matrix acting on the vector ${\boldsymbol\Phi}\equiv ({\boldsymbol{\phi}},\widetilde{\boldsymbol{\phi}})^T$. More precisely, in terms of the metric $G$ and antisymmetric tensor $B$, the generalized metric reads
\be
{\cal E}=\left(\begin{array}{cc}
G-B\cdot G^{-1}\cdot B & -B\cdot G^{-1} \\
G^{-1}\cdot B & G^{-1}\end{array}
\right)  \,.
\ee
For a given element $\sO\in O(2,2,\bZ)$ acting on ${\boldsymbol\Phi}$ as ${\boldsymbol\Phi}' = \sO {\boldsymbol\Phi}$, the parameters of the theory get transformed according to
\be\label{eq: O action}
\sO \, : \, \cE \, \to \,\cE'=(\sO^{-1})^{\intercal}\cE \sO^{-1}\;.
\ee 
In this formulation, the generators of the duality group \eqref{eq:duality c=2} admit the following four dimensional representation 
\begin{align}
\begin{pmatrix}
a&-b&0&0\\
-c&d&0&0\\
0&0&d&c\\
0&0&b&a
\end{pmatrix} \quad  &{\rm for} \quad \begin{pmatrix}
    a & b \\ c & d
\end{pmatrix} \in SL(2,\bZ)_\tau \nonumber\\
\begin{pmatrix}
d'&0&0&c'\\
0&d'&-c'&0\\
0&-b'&a'&0\\
b'&0&0&a'
\end{pmatrix}\quad &{\rm for} \quad \begin{pmatrix}
    a' & b' \\ c' & d'
\end{pmatrix} \in SL(2,\bZ)_\rho\\[5pt]
M=\begin{pmatrix}
0&0&1&0\\
0&1&0&0\\
1&0&0&0\\
0&0&0&1
\end{pmatrix}\in \bZ_2^M
\quad &, \quad 
I=\begin{pmatrix}
-1&0&0&0\\
0&1&0&0\\
0&0&-1&0\\
0&0&0&1
\end{pmatrix}\in \bZ_2^I \nonumber
\end{align}
As we will discuss further below in this section, this description of the theory enables to construct a matrix representation of topological manipulations, and more generally duality symmetries. Precisely, to a given defect $\cD$ implementing such a manipulation, one associates a $4\times 4$ matrix $\sD\in O(2,2,\bQ)$ acting on the fields as described around \eqref{eq: O action}. Notice that the coefficients are in $\bQ$ instead of $\bZ$. This is a consequence of the fact that, as we will verify in several examples below, duality symmetries usually involve topological manipulations, such as discrete gauging, which cannot be represented by a matrix in $O(2,2,\bZ)$.

\subsection{Discrete gauging and duality symmetries}\label{sec:discretegauging}
Similarly to the $c=1$ case, gauging a non-anomalous subgroup of $U(1)_{\bf n}^2 \times U(1)_{\bf w}^2$ connects two points in the toroidal branch and therefore can be represented as a $4\times 4$ matrix $\sigma$ acting on the 4-component vector $(\boldsymbol{\phi}^{\intercal}, \widetilde{\boldsymbol \phi}^{\intercal})$, or equivalently on the charge $(\bf n^{\intercal},\bf w^{\intercal})$, as
\be
\mat{\boldsymbol{\phi}'\\ \widetilde{\boldsymbol \phi}'} = \sigma \mat{\boldsymbol{\phi}\\\widetilde{\boldsymbol \phi}}\quad  \Longleftrightarrow \quad \mat{\boldsymbol{n}'\\{\boldsymbol w}'} = (\sigma^{-1})^{\intercal} \mat{\boldsymbol{n}\\{\boldsymbol w}}\,.
\ee
Dirac quantization condition after the gauging, implies that 
\be
{\bf n}'^{\intercal}{\bf w}' \in \bZ \quad \Longrightarrow \quad \sigma \in O(2,2,\bQ)\,.
\ee
Therefore any gauging of subgroups of  $U(1)_{\bf n}^2 \times U(1)_{\bf w}^2$ can be represented as a matrix $\sigma \in O(2,2,\bQ)$. This is a particular example of the matrix representation for topological manipulations introduced before. We emphasize that the map between gaugings and matrices in $O(2,2,\bQ)$ is not bijective: there can be different matrices representing the same gauging operation\footnote{Equivalently in the c=1, the group $O(1,1,\bQ) = \Big\{\mat{p/q & 0 \\ 0& q/p}\; , \; \mat{0 & p/q \\ q/p & 0}$\Big\} represents all the possible gauging of subgroups $\bZ_p \times \bZ_q \subset U(1)_n \times U(1)_w $ but the diagonal and off-diagonal matrices span the same set of operations.}. A duality symmetry defect $\cD$ then implies the existence of an $O(2,2,\bQ)$ matrix $\sD$ which leaves the theory parameters invariant. This condition can be written, using the generalized metric $\cal E$, as\footnote{This condition can be easily generalizable to CFTs with integer central charge $c > 2$.}
\be\label{eq: dualsymm on E}
(\sD^{-1})^{\intercal} \mathcal{E} \sD^{-1} - \mathcal{E} = 0\,.
\ee
At this point it is important to emphasize the distinction between the actual symmetry operation over the Hilbert space (or equivalently on the conformal primaries) implemented by $\cD$, which does not need to be invertible, and its matrix representation $\sD$ acting on the global charges, for which an inverse matrix can always be defined. 
The invariance of the theory parameters in \eqref{eq: dualsymm on E} is the manifestation of the self-duality of the theory under the gauging of a global symmetry subgroup $G$ parameterized by $\sD \in O(2,2,\bQ)$. Thoughout this paper, will show in several examples how the actual subgroup $G$ can be extracted from the matrix $\sD$ satisfying \eqref{eq: dualsymm on E}. On the contrary, when $\sD\in O(2,2,\bZ)$, this corresponds to an invertible symmetry which permutes the primary operators.  The action of the corresponding symmetry defect on the set of conformal primary operators is
\be\label{eq: action on ops}
\cD \,:\,\begin{cases}
 & V_{\bf n,\bf w} \rightarrow \sqrt{|G|} (-1)^{\alpha} V_{\bf n',\bf w'} \qquad \text{ if } n_i',w_i' \in \bZ\\
 &V_{\bf n,\bf w} \rightarrow \text{ non genuine op.} \qquad \text{ if } n_i',w_i' \not\in \bZ
\end{cases}
\ee
where $\bf n',\bf w'$ are the charges obtained upon acting with the associated matrix as
\be\label{eq: general charge map}
\begin{pmatrix}
    {\bf n}'\\{\bf w}'
\end{pmatrix}=(\sD^{-1})^{\intercal}\cdot \begin{pmatrix}
    {\bf n}\\{\bf w}
\end{pmatrix} \; ,
\ee
and the phase $\alpha$ comes by imposing the consistency of this action with locality (see appendix \ref{App:somedetails} for more details).

As mentioned before, matrices $\sD$ naturally decompose in terms of duality operations in $O(2,2,\bZ)$ and topological manipulations associated to gauging of discrete subgroups of the global symmetry, represented by matrices $\sigma \in O(2,2,\bQ)$. Let us momentarily concentrate on the latter. Given a matrix $\sigma \in O(2,2,\bQ)$ implementing a gauging of a subgroup $G \subset U(1)_{\bf n}^2 \times U(1)_{\bf w}^2$, it becomes instrumental to extract the form of such a subgroup of the global symmetry. In the following, we present an algorithmic way of achieving that. Consider the set of primary operators
\be \label{eq: inv set sigma}
S^\sigma\equiv \left\lbrace V_{{\bf n},{\bf w}}\, | \, \{({\bf n},{\bf w})\}\xrightarrow{(\sigma^{-1})^{\intercal}}  \{{\bf e}\} \right\rbrace
\ee
where ${\bf e}_a$ is the set of unital vector charges, namely $({\bf e}_a)_i = \delta_{i,a}$ ($i,a\in \{1,\ldots 4\}$). Such operators are the minimal gauge-invariant operators, both in the untwisted and twisted Hilbert space of the ungauged theory, which become genuine operators after the gauging\footnote{For instance, in the $c=1$ case a gauging of $\bZ_k \subset U(1)_n$ acts on the charges as $(n,w) \rightarrow (\frac{n}{k},kw)$. Therefore primaries with charge $(k,0) ,(0,\frac{1}{k})$ are mapped to $(1,0),(0,1)$ and they correspond to the invariant operators under the action of $\bZ_k$.}. A trial version of the subgroup $G$ is parametrized by five integers $\{k,a,b,c,d\}$ such that its action on a vertex operator reads
\be
G: \, V_{{\bf n},{\bf w}} \, \to \, V_{{\bf n},{\bf w}}' = e^{\frac{2\pi i}{k}(an_1+bn_2+cw_1+dw_2)} V_{{\bf n},{\bf w}} \,.
\ee
Being invarinat under the action of $G$, the operators in $S^\sigma$ satisfy the condition
\be\label{eq: invariant cond}
V_{{\bf n},{\bf w}}\in S^\sigma \, \Rightarrow \, V_{{\bf n},{\bf w}}=V_{{\bf n},{\bf w}}' \;.
\ee
Hence, given a matrix $\sigma\in O(2,2,\bQ)$, determining the set $S^\sigma$ allows to easily derive the action of the symmetry $G$, \ie{} determining $k,a,b,c,d$.\footnote{\label{ft: gcd solution} Generically, the solution is not unique. More precisely, the condition \eqref{eq: invariant cond} is equally satisfied by any subgroup $H\subset G$. However, there is always a solution for which ${\rm gcd}(k,a,b,c,d)=1$, which we identify with the maximal group.  } In order to illustrate this procedure, we apply it to the following example 
\be\label{eq: sigma Z2Z2}
\sigma = \left(\begin{array}{cccc}
1 & 1 & 0 & 0 \\
1 & -1 & 0 & 0 \\
0 & 0 &\tfrac12 & \tfrac12 \\
0 & 0 & \tfrac12 & -\tfrac12
\end{array}\right)
\quad , \quad 
(\sigma^{-1})^T = \left(\begin{array}{cccc}
\tfrac12 & \tfrac12 & 0 & 0 \\
\tfrac12 & -\tfrac12 & 0 & 0 \\
0 & 0 & 1 & 1 \\
0 & 0 & 1 & -1
\end{array}\right)
\ee
for which \eqref{eq: inv set sigma} corresponds to
\be
S^\sigma=\left\lbrace V_{1,1,0,0},V_{1,-1,0,0},V_{0,0,\tfrac12,\tfrac12},V_{0,0,\tfrac12,-\tfrac12}\right\rbrace\,.
\ee
Now, imposing the invariance condition given in \eqref{eq: invariant cond}, it is straightforward to verify that integer solutions for $\{a,b,c,d\}$ only exist for $k\in 2\bZ$. Finally, the maximal non-trivial solution (see footnote \ref{ft: gcd solution}) corresponds to
\be
k=2 \, , \, a=b= 1 \, , \, c=d=0 \, \Rightarrow \, G\equiv\bZ_2^D \subset U(1)_{n_1}\times U(1)_{n_2}  
\ee
We will encounter the matrix \eqref{eq: sigma Z2Z2} again when studying the quadri-critical point in the next section.

We want now to discuss the existence of non-invertible duality symmetries on the toroidal branch. Even if the condition \eqref{eq: dualsymm on E} in principle generates all the possible $\sD \in O(2,2,\bQ)$ implying a duality symmetry defect $\cD$, explicit solutions of such equation are generically hard to find. To this aim, we proceed to describe a particular method by which some solutions can be easily obtained.
In particular by performing some topological manipulation $\cT$, namely a combination of gaugings and dualities, we can connect the theory at $(\tau,\rho)$ to another point $(\widehat{\tau},\widehat{\rho})$ in the toroidal branch for which some duality symmetries are manifest. We denote the latter symmetries by $\widehat{\cD}$ and their associated matrix by $\widehat{\sD}$. Therefore the original theory will automatically enjoy a non-invertible symmetry defect $\cD$ corresponding to a matrix $\sD$ constructed as
\be\label{eq: composition of matrices}
\sD = \mathsf{T} \,\widehat \sD \,\mathsf{T}^{-1}\,,
\ee
where $\mathsf{T} \in O(2,2,\bQ)$ is the matrix representation of the topological manipulation $\cT$ used to connect the points $(\tau,\rho)$ and $(\widehat{\tau},\widehat{\rho})$ in the conformal manifold.
For instance, a very convenient choice of $(\widehat{\tau},\widehat{\rho})$ is one for which the theory factorizes in terms of two $c=1$ RCFT's, namely $\widehat{\tau}_1=\widehat{\rho}_1=0$. In particular, one can show that such a factorized theory enjoy duality symmetries coming from the separate gaugings of subgroups of  $U(1)_{n_1}\times U(1)_{w_1}$ and $U(1)_{n_2}\times U(1)_{w_2}$.\footnote{This is a trivial consequence of the existence of duality symmetries in any $c=1$ RCFT as shown in section \ref{sec:c=1}.}

Consider a generic $c=2$ RCFT with \be
\tau = \frac{p}{q} + i \tau_2 \quad,\quad \rho = \frac{p'}{q'} + i \rho_2\, \qquad , \qquad p,q,p',q' \in \bZ\,.
\ee
where $\tau_2$ and $\rho_2$ are determined in terms of the rationality condition reviewed previously in this section. For any $p,q\in \bZ$, we can always perform a gauging $\sigma^*$ that makes $\tau_1,\rho_1 \in \mathbb{Z}$. Indeed this can be achieved by the gauging of $\bZ_{n_i,w_i} \subset U(1)_{n_i,w_i}$with 
\be
\frac{w_1}{n_1} = \frac{pq'}{qp'} \quad , \quad \frac{w_2}{n_2} = q q'\,.
\ee
This gauging brings the theory at $(\tau,\rho)$ to a theory with 
\be
\tau' = qp' + i \frac{q^2p'}{p} \tau_2  \quad , \quad \rho' = pq' + i \frac{q^2p'}{p} \rho_2
\ee
which, by applying the dualities $T_{\rho}^{pq'}T_{\tau}^{qp'}$, is equivalent to the factorized point
\be
\widehat\tau = i \frac{q^2p'}{p} \tau_2 \quad , \quad \widehat\rho = i \frac{q^2p'}{p} \rho_2\,.
\ee
Duality symmetry defects $\widehat{\cD}$ of the factorized theory above are easily detected by the logic exposed in section \ref{sec:c=1} and the topological manipulation just described reads $\cT=T_{\rho}^{pq'}T_{\tau}^{qp'}\sigma^*$.

This simple computation shows that, because of \eqref{eq: composition of matrices}, \emph{any} $c=2$ RCFT enjoy duality symmetries which are more manifest in a factorzied point connected to the original RCFT via topological manipulations. We will illustrate this procedure when studying some particular examples in the next section. Before concluding, we emphasize that the duality symmetries which are manifest in the factorized point generically do not span the entire set of duality symmetries in a given RCFT. For instance there can be symmetries coming from non-diagonal gauging in the factorized point which are, however, more manifest if we look at different points connected by topological manipulations. We will encounter an example of this kind when we look at the bi-critical point in section \ref{sec: dualities 2-crit}.

\section{Duality symmetries at multicriticality}\label{sec: multicritical}

As mentioned in the previous section, the intricate structure of the bosonic $c=2$ conformal manifold allows for several special loci where various orbifold branches intersect. Moreover, the theories sitting at these multicritical loci are generically rational. Examples of this kind will be the focus of this section, putting particular emphasis on the interplay between non-invertible duality symmetries and several deformations. Regarding the latter, we will study the additional exactly marginal deformations arising at multicriticality, together with some examples of RG flows triggered by duality preserving relevant operators. 

We begin from the points on the conformal manifold of $c=2$ theories where the global symmetry is enhanced. These loci are given in~\eqref{eq:EnhancedPtsc2} and we focus on the following:
\begin{align}
    (\tau ,\rho) &= (i,i) \; : \qquad SU(2)^2 \times \overline{SU(2)}^2 \; , \nonumber \\[5pt]
    (\tau ,\rho) &= (e^{i 2 \pi/3},e^{i 2 \pi/3}) \; : \qquad SU(3) \times \overline{SU(3)} \; .
\end{align}
Orbifolding by the $\mathbb{Z}_2$ and $\bZ_3$ symmetries of the $SU(2)^2$ point and the $SU(3)$ point respectively, one obtains two multicritical points, where four and two branches of the conformal manifold interesect. See figure~\ref{fig: c=2 manifold} for a pictorial representation of the conformal manifold.  Therefore, these are respectively a quadri-critical and a bi-critical points. These two multicritical points are respectively located in the toroidal branch at $(\tau,\rho)=(i,2i)$ and $(\tau,\rho)=(\omega,\alpha)$ with $\omega=e^{\frac{2\pi i}{3}}$, $\alpha=\omega+i\sqrt{3}$. The reason behind this choice is the fact that they exemplify many of the properties of $c=2$ RCFTs with non-invertible duality symmetries. At the quadri-critical, point the theory is factorized into two copies of $c=1$, while at the bi-critical point the $B$-field takes non-zero value and the theory is genuinely $c=2$. 

\subsection{The quadri-critical point}

We begin by looking at the point $(\tau, \rho) = (i,i)$, featuring an $ SU(2)^2 \times \overline{SU(2)}^2$ enhanced global symmetry. We review some features of this point in Appendix \ref{app:enhancedc1}. The theory is factorized into two copies of $c=1$, as it can be seen already in the holomorphic decomposition of the fields 
\begin{align}
    \phi^a = \frac{1}{\sqrt{2}} \left( X^a + \overline X^a \right) \; , \quad
    \widetilde \phi^a = \frac{1}{\sqrt{2}} \left( X^a - \overline X^a \right) \; .
\end{align}
Each factor enjoys two $\mathbb{Z}_2$ symmetries, i.e. the shift $\phi^a \to \phi^a + \pi$ and the reflection $\phi^a \to - \phi^a$. On the chiral fields, these act as a shift by $\sqrt{2}\pi/2$ and a sign flip. One can then identify four (non-anomalous) $\bZ_2$ symmetries at $(\tau, \rho) = (i,i)$, which in the chiral basis take the following form  
\begin{align}\label{eq:Z2 autom}
{\mathbb Z}_2^s \; :& \quad \left(X^a, \; \overline X^a\right) \, \to \, \left(X^a + \sqrt{2}\pi/2 , \; \overline X^a + \sqrt{2} \pi/2\right)  \; , \nonumber \\[5pt]
{\mathbb Z}_2^{orb} \; :& \quad \left(X^a, \; \overline X^a\right) \, \to \, \left(-X^a, \; -\overline X^a\right)  \; , \nonumber\\[5pt]
{\mathbb Z}_2^{R_1} \; :& \quad \left(X^1, \; \overline X^1\right) \, \to \, \left(X^1 + \sqrt{2}\pi/2 , \; \overline X^1 + \sqrt{2} \pi/2\right)   \\
&\quad \left(X^2, \; \overline X^2\right) \, \to \, \left(-X^2 , \; - \overline X^2 \right) \; , \nonumber \\[5pt]
{\mathbb Z}_2^{R_{2}} \; :& \quad \left(X^1, \; \overline X^1\right) \, \to \, \left(-X^1 + \sqrt{2}\pi/2  , \; -\overline X^1 + \sqrt{2}\pi/2 \right ) \; , \nonumber \\
&\quad \left(X^2, \; \overline X^2\right) \, \to \, \left(X^2 + \sqrt{2}\pi/2, \;  \overline X^2+ \sqrt{2}\pi/2 \right)  \; . \nonumber
\end{align}
The $\bZ_2^s$ shift symmetry corresponds to a simultaneous translation in field space. The $\mathbb{Z}_2^{orb}$ implements a reflection on both fields, analogous to the $c=1$ case (see appendix~\ref{app:enhancedc1}). The remaining $\bZ_2$ symmetries implement combinations of the previous actions. For more details, see~\cite{Dulat:2000xj}\footnote{The reader should be aware that we are using a slightly different notation than in~\cite{Dulat:2000xj}.}. As usual, the orbifold by the $\bZ_2^s$ shift symmetry leads to another theory in the toroidal branch. The remaining $\bZ_2$ actions define three independent orbifold branches. Furthermore, the four actions just described are mapped to each other by the $SU(2)^2\times \overline{SU(2)}^2$ global symmetry. Therefore, performing the quotient by any of them leads to four different descriptions of the same point of the conformal manifold, where the toroidal and the three orbifold branches meet.

Note that the orbifold by the $\mathbb{Z}_2^s$ symmetry is implemented precisely by the matrix in Eq.~\eqref{eq: sigma Z2Z2}.
The resulting theory sits on the toroidal branch at $(\tau,\rho)=\left(i,i/2\right)$, and by acting with $S_{\rho} \, : \, \rho \to - 1/\rho$, the theory is mapped to $\left( \tau,\rho \right) = \left(i,2i\right)$. 
One therefore identifies the theory at $\left( \tau,\rho \right) = \left(i,2i\right)$ with the quadri-critical point. The overall map of the charges read
\be\label{eq:Z2mapSU2fourcrit}
n_1'=\frac{n_1+n_2}{2} \; , \: n_2'=\frac{n_1-n_2}{2} \; , \: w_1'= w_1+w_2\; , \: w_2'=w_1-w_2\; ,
\ee
where the non-primed symbols denote the original charges at the enhanced symmetry point.

The theory at the quadri-critical point is itself rational and, moreover, corresponds to a diagonal modular invariant since $\rho=2\tau$~\cite{Gukov:2002nw}. This property is also manifest by the fact that this theory can be regarded as a product of two rational $c=1$ theories at radius equal to $\sqrt{2}$. Even if many of the properties of the global symmetries featured at $(\tau,\rho)=(i,2i)$  naturally descend from its product structure, we will see that there are still some salient features pertaining to the $c=2$ theory, such as the non-Abelian fusion satisfied by the (non-invertible) duality symmetries. This is just a consequence of the fact that the latter symmetries descend from a non-Abelian subgroup of the $c=2$ duality group \eqref{eq:duality c=2} which is not a product of two $c=1$ dualities.

\subsubsection{Duality symmetry and fusion category}

As we have already discussed in the general case, RCFT points in the toroidal branch enjoy duality symmetries, obtained by composing gaugings and dualities in different ways. We now discuss some of the duality symmetries present in the quadri-critical point $(\tau,\rho) = (i, 2i)$. Since it is a factorized point (\ie{} it can be decomposed as a product of two $c=1$ theories) with $\rho_2,\tau_2 \in \bZ$, it generically hosts duality symmetries coming from diagonal gaugings, of the form\footnote{For sake of notational simplicity, from now on we will omit the symbol $\circ$ denoting composition dualities and topological manipulations. For instance $M\sigma_{2,1}\equiv M\circ \sigma_{2,1}$.}
\bea\label{eq: 4crit dual symmetries}
&\cD_A = M  \sigma_{2,1} \to  
    \sD_A=\mat{0&0&\tfrac12&0\\0&1&0&0\\ 2&0&0&0\\0&0&0&1}
\; , \; \cD_{B} = S_{\rho} M \tilde{\sigma}_{2,1} \to \sD_B=\mat{0&0&0&-\frac{1}{2}\\1&0&0&0\\ 0&-2&0&0\\ 0&0&1&0} \;\;\;
\eea
together with the invertible duality symmetries $S_{\tau}$ and $I$ that do not involve any gauging. The matrices above represent the action of the associated operators over the $({\boldsymbol\phi},\widetilde{\boldsymbol\phi})$. In \eqref{eq: 4crit dual symmetries} we introduced the following notation for the diagonal gauging
\begin{align}
\sigma_{N,M}&\equiv {\rm gauging } \; \bZ_N\times\bZ_M \subset U(1)_{n_1}\times U(1)_{w_1}\\
\widetilde\sigma_{N,M}&\equiv {\rm gauging } \; \bZ_N\times\bZ_M \subset U(1)_{n_2}\times U(1)_{w_2}\nonumber\,.
\end{align}
and their corresponding matrix representation reads
\be
\sigma_{N,M}\to\mat{\frac{N}{M}&0&0&0\\0&1&0&0\\ 0&0&\frac{M}{N}&0\\0&0&0&1} \quad , \quad
\widetilde\sigma_{N,M}\to\mat{1&0&0&0\\0&\frac{N}{M}&0&0\\ 0&0&1&0\\0&0&0&\frac{M}{N}} \,,
\ee

As emphasized before, this set of duality symmetries does not capture the full set of symmetries of this point; for instance duality symmetries coming from non-diagonal gaugings can be present\footnote{It was also recently noticed in \cite{Choi:2023vgk} that duality defects coming from gauging non-invertible symmetries can be present in $2$d CFT.}. It would be nice to develop a systematic method capable of identifying all duality symmetries, along the lines of \cite{Niro:2022ctq}.

Given a set of duality symmetries closed under fusion, we can discuss the fusion algebra and the underlying fusion category (see appendix \ref{App:TY} for some details on this type of categorical symmetries). Since they are symmetries coming from self-dualities, the fusion algebra will be graded by (a subgroup of) the underlying $O(2,2,\bZ)$ duality group at $c=2$. This grading fixes the generic structure of the fusion algebra to be
\be
\cD_{\alpha} \times \cD_{\beta} = \cD_{\alpha\beta} \times \cC_0 \,,
\ee
where $\cD_{\alpha\beta}$ is the composed duality symmetry whose matrix representation is obtained by multiplying the matrix representation of $\cD_{\alpha}$ and $\cD_{\beta}$ and $\cC_0$ is some combination of topological lines generating a subgroup of the invertible $U(1)^2_{\mathbf{n}}\times U(1)^2_{\mathbf{w}}$ symmetry. The occurrence of the gauging $\sigma_{\alpha}$ in the definition of $\cD_{\alpha}$ ensures that this topological defect absorbs all the lines generating the gauged symmetry $G_\alpha \subset U(1)^2_{\mathbf{n}}\times U(1)^2_{\mathbf{w}}$, namely
\be
\cD_{\alpha} \times \eta_\alpha = \eta_\alpha \times \cD_\alpha = \cD_\alpha \qquad \forall \alpha \in G_\alpha \subset U(1)^2_{\mathbf{n}}\times U(1)^2_{\mathbf{w}}\, .
\ee
The fusion of $\cD_\alpha$ with its orientation reversal can be uniquely fixed by imposing the consistency with the above property of absorption, finding
\be
\cD_\alpha \times \overline{\cD}_\alpha = \sum\limits_{\alpha \in G} \eta_\alpha \qquad G_\alpha\subset U(1)^2_{\mathbf{n}}\times U(1)^2_{\mathbf{w}}\,.
\ee
Let us now delve into the specific examples presented in \eqref{eq: 4crit dual symmetries}. From the gauging matrices used to define those defects we find
\be\label{eq: absorb}
\eta_{a,0}\widetilde \eta_{b,0}\times\mathcal{D}_{A} = \widetilde \eta_{b,0}\times\mathcal{D}_{A} \quad,\quad \eta_{a,0}\widetilde \eta_{b,0}\times\mathcal{D}_{B} =  \eta_{a,0}\times\mathcal{D}_{B} \,,
\ee
where $\eta_{a,0}$($\eta_{0,b}$) generate $\bZ_2 \subset U(1)_{n_1}$ ($\bZ_2 \subset U(1)_{w_1}$) while $\widetilde\eta_{a,0}$ ($\widetilde\eta_{0,b}$) generate $\bZ_2 \subset U(1)_{n_2}$ ($\bZ_2 \subset U(1)_{w_2}$). Here $a,b\in\bZ_2$.  Because of the grading structure we also need to include additional duality defects
\be
\cD_{C} = S_{\rho} \sigma_{2,1} \widetilde\sigma_{2,1} \quad,\quad \cD_{\tau C} = S_{\tau}S_{\rho} \sigma_{2,1} \widetilde\sigma_{2,1}\quad , \quad C\,,
\ee
where $C$ is the topological defect implementing charge conjugation and the new duality defects satisfy 
\be
 \eta_{a,0}\widetilde \eta_{b,0}\times\mathcal{D}_{C} = \mathcal{D}_{C} \quad , \quad  \eta_{a,0}\widetilde \eta_{b,0}\times\mathcal{D}_{\tau C} = \mathcal{D}_{\tau C} \,.
\ee 
Applying the generic rules described above we find the fusion between defects and their inverse to be 
\be\label{eq: inverse fusion}
 \mathcal{D}_{C} \times \overline{\mathcal{D}}_{C} = \sum\limits_{a,b=0,1} \eta_{a,0}\widetilde \eta_{b,0} \quad , \quad \mathcal{D}_{A} \times\overline{\mathcal{D}}_{A} = \sum\limits_{a=0,1} \eta_{a,0}\quad , \quad \mathcal{D}_{B} \times\overline{\mathcal{D}}_{B} = \sum\limits_{b=0,1} \widetilde \eta_{b,0}\,
\ee
while some fusion rules between different duality symmetries read
\be
\begin{split}
&\cD_{C}\times \cD_{A} =\cD_{B}\sum\limits_{a=0,1}\eta_{a,0}\;\;,\;\; \cD_{A}\times \cD_{C} =  S_{\tau} \cD_{A}\sum_{b=0,1}\widetilde\eta_{b,0}\\
&\mathcal{D}_A \times \mathcal{D}_{B} = S_{\tau} \sum_{a,b} \eta_{a,0}\widetilde \eta_{b,0} \;\;,\;\; \mathcal{D}_{B}\times \mathcal{D}_A =
 \cD_{C}  \,,
\end{split}
\ee
together with other ones that can be easily derived.
We notice that the category is non-commutative, as a consequence of the non-Abelian $O(2,2,\bZ)$ grading\footnote{Non-commutative categorical symmetries appeared before in the literature in $4$d (see e.g. \cite{Bashmakov:2022uek,Antinucci:2022cdi}) and also in $2$d $c=2$ theories (see e.g. \cite{Nagoya:2023zky}).}. In order to describe the non-commutative grading more explicitly, we introduce the group with five generators $\{s,d_c,d_a,i,c\}$ satisfying the following multiplication law
\begin{align}
s^2=d_c^2=c \quad , \quad d_a^2=c^2=i^2=1 \quad &, \quad isi=s^{3}\quad , \quad id_ci=d_c^{3}\nonumber\\  a s a = d_c \quad &, \quad a d_c a= s \;\;.
\label{eq: non com grading}
\end{align}
It can be easily verified that the fusion of the category described before is graded by the non-commutative group \eqref{eq: non com grading} by making the following assignment of (non-)invertible defects to the generators
\be
S_\tau \to s \;\; , \;\; \cD_C\to d_c \;\; , \;\; \cD_A\to d_a \;\; , \;\; I\to i \;\; , \;\; C\to c \;,
\ee
with the remaining defects corresponding to products of the ones above.

As already emphasized, the quadri-critical point is the product of two $c=1$ CFT at $R= \sqrt{2}$. Therefore the partition function can be written as
\be
Z_{c=2}[\tau=i,\rho=2 i] = \sum_{m,n \in \bZ_{4}}\chi_m \chi_n \overline{\chi}_{-m}\overline{\chi}_{-n} = Z_{c=1}[R=\sqrt{2}]Z_{c=1}[R=\sqrt{2}]
\ee
and we can trivially prove that the partition function of the $c=2$ theory at the quadri-critical point is self-dual under gauging of $\bZ_2 \subset U(1)_{n_1}$ and $\bZ_2 \subset U(1)_{n_2}$, implying the existence of the defects described above (see appendix \ref{App:TY}).

\subsubsection{Modular bootstrap at the quadri-critical point}

As a further check that the above structure is actually consistent, we compute the twisted partition functions and make use of modular covariance. We find that all the elements introduced above lead to well defined defect Hilbert spaces, hence complying with \eqref{eq: trace defect H space}. For sake of the extension of this exposition, we do not include all the examples but just a few illustrative ones and comment on the rest.

Regarding the defect $\cD_A$, it can be explicitly checked that its action emulates the action of $T$-duality in one of the sectors of the factorized theory, namely $\overline p_1\to-\overline p_1$, hence the computation works out exactly as at $c=1$ and $R=\sqrt{2}$ \cite{Thorngren:2021yso}. Similarly, target space inversion $I$ implements a reflection in one of the two sectors, hence clearly leading to a well defined defect on the product theory. 

We will then proceed to illustrate the computation for the non-invertible defects ${\cal D}_B$ and ${\cal D}_{C}$. Making use of \eqref{eq: general charge map} and \eqref{eq: momenta in terms of charges}, the action of $\cD_{B}$ maps 
\be\label{eq:DB mom map}
{\bf p}\to R_B{\bf p} \quad , \quad \overline{\bf p}\to \overline{R}_B\overline{\bf p}\quad , \quad  R_B=\begin{pmatrix}
    0&1\\-1&0
\end{pmatrix} \quad , \quad
  \overline{R}_B=\begin{pmatrix}
    0&1\\1&0
\end{pmatrix} \,.
\ee
In addition, it turns out that the phase $\alpha_B({\bf n},{\bf w})$ determined by equation \eqref{eq: c=2 phase} is trivial once we impose the transformed operator to be local\footnote{In fact, the action of ${\cal D}_B$ leads to the map of charges $\{n_1,n_2,w_1,w_2\}\to \{-2w_2,n_1,-\tfrac{n_2}{2},w_1\}$. Therefore, the resulting operator corresponds to a genuine local operator only for $n_2=2\bZ$. Moreover, plugging the previous map of charges into equation \eqref{eq: c=2 phase} leads to the solution $\alpha_B({\bf n},{\bf w})=n_2 w_2$ mod 2 for the phase entailing mutual locality. The latter clearly becomes trivial for $n_2$ even.}. Due to \eqref{eq:DB mom map}, when computing the partition function twisted by the insertion of $\cD_B$ along the spatial cycle, the sum gets contributions only from states with $p_1=p_2=0$, $\overline p_1=\overline p_2=2n$ ($n\in\bZ$).
Finally, in order to perform the sum over oscillator modes, we consider a basis over which $R_B$ ($\overline R_B$) act diagonally with eigenvalues $\pm i$ ($\pm 1$). Putting all together we get 
\begin{align}
Z_{(1,{\cD}_B)}&= \sqrt{2}\frac{q^{-\frac{1}{12}}\overline q^{-\frac{1}{12}}\sum_{n\in\bZ} \overline{q}^{4n^2}}{\prod_{m=1}^\infty (1+iq^{m})(1-iq^{m})(1+\bar q^{m})(1-\bar q^{m})}
= \sqrt{2}\frac{\vartheta_4(4\tau)\vartheta_4(2\bar\tau)\vartheta_3(8\bar\tau)}{\eta(2\tau)\eta(\bar\tau)^2}\nonumber\\ 
\xrightarrow{\tau\to-\frac{1}{\tau}} \,\, Z_{({\cD}_B,1)}&=\frac{\vartheta_2(\tau/4)\vartheta_2(\bar\tau/2)\vartheta_3(\bar\tau/8)}{4\eta(\tau/2)\eta(\bar\tau)^2}=
\frac{1}{|\eta(\tau)|^4}\left(q^{\frac{3}{32}}\overline{q}^{\frac{1}{16}}+2q^{\frac{3}{32}}\overline{q}^{\frac{1}{8}}+\ldots\right)
\;.
\end{align}
We therefore verify that $Z_{({\cD}_B,1)}$ has a consistent interpretation as a trace over the defect Hilbert space.

Now we turn to the defect $\cD_C$ implementing the following action 
\be\label{eq:DC mom map}
{\bf p}\to R_C{\bf p} \quad , \quad \overline{\bf p}\to \overline{R}_C\overline{\bf p}\quad , \quad  R_C=-\overline{R}_C=\begin{pmatrix}
    0&1\\-1&0
\end{pmatrix} 
\,.
\ee
In addition, the states get multiplied by the $\bZ_2$ phase $\alpha_C={\bf n}\cdot{\bf w}$. Given the transformation \eqref{eq:DC mom map}, the twisted partition function gets contribution from states in the identity multiplet $p_a=\overline p_a=0$ ($a=1,2$). Following similar steps as the ones described above, we get  
\begin{align}
Z_{(1,{\cD}_C)}&= 2\frac{q^{-\frac{1}{12}}\overline q^{-\frac{1}{12}}}{\prod_{m=1}^\infty (1+iq^{m})(1-iq^{m})(1+i\bar q^{m})(1-i\bar q^{m})}
= 2\frac{\vartheta_4(4\tau)\vartheta_4(4\bar\tau)}{|\eta(2\tau)|^2}\\
\xrightarrow{\tau\to-\frac{1}{\tau}} \,\, Z_{({\cD}_C,1)}&=\frac{\vartheta_2(\tau/4)\vartheta_2(\bar\tau/4)}{|\eta(\tau/2)|^2} =
\frac{4}{|\eta(\tau)|^4}\left[q^{\tfrac{3}{32}}\overline{q}^{\tfrac{3}{32}}+q^{\tfrac{3}{32}}\overline{q}^{\tfrac{3}{32}}\left(q^{\tfrac{1}{4}}+\overline{q}^{\tfrac{1}{4}}\right)+\ldots\right]
\;,\nonumber
\end{align}
hence corroborating that $\cD_C$ leads to a consistent defect Hilbert space. 

We end this subsection by considering the case of $\cD_{\tau C}$ whose action maps $p_a\to-p_a$ ($a=1,2$) while leaving the right movers invariant. The $\bZ_2$ phase again obtains $\alpha_{\tau C}={\bf n}\cdot{\bf w}$. By comparison with the $c=1$ case reviewed in section \ref{sec:c=1}, one notices that this action can be regarded as a composition of two (left) $T$-dualities on each $c=1$ factor (composed with a gauging of the corresponding ${\mathbb Z}_2$ subgroups), hence leading to
\begin{align}
Z_{(1,{\cal D}_{\tau C})}&
%=2 \left(\sum_{m\geq 0}(-1)^m\cV_{m^2}(\tau) \overline \chi_0(\overline \tau)\right)^2
=2\frac{\vartheta_4(2\tau)^2\vartheta_3(4\overline\tau)^2}{|\eta(\tau)|^4}\\ 
\xrightarrow{\tau\to-\frac{1}{\tau}} \,\, Z_{({\cal D}_{\tau C},1)}&=\frac{\vartheta_2(\tau/2)^2\vartheta_3(\overline\tau/4)^2}{4|\eta(\tau)|^4}=\frac{1}{|\eta(\tau)|^4}\left(q^{\frac18}+4q^{\frac18}\overline{q}^{\frac18}+4q^{\frac18}\overline{q}^{\frac14}\ldots\right)\nonumber
\end{align}
Note the important role played by the quantum dimension here. If it was absent, the final result would be
\be
\frac{\vartheta_2(\tau/2)^2\vartheta_3(\overline\tau/4)^2}{8|\eta(\tau)|^4}\sim\frac12 q^{\frac{1}{8}}+\ldots
\ee
clearly spoiling the interpretation of the twisted partition function as a trace over a Hilbert space. Of course, the positive outcome of the modular bootstrap analysis for ${\cal D}_C$ and ${\cal D}_{\tau C}$ naturally implies the consistency of the invertible defect implementing $S_\tau$.

\subsubsection{Marginal and relevant deformations at the quadri-critical point}

We proceed to study the interplay between the symmetry structure presented in the previous sections and the various marginal and relevant deformations of the theory. As a multicritical point, one expects finding additional independent exactly marginal deformations which span the orbifold branches, besides the ones of the form $\partial_\mu \phi^a\partial_\nu\phi^b$ which span the four dimensional toroidal branch. The strategy consists in constructing the set of marginal operators of the form $J^a_{i}\overline{J}_{j}^b$, where $a,b=1,2$ and $i,j=+,-$, and in terms of the chiral fields
\begin{align}
J_\pm^1 &= V_{\pm1,0,\pm1,0} = e^{\pm i\sqrt{2}X^1} \; , \qquad  \overline J_\pm^1= V_{\pm1,0,\mp1,0} = e^{\pm i\sqrt{2}\,\overline X^1} \; , \nonumber \\[5pt]
J_\pm^2 &= V_{0,\pm1,0,\pm1} = e^{\pm i\sqrt{2}X^2} \; , \qquad  \overline J_\pm^2= V_{0,\pm1,0,\mp1} = e^{\pm i\sqrt{2}\,\overline X^2} \; , \nonumber \\[5pt]
J_3^a &= \frac{i}{\sqrt{2}}\partial X^a \; , \qquad \overline J_3^a = \frac{i}{\sqrt{2}}\overline\partial \, \overline X^a \; .
\end{align}
At the enhanced symmetry point, they are all equivalent to $J_3^a \overline{J}_3^b$ by the $SU(2)^2\times \overline{SU(2)}^2$ global symmetry. We then construct invariant combinations under the $\mathbb{Z}_2$ automorphisms and charge conjugation. The action on the currents is 
\begin{align}\label{eq:Z2 autom currents}
{\mathbb Z}_2^s \; :& \; \left(J_3^a , \; J^a_\pm \right)\, \to \, \left(J_3^a , \; -J^a_\pm \right) \; , \nonumber \\[5pt]
{\mathbb Z}_2^{orb} \; :&  \; \left(J_3^a , \; J^a_\pm \right)\, \to \, \left(-J_3^a , \; J^a_\mp \right) \; , \nonumber\\[5pt]
{\mathbb Z}_2^{R_1} \; :& \; \left(J_3^1 , \; J^1_\pm \right)\, \to \, \left(J_3^1 , \; -J^1_\pm \right) \; , \nonumber \\
& \; \left(J_3^2 , \; J^2_\pm \right)\, \to \, \left(-J_3^2 , \; J^2_\mp  \right) \; , \nonumber \\[5pt]
{\mathbb Z}_2^{R_2} \; :& \; \left(J_3^1 , \; J^1_\pm \right)\, \to \, \left(-J_3^1 , \; J^1_\mp \right) \; , \nonumber \\
& \; \left(J_3^2 , \; J^2_\pm \right)\, \to \, \left(J_3^2 , \; -J^2_\pm \right) \; .
\end{align}
Given the $\mathbb{Z}_2$-invariant operators, the exactly marginal operators are further combinations of these, whose OPE with themselves has no simple poles~\cite{Ginsparg:1987eb}. Finally, the exactly marginal operators are mapped to the quadri-critical point by application of the map~\eqref{eq:Z2mapSU2fourcrit}. 

For the case of $\mathbb{Z}_2^{orb}$, the four exactly marginal operators
\begin{align}\label{eq:extmarginalZ2rot}
    \mathcal{O}_{ab}&=J_+^a\overline{J}_+^b + J_-^a\overline{J}_-^b + J_+^a\overline{J}_-^b + J_-^a\overline{J}_+^b \; , a\,b = 1,2 \; ,
\end{align}
are mapped to the quadri-critical points as
\begin{equation}
\bZ_2^{orb} \text{ branch}\begin{cases}
    \mathcal{O'}_{11} &= V_{0,0,+1,-1} + V_{+2,-2,0,0} + V_{0,0,-1,+1} + V_{-2,+2,0,0} \; , \nonumber \\
    \mathcal{O'}_{12} &= V_{2,0,0,-1} + V_{0,-2,+1,0} + V_{-2,0,0,+1} + V_{0,+2,-1,0} \; , \nonumber \\
    \mathcal{O'}_{21} &= V_{-2,0,0,-1} + V_{0,-2,-1,0} + V_{+2,0,0,+1} + V_{0,+2,+1,0} \; , \nonumber \\
    \mathcal{O'}_{22} &= V_{0,0,-1,-1} + V_{-2,-2,0,0} + V_{0,0,+1,+1} + V_{+2,+2,0,0} \; ,
\end{cases}
\end{equation}
where they span the 4-dimensional $\mathbb{Z}_2^{orb}$ orbifold branch.

Similarly, and considering the action in Eq.~\eqref{eq:Z2 autom}, the deformations into the $\bZ_2^{R_1}$ branch are generated by the following $\mathbb{Z}_2^{R_1}$-invariant exactly marginal operators 
\begin{equation}\label{eq:extmarginalZ2R}
\bZ_2^{R_1} \text{ branch}\begin{cases}
     \mathcal{O'}_A^{R_1} 
     &=% J^1_+\overline{J}^2_{+} - J^1_+\overline J^2_{-} + J^1_-\overline J^2_{-} - J^1_-\overline J^2_{+} \nonumber \\&\to \;
     V_{+2,0,0,-1} - V_{0,-2,+1,0} + V_{-2,0,0,+1} - V_{0,+2,-1,0} \; , \nonumber \\[7pt]
    \mathcal{O'}_B^{R_1} &= 
    %J^2_+\overline J^1_{+} - J^2_-\overline J^1_{+} + J^2_-\overline J^1_{-} - J^2_-\overline J^1_{+} \nonumber \\&\to \;
    V_{-2,0,0,-1} - V_{0,+2,+1,0} + V_{+2,0,0,+1} - V_{0,-2,-1,0} \; ,
\end{cases}
\end{equation}
while for $\mathbb{Z}_2^{R_2}$, we get the following
\begin{equation}\label{eq:extmarginalZ2TR}
\bZ_2^{R_2} \text{ branch}\begin{cases}
    \mathcal{O'}_A^{R_2} &= 
    %J^1_+\overline{J}^1_{+} - J^1_+\overline J^1_{-} + J^1_-\overline J^1_{-} - J^1_-\overline J^1_{+} \nonumber \\&\to \;
    V_{0,0,+1,-1} - V_{+2,-2,0,0} + V_{0,0,-1,+1} - V_{-2,+2,0,0} \; , \nonumber \\[7pt]
    \mathcal{O'}_B^{R_2} &=
    %J^2_+\overline J^2_{+} - J^2_+\overline J^2_{-} + J^2_-\overline J^2_{-} - J^2_-\overline J^2_{+} \nonumber \\&\to \;
    V_{0,0,-1,-1} - V_{-2,-2,0,0} + V_{0,0,+1,+1} - V_{+2,+2,0,0} \; .
\end{cases}
\end{equation}
Both the $\mathbb{Z}_2^{R_1}$ and $\mathbb{Z}_2^{R_2}$ orbifold branches are 2-dimensional. 

The duality defects discussed in the previous subsection are preserved by some combinations of the exactly marginal operators at the quadri-critical point\footnote{Note that, given their charges, none of these vertex operators acquires a phase under the action of the duality symmetries described in the previous subsection.}. For the four operators $\mathcal{O'}_{ab}$ parametrizing the $4$-dimensional $\bZ_2^{orb}$ orbifold branch, we get 
\begin{align}
    \mathcal{O'}_{11} + \mathcal{O'}_{12} \; , \quad \mathcal{O'}_{21} + \mathcal{O'}_{22} \qquad \text{preserved by }& \qquad \mathcal{D}_A \; , \nonumber \\[5pt]
    \mathcal{O'}_{11} + \mathcal{O'}_{21} \; , \quad \mathcal{O'}_{12} + \mathcal{O'}_{22} \qquad \text{preserved by }& \qquad \mathcal{D}_B \; , \nonumber \\[5pt]
    \mathcal{O'}_{11} + \mathcal{O'}_{22} \; , \quad \mathcal{O'}_{12} + \mathcal{O'}_{21} \qquad \text{preserved by }& \qquad S_{\tau} \; , \; I \; , \; \mathcal{D}_{C} \; , \nonumber \\[5pt]
    \mathcal{O'}_{11} \; , \; \mathcal{O'}_{12} \; , \; \mathcal{O'}_{21} \; , \; \mathcal{O'}_{22} \qquad \text{preserved by }& \qquad \mathcal{D}_{\tau C} \; .
\end{align}
In particular $\cD_{\tau C}$ is preserved along the full $\bZ_2^{orb}$ orbifold branch. In addition, there are two extra non-invertible symmetries $\cD_{A,B}$ which are preserved along certain $2$-dimensional submanifolds of this particular orbifold branch.

Regarding the $\mathbb{Z}_2^{R_1}$, $\mathbb{Z}_2^{R_2}$ branches the invariant combinations are
\begin{align}
    \mathcal{O'}_{A}^{R_1} + \mathcal{O'}_{B}^{R_1}\qquad \text{preserved by }& \qquad I \; , \; \mathcal{D}_{C} \; , \nonumber \\[5pt]
    \mathcal{O'}_{A}^{R_1} - \mathcal{O'}_{B}^{R_1} \qquad \text{preserved by }& \qquad S_{\tau} \; , \nonumber \\[5pt]
    \mathcal{O'}_{A}^{R_2} + \mathcal{O'}_{B}^{R_2} \qquad \text{preserved by }& \qquad S_{\tau} \; , \; I \; , \nonumber \\[5pt]
   \mathcal{O'}_{A}^{R_2} - \mathcal{O'}_{B}^{R_2} \qquad \text{preserved by }& \qquad \mathcal{D}_{C} \; .
\end{align}
Interestingly, we find that $\mathcal{D}_A$ and $\mathcal{D}_{\tau B} = S_{\tau}^{-1}MS_{\tau}\circ\widetilde{\sigma}_{2,1}$ act as
\begin{align}
    \mathcal{D}_A \; :& \quad \mathcal{O'}_A^{R_1} \to \mathcal{O'}_A^{R_2} \; , \nonumber \\[5pt]
    & \quad \mathcal{O'}_B^{R_1} \to \mathcal{O'}_B^{R_2} \; , \nonumber \\[5pt]
    \mathcal{D}_{\tau B} \; :& \quad \mathcal{O'}_A^{R_1} \to -\mathcal{O'}_A^{R_2} \; , \nonumber \\[5pt]
    & \quad \mathcal{O'}_B^{R_1} \to -\mathcal{O'}_B^{R_2} \; ,
\end{align}
thus they effectively exchange the two orbifold branches. 

\subsubsection*{Relevant deformations}
We now proceed with the analysis of the relevant deformations of the quadri-critical point. Classifying them by their dimension, we have
\begin{center}
\begin{tabular}{ |m{2cm}||m{10cm} | } 
 \hline
 $(h,\overline h)$ & $(\mathbf{n},\mathbf{w})$ \\ 
  \hline
$ \left( \frac{1}{8} \, , \, \frac{1}{8} \right)$& $(\pm 1,0,0,0) \; , \; (0,\pm 1,0,0)$ \\ [1ex] 

  $\left( \frac{1}{4} \, , \, \frac{1}{4} \right) $& $(\pm 1,\pm 1,0,0) \;,\; (\pm 1 , \mp 1,0,0)$ \\ [1ex] 

 $ \left( \frac{1}{2} \, , \, \frac{1}{2} \right) $& $(\pm 2,0,0,0)\;,\;(0,\pm 2 ,0,0)\;,\;(0,0,\pm1,0)\;,\;(0,0,0,\pm 1) $ \\ [1ex] 
  
 $ \left( \frac{5}{8} \, , \, \frac{5}{8} \right) $& $(0,\pm1,\pm1,0)\;,\;(\pm 2,\mp 1,0,0)\;,\; (\pm1,0,0,\pm1)\;,\;(\pm1,0,0,\mp1)$ $(\pm2,\pm1,0,0)\;,\;(\pm2,\mp1,0,0)\;,\;(\pm1,\pm2,0,0)\;,\;(\pm1,\mp2,0,0)$\\ [1ex] 
 \hline
\end{tabular}
\end{center}
Some non-invertible defects at the quadri-critical point commute with particular combinations of these operators. We thus proceed to analyse some representative cases of duality-preserving RG flows, triggered by some of the relevant deformations presented in the above table. Given the factorized structure quadri-critical theory, these RG flows can be studied by building up on the intuition acquired at $c=1$ (see \cite{Thorngren:2021yso} for some examples of the latter). This enables to check some non-trivial constraints that are a direct consequence of the presence of non-invertible duality symmetries. 

We begin by considering the case of the most relevant operators, that is with conformal weights $(h,\overline h)=\left(\tfrac18,\tfrac18\right)$. There are two possible independent charge conjugation invariant combinations, namely
\be
R_{1/8}^1=V_{1,0,0,0}+V_{-1,0,0,0}\sim \cos\phi^1 \quad , \quad R_{1/8}^2=V_{0,1,0,0}+V_{0,-1,0,0}\sim \cos\phi^2
\ee
Each of them preserve a $U(1)_m \times U(1)_w$ invertible symmetry participating in a mixed 't Hooft anomaly of the form
\be\label{eq:cont anomaly}
\cA_{3d} = \frac{i}{2\pi}\int AdB \qquad A\in U(1)_m \;,\; B \in U(1)_w\,.
\ee
Due to this anomaly the IR fixed point must be gapless. Moreover both $R_{1}$ and $R_2$ preserve  respectively $D_{\tau B}=S_\tau^{-1} S_\rho M \widetilde\sigma_{2,1}$ and $D_A=M\sigma_{2,1}$.\footnote{Note that $\cD_{\tau B} = S_{\tau}^{-1} \cD_{B}$ and therefore is not an independent defect.} Therefore the IR gapless theory must enjoy such non-invertible defect\footnote{The non-invertible symmetry cannot decouple from the gapless sector since its invertible subgroup partecipates in the 't Hooft anomalies generating the gapless mode.}. In this case, it is straightforward to show that the IR theory corresponds to the $c=1$ compact boson at $R = \sqrt{2}$, hence exactly satisfying all the symmetry constraints. 

Let us now comment on slightly more involved flows, triggered by the relevant operators with $(h,\overline h)=\left(\tfrac12,\tfrac12\right)$ of the form
\be\label{eq:rel 1/2}
R_{1/2}^1 \sim \cos(\widetilde \phi_1)+\cos(2\phi_2) \quad,\quad R_{1/2}^2 \sim \cos(\widetilde \phi_2) + \cos(2\phi_1)\,.
\ee
Both $R_{1/2}^{1,2}$ break $U(1)^{4}$ to a subgroup $U(1)\times U(1)\times \bZ_2$ with a $\bZ_2$ anomaly\footnote{This anomaly descends from the mixed anomaly \eqref{eq:cont anomaly} upon restricting the gauge bundles to the appropriate preserved $\bZ_2$ subgroups. More precisely, a $\bZ_2$ connection can be naturally embedded into a flat connection $A\in H^1(X,U(1))$ by means of a cocycle element $a\in H^1(X,\bZ_2)$ as
$A=\pi a$.
In addition, the uplift of $a$ to an integral cocycle allows to map the action of the external derivative over $A$ in terms of the Bockstein homomorphism $\beta \, : \, H^1(X,\bZ_2)\to H^2(X,\bZ_2)$, roughly $dA=2\pi \beta(a)$. A straightforward application of this reasoning to \eqref{eq:cont anomaly} leads to the mixed $\bZ_2$ anomaly in \eqref{eq: Z2 inv anomaly}.}
\be\label{eq: Z2 inv anomaly}
\cA_{3d} = \pi i \int a \cup \beta(b)
\ee
where $a,b \in H^1(X_3,\bZ_2)$ are discrete gauge fields for $\bZ_2 \times \bZ_2\subset U(1) \times U(1) \times \bZ_2$ and $\beta$ is the Bockstein homomorphism (see e.g. \cite{HatcherBook}). In the case of $R_{1/2}^1$, the discrete subgroup corresponds to a $\bZ_2 \times \bZ_2$ momentum and winding symmetries acting on $(\phi_2,\widetilde\phi_2)$, and vice versa for $R_{1/2}^2$.  
On the contrary, the relevant combination $R_{1/2}^1 + R_{1/2}^2$ preserves an anomaly free $\bZ_2 \times \bZ_2$. Therefore there are no continuous anomalies and the theory can be non-trivially gapped if we deform with $R_{1/2}^{1,2}$ or even trivially gapped if we consider the combination $R_{1/2}^1 + R_{1/2}^2$. However both deformations in \eqref{eq:rel 1/2} preserve the non-invertible symmetry generated by $\cD_{C}$ and the sum $R_{1/2}^1 + R_{1/2}^2$ preserves all the duality defects $\cD_A , \cD_{C}, \cD_{\tau C}$. Since both $\cD_A$ and $\cD_{\tau B}$ have a non-invertible anomaly, while $\cD_{C}$ is anomaly free (see appendix \ref{App:TY} and references therein)\footnote{In particular, we are referring to the anomalies coming from the TY($\mathbb{Z}_2)$ subcategory spanned by one of the three non-invertible defects together with the its corresponding $\bZ_2$ symmetry.}, the IR theory must match such anomalies. We proceed now to study the RG flows
\bea\label{eq: rel 1/2 RG}
&S = S_{c=2}[(i,2i)] + \mu\int  R_{1/2}^1\\
&S = S_{c=2}[(i,2i)] + \mu\int  (R_{1/2}^1+R_{1/2}^2)
\eea
and check explicitly the non trivial constraints coming from the non-invertible 't Hooft anomalies. In order to do that it is instructive to recall that a $c=1$ compact boson at radius $\sqrt{2}$ corresponds to the bosonization of two free Majorana fermions (equivalently a Dirac fermion). From this perspective, the relevant deformations $\cos 2\phi$ and $\cos\widetilde\phi$ are mapped to mass terms for the Majorana fermions. More precisely, in terms of the two Majora components $\chi_{1,2}$, the bosonization map reads
\be
\chi_1+i\chi_2 = e^{iX} \quad , \quad \overline\chi_1+i\overline\chi_2 = e^{i\overline X}
\ee
hence, recalling that $2\phi=X+\overline X$ and $\tilde\phi=X-\overline X$, one finds
\be
\cos2\phi\sim \chi_1\overline\chi_1-\chi_2\overline\chi_2 \quad , \quad \cos\widetilde\phi\sim \chi_1\overline\chi_1+\chi_2\overline\chi_2 \,.
\ee
These considerations, when applied to the theory at hand, imply 
\bea
& \cos(\widetilde\phi_1)+ \cos(2\phi_2) \quad\rightarrow\quad \chi_1\overline\chi_1+\chi_2\overline\chi_2+\chi_3\overline\chi_3 - \chi_4\overline\chi_4\\
&\cos(\widetilde \phi_1)+\cos(2\phi_2)+\cos(\widetilde \phi_2) + \cos(2\phi_1) \quad \rightarrow\quad \chi_1\overline\chi_1+\chi_3\overline\chi_3\,.
\eea
Therefore the IR theory of the RG flow triggered by $R^{1}_{1/2}$ is the bosonized version of a trivial theory, \ie{} a $\bZ_2 \times \bZ_2$  gauge theory, compatible with the anomaly free $TY(\bZ_2\times \bZ_2)$ symmetry and an anomalous $\bZ_2$ invertible symmetry. On the other hand, the RG triggered by the sum $R^{1}_{1/2}+R^{2}_{1/2}$ leads in the IR to two Majorana fermions in the fermionized theory. Upon bosonizing each of them separately, we land on the $c=1$ ${\rm Ising}^2$ theory. In the latter theory, the two preserved non-invertible defects are naturally realized by the Kramers-Wannier duality defect preserved by each factor.

\subsection{The bi-critical point}
Let us now delve into the more interesting case of the bi-critical point $(\tau,\rho) = (\omega,\alpha)$ where $\omega=e^{2\pi i /3}$ and $\alpha = -\frac{1}{2} + i 3\frac{\sqrt{3}}{2}$. Note that this theory cannot be regarded as a product of two $c=1$ theories. As discussed previously, we take the theory $(\tau,\rho) = (\omega,\omega)$ with enahnced global symmetry $SU(3)\times \overline{SU(3)}$ as the starting point for the analysis. In this theory, the holomorphic decomposition of the fields takes the form
\begin{align}
    \phi^1 = \frac{1}{\sqrt{2}}\left( X^1 + \overline X^1 \right) + \frac{1}{\sqrt{6}} \left( X^2 + \overline X^2 \right) \; , \quad 
    \phi^2 = \sqrt{\frac{2}{3}}\left( X^2 + \overline X^2 \right) \; ,
\end{align}
and enjoys the shift and rotational  $\mathbb{Z}_3$ symmetries determined by the following action on the chiral fields 
\begin{align}\label{eq:Z3 autom}
{\mathbb Z}_3^s \; :& \quad \left(X^1, \; X^2\right) \, \to \, \left(X^1 + 2\frac{\sqrt{2}}{3}\pi , \; X^2 \right) \; , \nonumber \\
& \quad \left(\overline X^1, \; \overline X^2\right) \, \to \, \left(\overline X^1 + \frac{\sqrt{2}}{3}\pi , \; \overline X^2 - \sqrt{\frac{2}{3}}\pi \right)  \; , \nonumber \\[5pt]
{\mathbb Z}_3^{orb} \; :& \quad \left(X^1, \; X^2\right) \, \to \, \left(\cos{\frac{2\pi}{3}} X^1 + \sin{\frac{2\pi}{3}} X^2, \; -\sin{\frac{2\pi}{3}} X^1 + \cos{\frac{2\pi}{3}} X^2\right)  \; ,\\
& \quad \left(\overline X^1, \; \overline X^2\right) \, \to \, \left(\cos{\frac{2\pi}{3}} \overline X^1 + \sin{\frac{2\pi}{3}}\overline X^2, \; -\sin{\frac{2\pi}{3}}\overline X^1 + \cos{\frac{2\pi}{3}}\overline X^2\right)  \; . \nonumber
\end{align}
and similarly for the antiholomorphic fields. Again, these two $\bZ_3$ actions can be mapped to each other by means of global $SU(3)\times \overline{SU(3)}$ rotations. Consequently, the corresponding orbifolds lead to two equivalent descriptions of the same theory, leading to a bi-critical point. 

In order to move to the bi-critical point, we gauge the $\mathbb{Z}_3^s$ symmetry. This operation implements the following map of charges
\begin{align}\label{eq: gauging from su3 to 2crit}
    n'_1 = \frac{2}{3}n_1 + \frac{1}{3} n_2 \; , \quad n'_2 = -\frac{1}{3}n_1 + \frac{1}{3}n_2 \; , \quad w'_1 = w_1 + w_2 \; , \quad w'_2 = - w_1 + 2 w_2 \; 
\end{align}
and $(\tau, \rho)=(\omega, \omega) \to (\omega, \omega/3)$,
subsequently performing the $SL(2,\mathbb{Z})$ transformation  $T_{\rho}^{-2}S_{\rho}$, finally landing at the point $(\tau, \rho)=(\omega, \alpha)$. For these two operations together, the overall map of the charges results
\begin{align}\label{eq:Z3mapSU3bicrit}
    n'_1 &= \frac{4}{3}n_1 + \frac{2}{3}n_2 + w_1 - 2 w_2 \; , \quad n'_2=-\frac{2}{3}n_1 + \frac{2}{3}n_2 + w_1 + w_2 \; , \nonumber \\[5pt]
    w'_1 &= \frac{1}{3}n_1 - \frac{1}{3}n_2 \; , \quad w'_2 = \frac{2}{3} n_1 + \frac{1}{3}n_2 \; . 
\end{align}
as it can be easily verified by conjugating the generalized metric with the corresponding matrix\footnote{Recall that, as described in section \ref{sec: c=2}, for a map implementing $({\bf n},{\bf w})\to (M^{-1})^{\intercal}({\bf n},{\bf w})$, the generalized metric transforms accordingly as $\cE\to (M^{-1})^{\intercal}\cE M^{-1}$.}.

\subsubsection{Duality symmetry and fusion category}\label{sec: dualities 2-crit}
As shown in section \ref{sec: c=2}, a generic RCFT can always be connected to a factorized point via the composition of dualities and gaugings on each factor. In this case, we choose to move to the factorized point at $(\tau,\rho)=(i\sqrt{3},i3\sqrt{3})$ by means of the following naminpulation
\be\label{eq: move 2c to fac}
T_{\rho}T_{\tau}\sigma_{\widetilde M  = 2} \, : \,(\omega,\alpha) \, \to \, (i\sqrt{3},i3\sqrt{3})\,.
\ee
Duality symmetries pertaining to the bi-critical theory can be found by composing the above map with the duality symmetries of the factorized point. Indeed, following the procedure described in section \ref{sec: c=2}, we find the following defects and their associated matrices in $O(2,2,\bQ)$
\be\label{eq:D1 D2}
\begin{split}
   & \mathcal{D}_1 \to \sD_1=\mat{0&\frac{1}{3}&\frac{1}{3}&0\\0&1&0&0\\3&-1&0&0\\-1&\frac{1}{3}&\frac{1}{3}&1} \quad,\quad 
    \mathcal{D}_{2} \to \sD_2=\frac{1}{9}\mat{-1&2&-4&-2\\-2&1&-2&-4\\-28&14&-1&-2\\14&-28&2&1}\,.
\end{split}
\ee
Another useful point at which additional duality symmetries become more manifest is the diagonal RCFT $(\tau = \omega, \rho = 3 \omega)$, which is connected to the bi-critical point via $T_{\rho}^{-1}$
\be\label{eq:Trho}
T_{\rho}^{-1} \, : \, (\omega,\alpha) \, \to \, (\omega,3\omega)\,.
\ee
At this point we have the duality symmetry coming from composing mirror symmetry $M$ and a gauging of $\bZ_3$ shift symmetry. However, this can be checked to coincide with $\cD_1$ when written in the bi-critical point. In addition, there is a new duality symmetry coming from the duality matrix $IMS_{\rho}S_{\tau}$. When mapped back to the bi-critical point, the action of such a symmetry becomes implemented by the following matrix
\be\label{eq:D3}
\cD_{3 } \to \sD_3=\mat{1&0&0&0\\ \frac{1}{3}&0&0&-\frac{1}{3}\\ -\frac{1}{3}&1&1&\frac{1}{3}\\ 1 &-3& 0 & 0}\in O(2,2,\bQ)\,,
\ee
Accordingly, it can be easily verified that these three operations leave the generalized metric invariant, that is
\be
(\sD_a^{-1})^\intercal \cE \sD_a^{-1}=\cE \quad , \quad a=1,2,3 \; .
\ee
Note that we are omitting the description of the actions implemented by the above defects in terms of a string of dualities and topological manipulations, just because these are quite involved and barely illuminating.   

We will focus on the three duality symmetries just described. Of course, we will identify new ones when consider their fusion products below. Before delving into those aspects, let us first identify the discrete gaugings leading to the self-dualitites implied by these non-invertible defects. We do that by following the arguments in Sec. \ref{sec:discretegauging}. We can interpret $\cD_{1,2,3}$ as performing the gauging of the following subgroups of the global symmetry
\bea
(\cD_1)& \quad \bZ_3^{(1)} \quad : \quad & V_{\bf n,\bf w} \rightarrow e^{\frac{2\pi i k}{3}(n_1+w_2)}V_{\bf n ,\bf w}\nonumber\\
(\cD_2)& \quad \bZ_3^{(2)} \times \bZ_9 \quad : \quad & V_{\bf n,\bf w} \rightarrow e^{\frac{2\pi i k_1}{9}(n_1-n_2+4(w_1+w_2))}e^{\frac{2\pi i k_2}{3}(n_2-w_1)}V_{\bf n ,\bf w}\,,\\
(\cD_3)& \quad \bZ_3^{(2)} \quad : \quad &  V_{\bf n,\bf w} \rightarrow e^{\frac{2\pi i k_2}{3}(n_2-w_1)}V_{\bf n ,\bf w}\,.\nonumber
\eea
These expressions should not be confused with the action of the corresponding non-invertible defects on the vertex operators. The latter will be implemented by the corresponding map of charges implied by the matrices in \eqref{eq:D1 D2}, \eqref{eq:D3}, together with the associated quantum dimensions and $\bZ_2$ phases, as discussed around \eqref{eq: action on ops}. We will be more explicit about these action when testing the defects against the modular bootstrap in the next subsection.  

We can now discuss the fusion between those defects. As matrices acting on the generalized metric, all $\cD_{1,2,3}$ square to the identity matrix. Therefore we get
\be
\cD_{1}\times \cD_{1} = \sum_{a \in\mathbb{Z}_3^{(1)}} \eta_a \quad ,\quad\cD_{2}\times \cD_{2} = \sum_{b \in\mathbb{Z}_3^{(2)}\times \bZ_9} \eta_b \quad ,\quad \cD_{3}\times \cD_{3} = \sum_{c \in\mathbb{Z}_3^{(2)}} \eta_c \,,
\ee
so that they separately generate a Tambara-Yamagami category with a $\bZ_2$ grading.
Moreover fusion between different defects generates new duality defects, arising from self-duality under gauging subgroups of $\bZ_3^{(1)}\times \bZ_3^{(2)}\times \bZ_9$. For instance 
\be\label{eq: fusion M tau rho}
\cD_1 \times \cD_{2} = \cC_0^{(1,2)} \cD_{4}
\ee
where $\cD_{4}$ is the duality defect constructed by gauging on half-space the symmetry $\bZ^{(1,2)D}_3 \times \bZ_9$, where 
\be
\bZ_3^{(1,2)D} \quad : \quad V_{\bf n,\bf w} \rightarrow e^{\frac{2\pi i k}{3}(n_1-n_2+w_1+w_2)}V_{\bf n ,\bf w}\,.
\ee
Comparing the symmetries on the l.h.s. and on the r.h.s. of the fusion \eqref{eq: fusion M tau rho} we can derive the form of $\cC_0^{(1,2)}$
\be
\cC_0^{(1,2)} = \sum_{a \in \bZ_3^{(1,2)AD}}\eta_a
\qquad, \qquad
\bZ_3^{(1,2)AD} = \frac{\bZ_3^{(1)}\times \bZ_3^{(2)}}{\bZ_3^{(1,2)D}}\,.
\ee
Similarly one can compute the other fusion rules, like
\be
\begin{split}
 \cD_{2} \times \cD_{3} = \cD_{5}\sum_{b \in \bZ_3^{AD}}\eta_b\,,
\end{split}
\ee
where $\cD_{5}$ comes from the half-gauging of 
\be
\begin{split}
    \bZ_3^{(2)}\times \bZ_9\supset\bZ_9^{D} \quad : \quad V_{\bf n,\bf w} \rightarrow e^{\frac{2\pi i k}{9}(-n_1+4n_2+2w_1+5w_2)}V_{\bf n ,\bf w}\,,
\end{split}
\ee
and
\be
\bZ_3^{AD} = \frac{\bZ_3^{(2)}\times \bZ_9}{\bZ_9^{D} }\,.
\ee
\subsubsection*{Self-duality under gauging}
The full set of non-invertible symmetries together imply that the RCFT at the bi-critical point is self-dual under the gauging of 
\be\label{eq: invertible 39 symm}
\begin{split}
     \bZ_3^{(1)} \quad &: \quad  V_{\bf n,\bf w} \rightarrow e^{\frac{2\pi i k}{3}(n_1+w_2)}V_{\bf n ,\bf w} \, ,\\
    \bZ_3^{(2)} \quad &: \quad  V_{\bf n,\bf w} \rightarrow e^{\frac{2\pi i k_2}{3}(n_2-w_1)}V_{\bf n ,\bf w} \, ,\\
    \bZ_9 \quad &: \quad  V_{\bf n,\bf w} \rightarrow e^{\frac{2\pi i k_1}{9}(n_1-n_2+4(w_1+w_2))}V_{\bf n ,\bf w}\,.
\end{split}
\ee
By using the closed form of the torus partition function for RCFTs, we proceed to explicitly check such self-dualities, thus confirming the above findings.

Following the general construction reviewed in App. \ref{app:PartFunctionc2}, the chiral algebra at the bi-critical point can be written as
\be\label{eq:bi-crit K chiral alg}
u(1)^2_{K}\times u(1)^2_{K} \quad , \quad K=\left(\begin{array}{cc}
6 & -3 \\ -3 & 6
\end{array}\right)\;.
\ee
Furthermore, the partition function for this RCFT  is given by a (diagonal) modular invariant combination of the chiral algebra characters 
\be
Z=\sum_{{\boldsymbol\lambda}\in{\cal D}}\chi_{\boldsymbol{\lambda}}\overline{\chi}_{\overline{\boldsymbol{\lambda}}} \, ,
\ee
where
\be
\chi_{\boldsymbol{\lambda}}=\frac{1}{\eta(\tau)^2}\sum_{{\bf l}\in{\mathbb Z}^2} q^{\frac12|{\boldsymbol\lambda}+K{\bf l}|^2_{K^{-1}}}
\quad , \quad 
\overline{\chi}_{\overline{\boldsymbol{\lambda}}}=\frac{1}{\eta(\overline\tau)^2}\sum_{{\bf r}\in{\mathbb Z}^2} \overline{q}^{\frac12|{\overline{\boldsymbol\lambda}}+K{\bf r}|^2_{K^{-1}}} \, .
\ee
The vector ${\boldsymbol\lambda}$ labelling each representation belongs to the lattice
\be
{\cal L}\equiv \left\lbrace {\boldsymbol\lambda}\in {\mathbb Z}^2 \, , \, {\boldsymbol\lambda}\sim {\boldsymbol\lambda}+K{\bf v} \, , \, {\bf v} \in {\mathbb Z}^2
 \right\rbrace \, .
\ee
The fact that the theory corresponds to a diagonal RCFT becomes evident by noticing that there is a duality frame in which $\rho=3\tau$ (see equation \eqref{eq:Trho}).

It turns to be useful to rewrite the action of the ${\mathbb Z}_3^{(1)}$ over the elements comprised in the different characters. This can be done straightforwardly by using the expressions \eqref{eq:gen charge map} to write the character labels in terms of the global charges. More precisely
\be
{\mathbb Z}_3^{(1)} \, : \,\, \chi_{\boldsymbol{\lambda}}\overline{\chi}_{\overline{\boldsymbol{\lambda}}} \, \to \, e^{i\frac{2\pi}{3}(\lambda_1+\overline\lambda_1)}
\chi_{\boldsymbol\lambda}\overline{\chi}_{\overline{\boldsymbol\lambda}}=e^{2\pi i\left({\boldsymbol\lambda}+\overline{\boldsymbol\lambda}\right)^T K^{-1}{\bf v}_3}
\chi_{\boldsymbol\lambda}\overline{\chi}_{\overline{\boldsymbol\lambda}} \, ,
\ee
with the generator of the ${\mathbb Z}_3$ sublattice ${\bf v}_3\equiv (2,-1)^T$. Armed with these tools, we can write the partition function twisted by $(\eta^a,\eta^b)$ along the temporal and spatial cycles respectively, where $\eta$ denotes the generator of ${\mathbb Z}_3^{(1)}$ 
\be
Z_{(\eta^a,\eta^b)}=\frac{1}{|{\cal L}|}\sum_{{\boldsymbol\lambda},{\boldsymbol\mu},\overline{{\boldsymbol\mu}}\in{\cal L}} e^{i4\pi a {\boldsymbol\lambda}^TK^{-1}{\bf v_3}}e^{i2\pi b({\boldsymbol\mu}+\overline{\boldsymbol\mu})^T K^{-1}{\bf v_3}}e^{-i2\pi {\boldsymbol\lambda}^TK^{-1}({\boldsymbol\mu}-\overline{\boldsymbol\mu})}\chi_{{\boldsymbol\mu}}\overline{\chi}_{\overline{\boldsymbol\mu}} \; .
\ee
Summing over ${\boldsymbol\lambda}$ and using the orthogonality relation \eqref{eq:orthogonal} leads to
\be
Z_{(\eta^a,\eta^b)}=\sum_{{\boldsymbol\mu}\in{\cal L}}e^{i4\pi b({\boldsymbol\mu}-a{\bf v_3})^TK^{-1}{\bf v_3}}\chi_{{\boldsymbol\mu}}\overline{\chi}_{{\boldsymbol\mu}-2a{\bf v_3}}\;.
\ee
The partition function of the gauged theory then obtains
\be
\frac13\sum_{a,b\in {\mathbb Z}_3}Z_{(\eta^a,\eta^b)} \; .
\ee
Now, the sum over $b$ imposes the constraint $a=2\mu_1$ mod 3, leading to
\be
\frac13\sum_{a,b\in {\mathbb Z}_3}Z_{(\eta^a,\eta^b)}=\sum_{{\boldsymbol\mu}\in{\cal L}}\chi_{{\boldsymbol\mu}}\overline{\chi}_{C{\boldsymbol\mu}}=\sum_{{\boldsymbol\mu}\in{\cal L}}\chi_{{\boldsymbol\mu}}\overline{\chi}_{{\boldsymbol\mu}} \quad , \quad C=\left(\begin{array}{cc}
-1 & 0 \\
1 & 1
\end{array}\right) \, ,
\ee
hence establishing the self-duality of the theory under gauging ${\mathbb Z}_3^{(1)}$. In the last equality, we used that the matrix $C$ preserves the lattice pairing, namely
\be
C^TK^{-1}C =K^{-1}\,.
\ee
A completely analogous computation shows that the theory at $(\tau,\rho)=(\omega,\alpha)$ is self-dual upon gauging ${\mathbb Z}_3^{(2)}$ in \eqref{eq: invertible 39 symm}. 

For sake of completeness, let us briefly comment on the computation that shows the self-duality of the theory under gauging the ${\mathbb Z}_9$ in \eqref{eq: invertible 39 symm}. Its action on the characters is
\be
{\mathbb Z}_9 \, : \,\, \chi_{\boldsymbol\lambda}\overline{\chi}_{\overline{\boldsymbol\lambda}} \, \to \, e^{i\frac{2\pi}{9}(2\lambda_1+\lambda_2-\overline\lambda_1-2\overline\lambda_2)}
\chi_{\boldsymbol\lambda}\overline{\chi}_{\overline{\boldsymbol\lambda}}=e^{2\pi i\left({\boldsymbol\lambda}^TK^{-1}{\bf v}_9-\overline{\boldsymbol\lambda}^TK^{-1}\overline{\bf v}_9\right)}
\chi_{\boldsymbol\lambda}\overline{\chi}_{\overline{\boldsymbol\lambda}} \, ,
\ee
with ${\bf v}_9\equiv(1,0)^T$ and $\overline{\bf v}_9\equiv(0,1)^T$. Therefore the twisted partition function formally reads
\be
Z_{(\eta^a,\eta^b)}=\frac{1}{|{\cL}|}\sum_{{\boldsymbol\lambda},{\boldsymbol\mu},\overline{{\boldsymbol\mu}}\in{\cL}} e^{i2\pi a {\boldsymbol\lambda}^TK^{-1}({\bf v}_9-\overline{\bf v}_9)}e^{i2\pi b\left({\boldsymbol\mu}^TK^{-1}{\bf v}_9-\overline{\boldsymbol\mu}^T K^{-1}\overline{\bf v}_9\right)}e^{-i2\pi {\boldsymbol\lambda}^TK^{-1}({\boldsymbol\mu}-\overline{\boldsymbol\mu})}\chi_{{\boldsymbol\mu}}\overline{\chi}_{\overline{\boldsymbol\mu}} \, .
\ee
The sum over ${\boldsymbol\lambda}$ now sets $\overline{\boldsymbol\mu}={\boldsymbol\mu}-a({\bf v}_9-\overline{\bf v}_9)$. Finally, taking the orbifold and summing over $b\in {\mathbb Z}_9$ fixes $a=\mu_1-\mu_2$ mod 9 hence proving the claimed self-duality
\be
\frac19\sum_{a,b\in {\mathbb Z}_3}Z_{(\eta^a,\eta^b)}=\sum_{{\boldsymbol\mu}\in{\cL}}\chi_{{\boldsymbol\mu}}\overline{\chi}_{C'{\boldsymbol\mu}}=\sum_{{\boldsymbol\mu}\in{\cL}}\chi_{{\boldsymbol\mu}}\overline{\chi}_{{\boldsymbol\mu}} \quad , \quad C'=\left(\begin{array}{cc}
0 & 1 \\
1 & 0
\end{array}\right) \, .
\ee
where we made use of the fact that the matrix $C'$ also preserves the lattice pairing.

\subsubsection{Modular bootstrap at the bi-critical point}

As with the previous example, we now proceed to run the modular bootstrap analysis to confirm that the symmetry operations introduced above define consistent topological defects. 

Let us begin with ${\cal D}_1$ which, by plugging its matrix representation $\sD_1$ into \eqref{eq: general charge map}, leads to the following map of the $U(1)$ charges
\be\label{eq:D1 charge map}
n_1\to 3w_1-w_2 \,\, , \,\, n_2\to \frac{n_1}{3}+n_2-w_1+\frac{w_2}{3} \,\, , \,\,
w_1\to \frac{n_1+w_2}{3} \,\, , \,\, w_2\to w_2 \, .
\ee
Before continuing, let us determine the phase associated to the action of ${\cal D}_1$ on vertex operators. As explained in Appendix \ref{App:somedetails}, mutual locality of correlation functions under the action \eqref{eq:D1 charge map} demands
\begin{align}\label{eq:D1 phase}
\alpha_1({\bf n}+{\bf n}',{\bf w}+{\bf w}')+&\alpha_1({\bf n},{\bf w})+\alpha_1({\bf n}',{\bf w}')=\\
&\frac{n_1+w_2}{3}(3w'_1-w'_2)+\frac{n'_1+w'_2}{3}(3w_1-w_2) -w_1w_2'-w_1'w_2\,\, {\rm mod} \,\, 2 \, .\nonumber
\end{align}
Note that, in order to recast the phase to the above form, we imposed $n_1+w_2\in 3{\mathbb Z}$ and $n'_1+w'_2\in 3{\mathbb Z}$, hence rendering the resulting vertex operators genuine. As such, there are multiple solutions, each of them differing by stacking ${\cal D}_1$ with a particular ${\mathbb Z}_2$ subgroup of the invertible symmetry. For a reason that will be clear momentarily, we choose the following solution
\be\label{eq:D1 phase solution}
\alpha_1({\bf n},{\bf w})=\frac{n_1+w_2}{3}(3w_1-w_2)-w_1(w_1+w_2)
\ee
Having fixed the phase in the action of the ${\cal D}_1$ defect, we proceed to run the modular bootstrap. The map \eqref{eq:D1 charge map} translates in the momentum basis to
\be
(p_1,p_2,\overline p_1,\overline p_2) \, \to \, (p_1,p_2,-\overline p_1,\overline p_2)
\ee
hence implying that only states with $\overline p_1=0$ contribute to the twisted partition function $Z_{(1,{\cal D}_1)}$. In particular, this implies $n_1=3w_1-w_2$. Moreover, \eqref{eq:D1 phase solution} trivializes in ${\mathbb Z}_2$ for these values of the charges, which is the main reason for choosing this particular solution. 
In terms of the chiral algebra characters, and using the conversion formulas \eqref{eq:gen charge map}, the states contributing to the twisted partition function have ${\boldsymbol \lambda}=(0,\lambda_2)^T$ and ${\bf r}=(r,2r)^T$ with $r\in{\mathbb Z}$. The basis of independent right moving characters then reads
\be
{\boldsymbol \lambda}+K\cdot{\bf r}=\left(0\, , \, \lambda_2 + 9r\right)^T \,.
\ee
Therefore, the independent values of $\lambda_2$ span a ${\mathbb Z}_9$. Turning to the holomorphic blocks, the scaling dimensions are given by (${\bf l}=(l_1,l_2)^T$)
\be\label{eq: K diagonal}
\frac12({\boldsymbol \lambda}+K{\bf l})^TK^{-1} ({\boldsymbol \lambda}+K{\bf l}) =\frac{3}{4}\left(l_++\frac{\lambda_2}{3}\right)^2 + \frac{9}{4}\left(l_-+\frac{\lambda_2}{9}\right)^2 \; . 
\ee
where we diagonalized the matrix $K$ in terms of $l_+
\equiv l_1+l_2$ and $l_-
\equiv -l_1+l_2$. Consistently with the fact that the theory at the bi-critical point $(\tau,\rho)=(\omega,\alpha)$ does not factorize, the $l_\pm$ variables are not free within ${\mathbb Z}$. In particular, they must satisfy either $(l_+, l_-)= (2l,2l')$ or $(l_+, l_-)= (2l+1,2l'+1)$ for $l,l'\in{\mathbb Z}$ \footnote{This can be easily verified by noticing that $l_1=\frac{l_+-l_-}{2}$ and $l_2=\frac{l_+-l_-}{2}$ have integer solutions only for $l_\pm=$even or $l_\pm=$odd. }. We then need to sum over both the even and odd sectors. 
In addition, states with an odd number of right-moving oscillator modes come with a $(-1)$ phase, as usual.
Summing over oscillator modes and plugging the solution described above, the twisted partition function results
\begin{align}
Z_{(1,{\cal D}_1)}&=\frac{\sqrt{3}}{|\eta(\tau)|^4} \sum_{\lambda_2\in{\mathbb Z}_9}\left\lbrace\left(\sum_{l\in{\mathbb Z}}q^{3\left(l+\frac{\lambda_2}{6}\right)^2}\sum_{l'\in{\mathbb Z}}q^{9\left(l'+\frac{\lambda_2}{18}\right)^2}+
\sum_{l\in{\mathbb Z}}q^{3\left(l+\frac{\lambda_2+3}{6}\right)^2}\sum_{l'\in{\mathbb Z}}q^{9\left(l'+\frac{\lambda_2+9}{18}\right)^2}\right)
\right. \nonumber\\
&\qquad \qquad \qquad \qquad 
\left. \times\sum_{r}\overline{q}^{9\left(r+\frac{\lambda_2}{9}\right)^2}\right\rbrace\vartheta_4(2\overline\tau) \; ,\\
&=\frac{\sqrt{3}}{|\eta(\tau)|^4} \sum_{\lambda_2\in{\mathbb Z}_9}\left\lbrace\left(\vartheta\left[{\tfrac{\lambda_2}{6}\atop 0}\right](6\tau)\, \vartheta\left[{\tfrac{\lambda_2}{18}\atop 0}\right](18\tau)+
\vartheta\left[{\tfrac{\lambda_2+3}{6}\atop 0}\right](6\tau)\,\vartheta\left[{\tfrac{\lambda_2+9}{18}\atop 0}\right](18\tau)\right) \right. \nonumber \\
&\qquad \qquad \qquad \qquad 
\left. \times \vartheta\left[{\tfrac{\lambda_2}{9}\atop 0}\right](18\overline\tau)\right\rbrace \vartheta_4(2\overline\tau) \; .
\end{align}
Performing the modular $S$ transformation $\tau\to-1/\tau$ and using the transformation properties listed in appendix \ref{app: modular} one gets
\begin{align}\label{eq: D1 twist pf}
Z_{({\cal D}_1,1)}&=\frac{1}{36}\frac{1}{|\eta(\tau)|^4}\sum_{\lambda_2\in{\mathbb Z}_9}\left\lbrace\left(\vartheta\left[{0\atop-\tfrac{\lambda_2}{6}}\right]\left(\tfrac{\tau}{6}\right)\, \vartheta\left[{0\atop-\tfrac{\lambda_2}{18}}\right]\left(\tfrac{\tau}{18}\right)+\vartheta\left[{0\atop-\tfrac{\lambda_2+3}{6}}\right]\left(\tfrac{\tau}{6}\right)\, \vartheta\left[{0\atop-\tfrac{\lambda_2+9}{18}}\right]\left(\tfrac{\tau}{18}\right)\right) \right. \nonumber \\
&\qquad \qquad \qquad \qquad 
\left. \times \vartheta\left[{0\atop-\tfrac{\lambda_2}{9}}\right]\left(\tfrac{\overline\tau}{18}\right)\right\rbrace\vartheta_2\left(\tfrac{\overline\tau}{2}\right) \; ,\\
&\approx \frac{1}{|\eta(\tau)|^4}\left(\overline{q}^{\frac{1}{16}}+4q^{\frac{1}{9}}\overline{q}^{\frac{13}{144}}+2q^{\frac{1}{9}}\overline{q}^{\frac{25}{144}}+2q^{\frac{1}{3}}\overline{q}^{\frac{1}{16}}+\ldots\right) \; . \nonumber
\end{align}
In going to the last line, we performed the sum over $\lambda_2$. As it becomes evident from the first orders in the expansion above, the defect Hilbert space has a consistent expansion in terms of Virasoro characters.

We now proceed to bootstrap the defect ${\cal D}_2$. We first notice that the right hand side of the equation \eqref{eq: c=2 phase} determining the phase $\alpha_2$ is already in $2{\mathbb Z}$ once restricted to genuine operators\footnote{More precisely, ${\cal D}_2$ implements an orbifold by the ${\mathbb Z}_3^{(1)}\times {\mathbb Z}_9$ symmetry in \eqref{eq: invertible 39 symm}. Hence, by restricting to genuine local operators we mean
\be
n_1-n_2+4(w_1+w_2)\in 9{\mathbb Z} \quad , \quad 
n_2-w_1\in 3{\mathbb Z}  \,. \nonumber
\ee}. We can therefore set $\alpha_2({\bf n},{\bf w})=0$. In addition, the action of ${\cal D}_2$ flips the sign of both $p_1$ and $p_2$. The subset of states contributing to $Z_{(1,{\cal D}_2)}$ thus have $p_1=p_2=0$, which in terms of the characters reads $\lambda_1=\lambda_2=l_1=l_2=0$. By manipulations analogous to the ones described above, one obtains
\begin{align}
Z_{(1,{\cal D}_2)}&=\frac{\sqrt{27}}{|\eta(\tau)|^4}\vartheta_4(2\tau)^2\left(\vartheta_3(6\overline\tau)\vartheta_3(18\overline\tau)+\vartheta_2(6\overline\tau)\vartheta_2(18\overline\tau)\right)\nonumber\\
\label{eq:mod boots D2}
\xrightarrow{\tau\to-\frac{1}{\tau}}Z_{({\cal D}_2,1)}& =\frac{1}{4|\eta(\tau)|^4}\vartheta_2\left(\frac{\tau}{2}\right)^2\left(\vartheta_3\left(\frac{\overline\tau}{6}\right)\vartheta_3\left(\frac{\overline\tau}{18}\right)+\vartheta_4\left(\frac{\overline\tau}{6}\right)\vartheta_4\left(\frac{\overline\tau}{18}\right)\right)\nonumber\\
&\approx \frac{1}{|\eta(\tau)|^4}\left(2q^{\frac{1}{8}}+12q^{\frac{1}{8}}\overline q^{\frac{1}{9}}+12q^{\frac{1}{8}}\overline q^{\frac{1}{3}}+\ldots\right) \,. 
\end{align}

As a final example, consider the defect ${\cal D}_5$ which, in particular, implements the orbifold by the ${\mathbb Z}_9$ gauging in \eqref{eq: invertible 39 symm}. For the phase required by mutual locality, we choose the following solution 
\be
\alpha_5({\bf n},{\bf w})=n_2(w_1+w_2)+w_1^2+w_2^2
\ee
which turns out to be trivial when evaluated on the states contributing to the twisted partition function (recall that all possible solutions differ just by stacking with an invertible ${\mathbb Z}_2$ symmetry defect). The action of ${\cal D}_5$ amounts to the following 
\be
{\bf p}\to R\cdot {\bf p} \quad , \quad R=-\frac{1}{2}\left(\begin{array}{cc}1&\sqrt{3}\\ \sqrt{3}&-1\end{array}\right)
\ee
Note that $R^2=1$, hence defining a ${\mathbb Z}_2$ action in momentum space (equivalently, ${\cal D}_5 : {\bf X}\to R\cdot {\bf X}$). In terms of the characters, the states contributing to the trace have $\lambda_2=-2\lambda_1$ $l_1=0$ and their holomorphic scaling dimension read
\be
\frac12 \left|({\boldsymbol{\lambda}+K{\bf l}})\right|^2_{K^{-1}} \,\, \xrightarrow[l_1=0 \, , \, l_2=l] {\lambda_2=-2\lambda_1} \,\, 3\left(l-\frac{\lambda_1}{3}\right)^2 \; .
\ee
It becomes evident from the expression above that the independent representations contributing to the trace are labelled by $\lambda_1$ taking values in ${\mathbb Z}_3$.  
On the other hand, the right moving sector can be dealt with by means of analogous manipulations as in \eqref{eq: K diagonal}. 

In order to perform the sum over oscillator modes, we notice that the matrix $R$ is no more than a rotation in field space, followed by flip of one of the momenta. This can be verified by performing the following change of basis on the holomorphic fields
\be
{\bf X}\, \to \,{\bf X}'= \cB \, {\bf X} \quad , \quad 
\cB =\left(\begin{array}{cc}
\tfrac{\sqrt{3}}{2}& -\tfrac12 \\
\frac12 & \tfrac{\sqrt{3}}{2}
\end{array}\right)
\ee
for which the action of $R$ reduces to $X_1'\to -X_1'$. 
Hence, by going to the appropriate basis, the summand acquire the usual sign factor depending on one of the oscillator numbers. Putting all together we end up with
\begin{align}
Z_{(1,{\cal D}_5)}&=\frac{\sqrt{9}}{|\eta(\tau)|^4} \sum_{\lambda_1\in{\mathbb Z}_3}\left\lbrace\left(\vartheta\left[{-\tfrac{\lambda_1}{6}\atop 0}\right](6\overline\tau)\, \vartheta\left[{-\tfrac{\lambda_1}{6}\atop 0}\right](18\overline\tau)+
\vartheta\left[{\tfrac{-\lambda_1+3}{6}\atop 0}\right](6\overline\tau)\,\vartheta\left[{\tfrac{-\lambda_1+3}{6}\atop 0}\right](18\overline\tau)\right) \right. \nonumber \\
&\qquad \qquad \qquad \qquad 
\left. \times \vartheta\left[{-\tfrac{\lambda_1}{3}\atop 0}\right](6\tau)\right\rbrace \vartheta_4(2\tau)
\end{align}
Finally, upon performing a modular $S$ transformation we obtain the trace over the twisted Hilbert space
\begin{align}\label{eq: D5 modular bootstrap}
Z_{(1,{\cal D}_5)}&=\frac{1}{12|\eta(\tau)|^4} \sum_{\lambda_1\in{\mathbb Z}_3}\left\lbrace\left(\vartheta\left[{0\atop \tfrac{\lambda_1}{6}}\right]\left(\frac{\overline\tau}{6}\right)\, \vartheta\left[{0\atop \tfrac{\lambda_1}{6}}\right]\left(\frac{\overline\tau}{18}\right)+
\vartheta\left[{0\atop \tfrac{\lambda_1-3}{6}}\right]\left(\frac{\overline\tau}{6}\right)\,\vartheta\left[{0\atop \tfrac{\lambda_1-3}{6}}\right]\left(\frac{\overline\tau}{18}\right)\right) \right. \nonumber \\
&\qquad \qquad \qquad \qquad 
\left. \times \vartheta\left[{0\atop -\tfrac{-\lambda_1}{3}}\right]\left(\frac{\tau}{6}\right)\right\rbrace \vartheta_2\left(\frac{\tau}{2}\right)\\
&\approx \frac{1}{|\eta(\tau)|^4}\left(q^{\frac{1}{16}}+2q^{\frac{1}{16}}\overline q^{\frac{1}{9}}+4q^{\frac{7}{48}}\overline q^{\frac{1}{9}}+6 q^{\frac{7}{48}}\overline q^{\frac{1}{3}}+\ldots\right) \,. \nonumber
\end{align}

\subsubsection*{Categorical Data}
As explained in \cite{Thorngren:2021yso} (see also appendix \ref{App:TY}) the computation of the twisted Hilbert space enables to determine the extra categorical data of the Tambara-Yamagami categorical symmetry, \ie{} the non-degenerate bi-character $\chi(a,b)$ and the Frobenius-Shur indicator $\epsilon$. As in the analysis of the modular bootstrap, we focus on the duality defects $\cD_{1,2}$ and $\cD_5$. The data corresponding to the remaining defects can be derived in a similar manner.

Following the analysis reviewed in appendix \ref{App:TY}, the F-symbols of the categorical symmetry imply selection rules for the spins of the states in the defect Hilbert space. In particular we get
\be
e^{4\pi i s}|\psi \rangle = \frac{\epsilon}{\sqrt{\bA}}\sum\limits_{a \in G} \widehat{\eta}_a|\psi\rangle\,,
\ee
where $\widehat{\eta}_a$ are the invertible symmetry operators generating $G \subset U(1)_{\bf n}^2 \times U(1)_{\bf w}^2$ acting on the Hilbert space twisted by $\cD$, with a particular resolution of the $4$-valent junction (see the discussion around Eq. \eqref{eq: spin sel rules})\footnote{The two possible resolutions are related by the bi-character.} and $|\psi \rangle$ are eigenstates of $\widehat{\eta}_a$ in such twisted Hilbert space. By performing F-moves, it is straightforward to check that 
\be
\widehat{\eta}_a \widehat{\eta}_b = \chi(a,b) \widehat{\eta_{ab}}
\ee
so that the group generated by $\widehat{\eta}_a$ is generically an extension of $G$. Let us now start by applying these constraints on $\cD_1$. In this case the abelian group $G = \bZ_3^{(1)}$ and the possible bi-characters are 
\be
\chi_r(a,b) = e^{\frac{2\pi i}{3}rab} \qquad r= \pm 1\,.
\ee
Therefore we get 
\be
e^{4\pi i s} = \epsilon e^{\pm \frac{i\pi}{3}n^2\mp \frac{\pi i}{4}} \qquad r = \pm 1\,,
\ee
which is compatible with the expansion \eqref{eq: D1 twist pf} only for the values $r=1 , \epsilon=1$.

In the case of $\cD_5$, the abelian group is $\bZ_9$ and the possible non-degenerate bi-characters are
\be
\chi_r(a,b) = e^{\frac{2\pi i}{9}rab} \qquad \text{gcd}(9,r)=1\,,
\ee
which imply
\be
e^{4\pi i s} = \epsilon e^{r \frac{\pi in^2}{9}}e^{-r\frac{\pi i}{4}}\,.
\ee
Such spin selection rules are compatible with \eqref{eq: D5 modular bootstrap} if $\epsilon=1,r=-1$.

The combined duality defect $\cD_2$, coming from the self-duality under $\bZ_3\times \bZ_9$ gauging, will therefore have a diagonal bi-character
\be
\chi(a,b) = \exp{\left[{2\pi i \mat{a_1 & a_2} \chi
\mat{b_1 \\ b_2}} \right]}\quad,\quad \chi = \mat{1/3 & 0 \\ 0 & -1/9}\,,
\ee
where $\mathbf{a},\mathbf{b} \in \bZ_3\times \bZ_9 $ and trivial Frobenius-Shur indicator.

\subsubsection{Marginal and relevant deformations at the bi-critical point}

As in the easier case of the quadri-critical point, in order to construct the marginal deformations spanning the orbifold branch in the bi-critical point, it is instructive to analyze symmetries and marginal operators of the $SU(3)$ enhanced symmetry point $(w,w)$. Following the discussion done in appendix~\ref{app:enhancedc2}, we find the following holomorphic currents 
\begin{align}
    I_{\pm} &= \pm V_{\pm1,\mp1,\pm1,0} = \pm e^{\pm i \sqrt{2}X^1} \; , \quad \overline I_{\pm} = \pm V_{\pm1,0,\mp1,0} = \pm e^{\pm i \sqrt{2} \, \overline X^1} \nonumber \\[5pt]
    K_{\pm} &= \mp V_{\pm1,0,\pm1,\pm1} = \mp e^{\pm i \frac{1}{\sqrt{2}}X^1 \pm i \sqrt{\frac{3}{2}}X^2} \; , \quad \overline K_{\pm} = \mp V_{0,\pm1,\mp1,\mp1} = \mp e^{\pm i \frac{1}{\sqrt{2}} \overline X^1 \pm i \sqrt{\frac{3}{2}} \overline X^2} \; , \nonumber \\[5pt]
    U_{\pm} &= \pm V_{0,\pm1,0,\pm1} = \pm e^{\mp i \frac{1}{\sqrt{2}}X^1 \pm i \sqrt{\frac{3}{2}}X^2} \; , \quad \overline U_{\pm} = \pm V_{\mp1,\pm1,0,\mp1} = \pm e^{\mp i \frac{1}{\sqrt{2}} \overline X^1 \pm i \sqrt{\frac{3}{2}} \overline X^2} \; , \nonumber \\[5pt]
    J^3 &= i \sqrt{2} \partial X^1 \; , \quad  \overline J^3 = i \sqrt{2}\, \overline \partial \, \overline X^1 \nonumber \\[5pt]
    J^8 &= i \sqrt{2} \partial X^2 \; , \quad  \overline J^8 = i \sqrt{2}\, \overline \partial \, \overline X^2 \; ,
\end{align}
which span the $su(3)$ algebra, along with a similar set of antiholomorphic currents for $\overline{su(3)}$. The two relevant $\mathbb{Z}_3$ symmetries are \cite{Dulat:2000xj}: 
\begin{align}
    \mathbb{Z}_3^{s} \; :& \qquad I_{\pm} \; \to \; \omega^{\pm 2} I_{\pm} \; , \quad K_{\pm} \; \to \; \omega^{\pm 1} K_{\pm} \; , \quad U_{\pm} \; \to \; \omega^{\pm 2} U_{\pm} \; , \nonumber \\[5pt]
    \mathbb{Z}_3^{orb} \; :& \qquad I_{\pm} \; \to \; U_{\pm} \; , \quad K_{\pm} \; \to \; I_{\mp} \; , \quad U_{\pm} \; \to \; K_{\mp} \; .
\end{align}
At this point, the various marginal operators are all related to $\{J_3\overline{J}_3, J_3 \overline{J}_8, J_8 \overline{J}_3,J_8\overline{J}_8\}$ by means of the enhanced $su(3)$ chiral algebra, and the invariant combinations under both $\mathbb{Z}_3^{s}$ and $\mathbb{Z}_3^{orb}$ are
\begin{align}
    \mathcal{O}_1 &= \mathcal{O}_A + \mathcal{O}_B + \mathcal{O}_C \; , \nonumber \\[5pt]
    \mathcal{O}_2 &= \mathcal{O}_{A\omega} + \mathcal{O}_{B\omega} + \mathcal{O}_{C\omega} \; 
\end{align}
where
\begin{align}
    \mathcal{O}_A &= I_+ \overline{I}_+ + U_+ \overline{U}_+ + K_-\overline{K}_- + h.c. = V_{2,-1,0,0} + V_{-1,2,0,0} + V_{1,1,0,0} + h.c. \; , \nonumber \\[5pt]
    \mathcal{O}_B &= I_+ \overline{K}_- + U_+ \overline{I}_+ + K_-\overline{U}_+ + h.c. = V_{1,-2,2,1} + V_{1,1,-1,1} + V_{2,-1,1,2} + h.c. \; , \nonumber \\[5pt]
    \mathcal{O}_C &= I_+ \overline{U}_+ + U_+ \overline{K}_- + K_-\overline{I}_+ + h.c. = V_{0,0,1,-1} + V_{0,0,1,2} + V_{2,0,0,1} + h.c.\; , \nonumber \\[5pt]
   \mathcal{O}_{A\omega} &= \omega^2 I_+ \overline{I}_+ + \omega U_+ \overline{U}_+ + K_-\overline{K}_- + h.c. \; , \nonumber \\[5pt]
    \mathcal{O}_{B\omega} &= \omega I_+ \overline{K}_- + U_+ \overline{I}_+ + \omega^2 K_-\overline{U}_+ + h.c. \; , \nonumber \\[5pt]
    \mathcal{O}_{C\omega} &= I_+ \overline{U}_+ + \omega^2 U_+ \overline{K}_- + \omega K_-\overline{I}_+ + h.c. \;.
\end{align} 
Finally, at the bi-critical point this set of operators is mapped to 
\begin{align}
    \mathcal{O'}_A &= V_{2,-2,1,1} + V_{0,2,-1,0} + V_{2,0,0,1} + h.c. \; , \nonumber \\[5pt]
    \mathcal{O'}_B &= V_{0,1,1,0} + V_{1,0,0,-1} + V_{-1,1,1,1} + h.c. \; , \nonumber \\[5pt]
    \mathcal{O'}_C &= V_{3,0,0,0} + V_{3,-3,0,0} + V_{0,3,0,0} + h.c. \; ,
\end{align}
and similarly for $\mathcal{O'}_{A\omega},\mathcal{O'}_{B\omega},\mathcal{O'}_{C\omega}$. Therefore we have the two exactly marginal operators $\mathcal{O'}_1,\mathcal{O'}_2$ that span the $\mathbb{Z}_3^{orb}$ branch. However in this case all the non-invertible duality symmetries are broken by all these marginal deformations.

\subsubsection*{Relevant deformations}
The relevant operators at the bi-critical point are the following:
\begin{center}
\begin{tabular}{ |m{2cm}||m{11cm} | } 
 \hline
 $(h,\overline h)$ & $(\mathbf{n},\mathbf{w})$ \\ 
  \hline
$ \left( \frac{1}{9} \, , \, \frac{1}{9} \right)$& $(\pm1,0,0,0) \;,\;(0,\pm 1,0,0) \; , \; (\pm 1,\mp 1,0,0)$ \\ [1ex] 

  $\left( \frac{1}{3} \, , \, \frac{1}{3} \right) $& $(\pm 1,\pm 1,0,0) \;,\; (\pm 2 , \mp 1,0,0)\;,\;(\pm1,\mp2,0,0)$ \\ [1ex] 

 $ \left( \frac{4}{9} \, , \, \frac{4}{9} \right) $& $(\pm 2,0,0,0)\;,\;(0,\pm 2 ,0,0)\;,\;(\pm2,\mp2,0,0) $ \\ [1ex]  
 $ \left( \frac{7}{9} \, , \, \frac{7}{9} \right) $& $(0,0,0,\pm1)\,,\,(\pm1,0,0,\pm1)\, , \, (0,\pm1,\mp1,0)\, , \,(0,0,\pm1,\pm1)$  
 $(0,0,\pm1,0)\, , \, (\pm1,\mp1,\pm1,\pm1)\, , \,(\pm2,\pm1,0,0)\, , \,(\pm1,\pm2,0,0)$ $(\pm3,\mp1,0,0)\, , \,(\pm1,\mp3,0,0)\, , \,(\pm3,\mp2,0,0)\, , \,(\pm2,\mp3,0,0)$\\ [1ex] 
 \hline
\end{tabular}
\end{center}
As in the quadri-critical point, we can find combinations that preserve some non-invertible duality symmetry, therefore leading to interesting constraints on the corresponding RG flow. In this case however, it is more challenging to determine the IR theory resulting from these duality preserving RG flows. Therefore all the possible symmetry constraints which we will describe in the following may lead to useful predictions for the low energy dynamics.

As before, let us give some representative examples which will serve to highlight some interesting features regarding these RG flows. A simple relevant deformation which preserves $\cD_1$ is
\be
R_{1/9} \sim \cos(\phi_2)\,.
\ee
Such a deformation preserves $U(1)_{n_1} \times U(1)_{w_1}\times U(1)_{w_2}$, with the first two factors participating in a continuous mixed 't Hooft anomaly; therefore we conclude that the IR theory is gapless. Because of the presence of $\cD_1$ we also deduce that such gapless theory must enjoy a non-invertible TY$(\bZ_3)$ symmetry. A very natural candidate for this IR theory is therefore the $R=\sqrt{3}$ $c=1$ CFT. However, as opposite to the quadri-critical case, since the bi-critical point is not factorized, we cannot immediately the IR fate of this process. In order to determine it, we can however perform a topological manipulation which brings the theory to a factorized point, perturbed by a genuine relevant operator. As explained around \eqref{eq: move 2c to fac} we have
\be
T_{\rho}T_{\tau}\sigma_{\widetilde M  = 2} \, : \,(\omega,\alpha) \, \to \, (i\sqrt{3},i3\sqrt{3})\,,
\ee
and, by this manipulation, the relevant deformation is mapped to
\be\label{eq: transformed cos}
T_{\rho}T_{\tau}\sigma_{\widetilde M  = 2} \, : \, \cos(\phi_2)|_{(\omega,\alpha)} \, \to \, \cos(2\phi_2)|_{(i\sqrt{3},i3\sqrt{3})}\,.
\ee
Therefore, at $(i\sqrt{3},i3\sqrt{3})$, we find that the sector described by $\phi_2$ flows to a gapped phase with two degenerate vacua, \ie{} the $\bZ_2 \in U(1)_{n_2}$ is spontaneously broken. The reason for this is that the $\bZ_2$ shift symmetry which is preserved by the deformation \eqref{eq: transformed cos} at the factorized point is itself involved in a mixed anomaly with a $\bZ_2$ subgroup of the winding symmetry. We then conclude that the IR theory is described by
\be
(i\sqrt{3},i3\sqrt{3}) + \cos(2\phi_2) \;\xrightarrow[]{\text{RG flow}} \;\left(c=1 \; R=\sqrt{3} \right)\;\times \left(\text{ 2 vacua}\right)
\ee
In order to understand the IR of the bi-critical point we can now perform the inverse topological manipulation 
\be
\sigma_{\widetilde N = 2} T_{\rho}^{-1}T_{\tau}^{-1} (i\sqrt{3},i3\sqrt{3})= (\omega,\alpha)\,
\ee
which maps the two vacua mentioned above to a single trivial ground-state. Therefore we get
\be
 (\omega,\alpha)+ \cos(\phi_2) \;\xrightarrow[]{\text{RG flow}} \;\left(c=1 \; R=\sqrt{3} \right)
\ee
in agreement with the expectations inferred from the symmetries. 

We now discuss a slightly more involved example, constructed by considering the combination of relevant deformations\footnote{Because these three deformation have no the same scaling dimensions, there is a range in the parameter space where $\cos(\phi_2)$ dominates and the IR theory is approximately the one described before. However in a generic point of the phase diagram, both deformations need to be taken into account.}
\be
R_{1/9, 7/9}\sim \cos(\phi_2) + \cos(3\phi_1-\phi_2) + \cos(\widetilde{\phi}_1)\,.
\ee
While $\cos(\phi_2)$ preserves the non-invertible symmetry $\cD_1$, the relevant operator $\cos(3\phi_1-\phi_2) + \cos(\widetilde{\phi}_2)$ preserves only the diagonal symmetry generated by $\widetilde{\cD}_1 \equiv \widetilde{\eta}_1\cD_1$ ($\widetilde{\eta}_1 \in \bZ_2^{(w_2)}$) due to the phase \eqref{eq:D1 phase solution} produced by the action of $\cD_1$. The full set of symmetries preserved by $R_{1/9, 7/9}$ is then 
\bea
&\bZ_3^{(1)}\times U(1)_{w_2}\times \widetilde{\cD}_1 \; \Longrightarrow \; TY(\bZ_3^{(1)}) \times U(1)
\eea
By looking just at the invertible part of the symmetry structure, this RG flow is compatible with a trivially gapped phase. However the presence of an anomalous TY$(\bZ_3)$ implies that the theory is either gapless or gapped with a spontaneoulsy broken non-invertible symmetry \cite{Antinucci:2023ezl, Apte:2022xtu,Cordova:2023bja}.  

\section{Duality symmetries of $SU(3)_1$ WZW}\label{sec: WZW}
We now want to comment about the (non-invertible) symmetry structure of the $c=2$ theory at $(\tau =\omega, \rho = \omega)$, where the chiral algebra enhances to $SU(3) \times \overline{SU(3)}$ (see appendix \ref{app:enhancedc2}) and the theory is equivalent to $SU(3)_1$ Wess-Zumino-Witten model (WZW). As opposed to the $c=1$ case, this theory also enjoys non-invertible duality defects exactly as the other rational points of the toroidal branch\footnote{Also the $SU(2) \times SU(2)$ point enjoys non-invertible duality defects as already emphasized in \cite{Ji:2019ugf}. Here we focus on the $SU(3)$ point since it cannot be expressed as product of $c=1$ theories. We also emphasize that these non-invertible symmetries are not part of the Verlinde lines of WZW since for $SU(N)_1$ all the Verlinde lines are invertible.}. An easy way to express such symmetries is by combining the ones discussed at the bi-critical point in section \ref{sec: dualities 2-crit} with the duality and gauging operations which connect the bi-critical point with the $SU(3)_1$ WZW (see the discussion around \eqref{eq: gauging from su3 to 2crit}). We then find that the non-invertible defect $\cD_1$ leads to the following one at the $SU(3)$ point 
\be
\begin{split}\label{eq: SU3 D1}
    \cD_1^{SU(3)} \, \to \, \sD_1^{SU(3)}=\frac{1}{3}\mat{-1& 2 & -4& -2\\ -2& 1& -2& -4\\ -4& 2& -1& -2\\ 2& -4& 2& 1} \; ,
\end{split}
\ee
where, for sake of simplicity, we chose to display only its representation in terms of an $O(2,2,\bQ)$ matrix. It is straightforward to verify that such matrix leaves invariant the generalized metric at $(\tau,\rho)=(\omega,\omega)$.

By similar methods, we can further evaluate the duality symmetries descending from the non-invertible defects $\cD_2$ and $\cD_3$ at the bi-critical point. However, we find that, at the $SU(3)_1$ WZW theory, these defects reduce to the one in \eqref{eq: SU3 D1} stacked with invertible defects. Explicitly we find
\be
\cD_2^{SU(3)} = -M \cD_1^{SU(3)} \quad , \quad \cD_3^{SU(3)} = M_3\cD_1^{SU(3)}\quad ,\quad M_3 = T_{\tau}^{-1}S_{\tau}T_{\rho}^{-1}S_{\rho} 
\ee
where $M$ is mirror symmetry and both $M$ and $M_3$ are invertible symmetries of this point\footnote{Another invertible symmetries of the $SU(3)$ point are $S_{\rho}T_{\rho}$ and $S_{\tau}T_{\tau}$. It is easy to check that they do note generate other non-trivial non-invertible symmetries when translated to the bi-critical point. Indeed $S_{\tau}T_{\tau}$ is still an invertible symmetry of the bi-critical point while $S_{\tau}T_{\tau}S_{\rho}T_{\rho} = M_3^2$.}. Therefore we have just one independent non-invertible defect, say $\cD_1^{SU(3)}$. Therefore, we have found that the $SU(3)_1$ enhanced symmetry point at $(\omega,\omega)$ has a TY$(\bZ_3)_{\chi, \epsilon}$ non-invertible symmetry. 

The presence of this defect implies that the theory at $(\tau = \omega, \rho = \omega)$ is self-dual under gauging the following $\bZ_3$ subgroup of the global symmetry
\be\label{eq: SU(3) Z3}
\bZ_3 \; : \; V_{\bf{n},\bf{w}} \rightarrow e^{\frac{2\pi i }{3} k (2n_1+n_2+2w_1+2w_2)}V_{\bf{n},\bf{w}} \,.
\ee

Now we proceed to prove the self-duality under gauging the $\bZ_3$ symmetry in \eqref{eq: SU(3) Z3} by explicitly computing the orbifold partition function. The chiral algebra at the $SU(3)$ point is characterized by a $K$-matrix of the form
\be\label{eq: Kmatrix SU(3)}
K=\begin{pmatrix}
    2&-1\\-1&2
\end{pmatrix} \; ,
\ee
and the partition function is given by the diagonal modular invariant
\be
Z_{(1,1)}=\sum_{\boldsymbol{\lambda}\in\cL} \chi_{\boldsymbol{\lambda}}\overline{\chi}_{\boldsymbol{\lambda}} \; ,
\ee
where the characters and the lattice $\cL$ are defined as before, but now with the $K$-matrix introduced in \eqref{eq: Kmatrix SU(3)}.
Furthermore, note that
\be
\bZ_3 \, : \, \chi_{\boldsymbol{\lambda}}\overline{\chi}_{\overline{\boldsymbol{\lambda}}} \, \to \, e^{2\pi i {\overline{\boldsymbol{\lambda}}}^TK^{-1}\overline{{\bf v}}_3}\chi_{\boldsymbol{\lambda}}\overline{\chi}_{\overline{\boldsymbol{\lambda}}}
\ee
with $\overline{{\bf v}}_3\equiv (0,-1)^T$. Denoting by $\eta$ to the generator of the above $\bZ_3$ action, the twisted partition function results
\be
Z_{(\eta^a,\eta^b)}=\frac{1}{|\cL|}\sum_{\boldsymbol{\lambda},\boldsymbol{\mu},\overline{\boldsymbol{\mu}}\in\cL}
e^{2\pi i a {\boldsymbol{\lambda}}^TK^{-1}\overline{{\bf v}}_3}
e^{2\pi i b{\overline{\boldsymbol{\mu}}}^TK^{-1}\overline{{\bf v}}_3}
e^{-i2\pi {\boldsymbol\lambda}^TK^{-1}({\boldsymbol\mu}-\overline{\boldsymbol\mu})}
\chi_{{\boldsymbol\mu}}\overline{\chi}_{\overline{\boldsymbol\mu}}
\ee
Following the same steps as in the previous sections, summing over $\boldsymbol{\lambda}$ and subsequently over $a,b\in\bZ_3$, we obtain
\be
\frac13\sum_{a,b\in\bZ_3}Z_{(\eta^a,\eta^b)}=
\sum_{\boldsymbol{\mu}\in\cL} \chi_{\boldsymbol{\mu}}\overline{\chi}_{\overline{C}\boldsymbol{\mu}} 
\quad , \quad 
\overline{C}=\left(\begin{array}{cc}
1 & 0\\
-1 & -1
\end{array}\right)\, .
\ee
Again, it can be easily verified that $\overline{C}$ preserves the pairing, namely $\overline{C}K\overline{C}^T=K$, hence $\overline{K}_{\overline{C}\boldsymbol{\mu}}\sim \overline{K}_{\boldsymbol{\mu}}$, leading to the claimed self-duality.  

Just for completeness, let us remark that the topological defect $\cD_1^{SU(3)}$ induces a consistent defect Hilbert space, as it complies with the conditions imposed by the modular bootstrap. First, the right hand side of equation \eqref{eq: c=2 phase} trivializes when evaluated on invariant states under \eqref{eq: SU(3) Z3} and consequently we set $\alpha({\bf n},{\bf w})=0$. Furthermore, the action of $\cD_1^{SU(3)}$ amounts to take $p_1\to-p_1$ and $p_2\to-p_2$, hence the twisted partition function only accounts for the contribution of states with $p_1=p_2=0$. The computation is almost identical to the one depicted around \eqref{eq:mod boots D2} and we will not include the details here. Performing the modular $S$ transformation, the trace over the defect Hilbert space is then of the form
\begin{align}
Z_{(\cD_1^{SU(3)},1)}&=\frac{1}{4|\eta(\tau)|^4}\vartheta_2\left(\tfrac{\tau}{2}\right)^2\left(\vartheta_3\left(\tfrac{\bar\tau}{6}\right)\vartheta_3\left(\tfrac{\bar\tau}{2}\right)+\vartheta_4\left(\tfrac{\bar\tau}{6}\right)\vartheta_4\left(\tfrac{\bar\tau}{2}\right)\right)\nonumber\\
&=\frac{2}{|\eta(\tau)|^4}\left(q^{\tfrac18}+6q^{\tfrac18}\bar q^{\tfrac13}+2q^{\tfrac58}+12q^{\tfrac58}\bar q^{\tfrac13} + \ldots \right)
\end{align}

Let us provide an additional interpretation of the above self-duality. The theory at hand has an alternative description as a $SU(3)_1$ Wess-Zumino-Witten (WZW) model. There are three integrable highest weight representations of the $SU(3)_1$ affine algebra corresponding to the following pair of fundamental weights 
\be
(\lambda_1,\lambda_2)\in\{(0,0),(1,0),(0,1)\}=\{{\bf 1},{\bf 3},{\bf \bar 3}\}
\ee
where ${\bf 1}$ and (${\bf \bar 3}$)${\bf 3}$ denote respectively the singlet and (anti-)fundamental representations of $SU(3)$. 
Note that the above labels precisely account for the three independent vectors of the lattice $\cL$ (indeed $|K|=3$ for this theory). Moreover, the characters associated to the representations ${\bf 3}$ and ${\bf \bar 3}$ are identical as functions of $q$, as it can be checked by explicit computation. From this perspective, the $\bZ_3$ subgroup \eqref{eq: SU(3) Z3} is no more than the center of $SU(3)$ generated by the Verlinde lines of the theory. The corresponding gauging implements the self-duality $SU(3)_1\sim \frac{SU(3)_1}{\bZ_3}$. As a matter of fact, the partition functions for these models
\be
Z_{SU(3)_1}=|\chi_{\bf 1}|^2+|\chi_{\bf 3}|^2+|\chi_{\bf \bar 3}|^2
\quad , \quad
Z_{\frac{SU(3)_1}{\bZ_3}}=|\chi_{\bf 1}|^2+\chi_{\bf 3}\overline\chi_{\bf \bar 3}+\chi_{\bf \bar 3}\overline\chi_{\bf 3}
\ee
are identical as functions of $(\tau,\bar\tau)$.

~

Regarding the set of relevant deformations at this point, we get
\begin{center}
\begin{tabular}{ |m{2cm}||m{10cm} | } 
 \hline
 $(h,\overline h)$ & $(\mathbf{n},\mathbf{w})$ \\ 
  \hline
$ \left( \frac{1}{3} \, , \, \frac{1}{3} \right)$& $(\pm1,0,0,\pm1)\,,\,(\pm 1,0,0,0)\,,\,(\pm 1,\mp 1,\pm1,\pm1)\,,\,(\pm1,\mp1,0,0)$ $(0,\pm1,0,0)\,,\,(0,\pm1,\mp1,0)\,,\,(0,0,\pm1,\pm1)\,,\,(0,0,\pm1,0)$ $(0,0,0,\pm1)$
\\ [1ex] 
 \hline
\end{tabular}
\end{center}
Contrary to the multicritical points, now there is no relevant deformation which preserves the non-invertible duality symmetry since any of the above operators is send to a non genuine one under the action of $\cD^{SU(3)}_{1,2,3}$.

\section*{Acknowledgments}
We would like to thank Riccardo Argurio, Luigi Tizzano, Christian Copetti, Andrea Antinucci and Giovanni Rizi for a careful reading of the manuscript and for giving us precious comments. The work of S.M. is supported by "Fondazione Angelo Della Riccia" and by funds from the Solvay Family. J.A.D. is a Postdoctoral Researcher of the F.R.S.-FNRS (Belgium). The research of J.A.D. and G.G is supported by IISN-Belgium (convention 4.4503.15) and through an ARC advanced project. The work of O. H. was supported by the FWO-Vlaanderen through the project G006119N and by the Vrije Universiteit Brussel through the Strategic Research Program ``High-Energy Physics''.

\appendix 

\section{Details on compact bosons}\label{App:somedetails}

In this section we expose some conventions and useful formulas that are used in the main text. We adopt the conventions in which a periodic free holomorphic field $X$ satisfies
\be
X(z)\sim X(z)+2\sqrt{2}\pi R \quad , \quad \langle X(z)X(w)\rangle=-\log (z-w)
\ee
Along the toroidal branch of a $c=D$ theory of free periodic scalar fields, a generic vertex operator is written as
\be
V_{{\bf n},{\bf w}}(z,\overline z)\equiv e^{i{\bf n}.{\boldsymbol \phi}(z,\overline z)+i{\bf w}.\widetilde{\boldsymbol \phi}(z,\overline z)}=e^{i{\bf p}.{\bf X}(z)+i\overline{\bf p}.\overline{\bf X}(\overline z)}
\ee
where we defined the $D$-dimensional vectors
\begin{align}
{\boldsymbol \phi}=(\phi_1,\ldots, \phi_D) \quad &, \quad \widetilde{\boldsymbol \phi}=(\widetilde\phi_1,\ldots, \widetilde\phi_D) \, , \nonumber\\
{\bf X}=(X_1,\ldots, X_D) \quad &, \quad \overline{\bf X}=(\overline X_1,\ldots, \overline{X}_D) \, , \\
{\bf n} =(n_1,\ldots, n_D) \quad &, \quad {\bf w}=(w_1,\ldots , w_D) \,, \nonumber
\end{align}
where $\{n_1\}$, $\{w_i\}$ denote the charges under the $U(1)_n^D\times U(1)_w^D$ global symmetry. Of course, operators along the orbifold branch can be written in a similar fashion, though arbitrary charges are not generically gauge invariant. The exact relation between the fields $({\boldsymbol \phi}, \widetilde{\boldsymbol \phi})$ and $({\bf X},\overline{\bf X})$ (or equivalently between $({\bf n},{\bf w})$ and $({\bf X},\overline{\bf X})$) depends on the details of the theory in consideration and we will be explicit in the examples considered in this paper. 

By omitting the antiholomorphic part for notational simplicity, the operator product expansion of two vertex operators reads
\be\label{eq: gen OPE}
V_{{\bf n},{\bf w}}(z,\overline z)V_{{\bf n}',{\bf w}'}(w,\overline w)\sim (-1)^{{\bf n}\cdot {\bf w}'} (z-w)^{{\bf p}.{\bf p}'}
:e^{i\sum_{m}^\infty (z-w)^m({\bf p}\cdot \partial{\bf X}(w))^m }::e^{i({\bf p}+{\bf p}')\cdot {\bf X}}(w):
\ee
Let us briefly comment on the $(-1)^{{\bf n}\cdot {\bf w}'}$ phase arising in the above OPE. The reason for it is due to imposing locality on the correlation functions of the theory. Take for instance a correlation function of the form
\be\label{eq:gen correlator}
\langle V_{{\bf n},{\bf w}}(z,\overline z)V_{{\bf n}',{\bf w}'}(w,\overline w) \ldots \rangle
\ee
where the $\ldots$ denote possible additional insertions at points away from $z$ and $w$. By locality, we mean that the above correlation function should be single valued, {\it i.e.} invariant under taking a point, say $w$, along a loop around $z$ and back to its original position. This turns out to be non-trivial by the fact that momentum and winding modes are not mutually local, due to the mixed anomaly. The operation just described is achieved by exchanging the insertions and then analytically continuing to a function of $(z-w)$ ($(\overline z-\overline w)$). For sake of simplicity, let us omit the additional insertions in \eqref{eq:gen correlator} and focus on the two-point function. At the level of the OPE, one obtains
\begin{align}
V_{{\bf n},{\bf w}}(w,\overline w)V_{{\bf n}',{\bf w}'}(z,\overline z)&\sim (-1)^{{\bf n}'\cdot{\bf w}}(w-z)^{{\bf p}.{\bf p}'}(\overline w-\overline z)^{\overline{\bf p}.\overline{{\bf p}}'}(\cdots) \nonumber\\
&\sim (-1)^{{\bf n}'\cdot{\bf w}}(-1)^{{\bf n}\cdot{\bf w}'+{\bf n}'\cdot{\bf w}}(z-w)^{{\bf p}.{\bf p}'}(\overline z-\overline w)^{\overline{\bf p}.\overline{{\bf p}}'}(\cdots)
\end{align}
where $(\cdots)$ denote normal ordered operators which do not affect the present argument. In going to the second line, we analytically continued by taking $(w-z)\to e^{i\pi}(z-w)$ ($(\overline w-\overline z)\to e^{-i\pi}(\overline z-\overline w)$) and then expressed the resulting phase in terms of the mutual spin\footnote{In fact, regardless of the particular relation between $({\bf p}, \overline{\bf p})$ and $({\bf n},{\bf w})$, the $2D$-dimensional momentum lattice corresponding to this kind of theories always satisfies that the mutual spin ${\bf p}.{\bf p}'-\overline{\bf p}.\overline{{\bf p}}'={\bf n}\cdot{\bf w}'+{\bf n}'\cdot{\bf w}$. Notice that we are assuming that both insertions are genuine local bosonic operators, hence ${\bf n}\cdot{\bf w}',{\bf n}'\cdot{\bf w}\in {\mathbb Z}$}. The correlation function is then invariant
\be
\langle V_{{\bf n},{\bf w}}(w,\overline w)V_{{\bf n}',{\bf w}'}(z,\overline z)\rangle =\langle V_{{\bf n},{\bf w}}(z,\overline z)V_{{\bf n}',{\bf w}'}(w,\overline w)\rangle \,,
\ee
consistently with locality.
More generally, it can be proven that the overall phase in \eqref{eq: gen OPE} is necessary and sufficient to ensure the locality of a generic correlation function (see \cite{Fuchs:2007tx} and \cite{Runkel:2022fzi} for a recent discussion). 

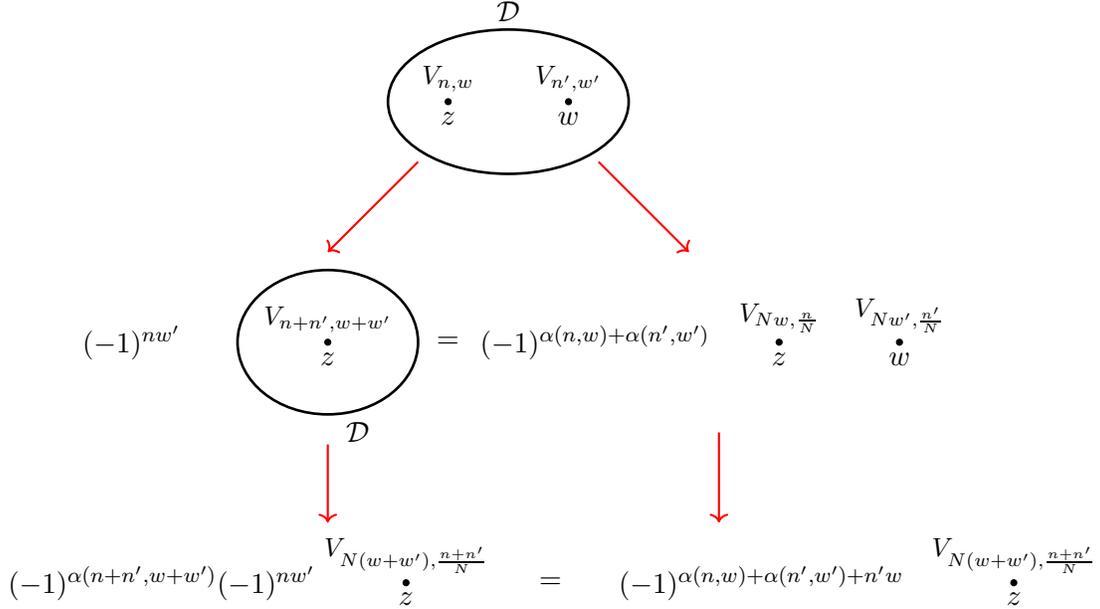
\begin{figure}
   \begin{tikzpicture}[scale=0.8]
   	\begin{scope}[shift={(-11.5,0)}]
	\node[] at (-5.5,2.5){$\quad (-1)^{nw'}$};
 		  	\draw[line width = 1] (-2,2.5) ellipse [x radius = 1.5,y radius = 1.2];
			\filldraw [black] (-2, 2.5) circle [radius = 0.05] node[below, black] {$z$}node[above, black] {\small\small$V_{n+n',w+w'}$};
			\node[] at(-1.5,1.) {$\mathcal{D}$} ;	
			\node[] at (0.,2.5){$= $};
			 \draw[line width=0.8, color=red,->] (-2,0.8)--(-2,-0.5);
	\end{scope}	
 
	\begin{scope}[shift={(-9.5,-4.)}]
	\node[] at (-7.,2.5){$\quad (-1)^{\alpha(n+n',w+w')}(-1)^{nw'}$};
			\filldraw [black] (-2.7, 2.5) circle [radius = 0.05] node[below, black] {$z$}node[above, black] {\small\small$V_{N(w+w'),\frac{n+n'}{N}}$};
			\node[] at (-0.3,2.5){$= $};
	\end{scope}	
 
 		\begin{scope}[shift ={(-7.5,4)}]
		\draw[line width = 1] (-3,2.5) ellipse [x radius = 2,y radius = 1.2];
		\filldraw [black] (-2, 2.5) circle [radius = 0.05] node[below, black] {$w$} node[above, black] {\small$V_{n',w'}$};
		\filldraw [black] (-4, 2.5) circle [radius = 0.05] node[below, black] {$z$} node[above, black] {\small$V_{n,w}$};	
		\node[] at(-3,4.) {$\mathcal{D}$} ;
		 \draw[line width=0.8, color=red,->] (-4.5,1.5)--(-6,0);
		  \draw[line width=0.8, color=red,->] (-1.5,1.5)--(0.,0);	
		\end{scope}				
		\begin{scope}[shift={(-3.5,0)}]

 			\node[] at (-5.8,2.5){$\quad (-1)^{\alpha(n,w)+\alpha(n',w')}$};
			\filldraw [black] (-0.5, 2.5) circle [radius = 0.05] node[below, black] {$w$} node[above, black]{\small$V_{Nw',\frac{n'}{N}}$};
			\filldraw [black] (-2.5, 2.5) circle      [radius = 0.05] node[below, black] {$z$}    node[above, black]  {\small$V_{Nw,\frac{n}{N}}$};	
			 \draw[line width=0.8, color=red,->] (-3.5,1)--(-3.5,-0.5);	
  
			\end{scope}			
			\begin{scope}[shift={(-7.5,-4.)}]
			\node[] at (1.2,2.5){$(-1)^{\alpha(n,w)+\alpha(n',w')+n'w}$};
			\filldraw [black] (5.4, 2.5) circle [radius = 0.05] node[below, black] {$z$} node[above, black] {\small$V_{N(w+w'),\frac{n+n'}{N}} $};			
		\end{scope}		 
  \end{tikzpicture}
    \caption{Constraint on the phase factor $\alpha$ coming from mutual locality, combined with the action of the non-invertible symmetry.}\label{fig: mutual phase}
\end{figure}

Moreover, the presence of this phase also affects the action of some topological operators implementing dualities that exchange momentum and winding modes. We illustrate this for the case of the $T$-duality defect at $c=1$ in Figure \ref{fig: mutual phase}, leading to the following constraint
\be
\alpha(n+n',w+w')+\alpha(n+n',w+w')+\alpha(n+n',w+w')=nw'+n'w \,\, {\rm mod} \,\, 2
\ee
A possible solution of the above is $\alpha(n,w)=nw$,
hence obtaining the action \eqref{eq:c=1 duality action}.

The above argument can be generalized for any duality defect ${\cal D}$ at $c=2$. For this purpose, let us denote by $D{\bf n}$ and $D{\bf w}$ respectively to momentum and winding charges transformed by the action of ${\cal D}$. These can be easily read from resorting to the matrix representation of the duality symmetry action, as we explain in the main text. Again, we denote by $\alpha({\bf n},{\bf w})$ to the potential phase generated by the action of ${\cal D}$ on $V_{{\bf n},{\bf w}}$. Now, the argument depicted in Figure \ref{fig: mutual phase} leads to 
\be\label{eq: c=2 phase}
\alpha({\bf n}+{\bf n}',{\bf w}+{\bf w}')+\alpha({\bf n},{\bf w})+\alpha({\bf n}',{\bf w}')={\bf n}\cdot{\bf w}'+(D{\bf n})\cdot (D{\bf w}') \,\, {\rm mod} \,\, 2
\ee

\subsection{The self-dual point at $c=1$}\label{app:enhancedc1}

As the dynamics of one of the $c=2$ multicritical theories studied in this paper, namely the quadri-critical point, keeps a close relation with the self-dual point at $c=1$, we briefly review some features about the latter theory in this subsection. 

The $c=1$ theory at radius $R=1$ is characterized by the presence of additional spin one conserved currents, that enhance the chiral algebra to $su(2) \times \overline{su(2)}$. More precisely, the holomorphic currents
\be
J_1(z)=\frac{1}{2}\left(e^{i\sqrt{2}X(z)}-e^{- i\sqrt{2}X(z)}\right) \, , \,  J_2(z)=\frac{1}{2i}\left(e^{i\sqrt{2}X(z)}+e^{-i\sqrt{2}X(z)}\right) \, , \, J_3(z)=
\frac{i}{\sqrt{2}}\partial X(z) \nonumber
\ee
satisfy the $su(2)_1$ chiral algebra
\be\label{eq:chiralcurrentsOPE}
J_i(z)J_j(w)=\frac12 \frac{\delta_{ij}}{(z-w)^2}+\frac{i\epsilon_{ijk}J_k(w)}{z-w}+\ldots
\ee
with the totally antisymmetric tensor satisfying $\epsilon_{123}=1$. Equivalently, in the complexified basis
\be
J_\pm(z)=e^{\pm i\sqrt{2}X(z)} \, \Rightarrow \, J_+(z)J_-(w)= \frac{1}{(z-w)^2}+\frac{2iJ_3(w)}{z-w}+\ldots
\ee

The key observation is that at $R=1$ there are two $\mathbb{Z}_2$ symmetries, the shift $\mathbb{Z}_2^s$ and the orbifold $\mathbb{Z}_2^{orb}$ which act on the fields and currents as 
\begin{align}
{\mathbb Z}_2^{s} \; &: \qquad X\to X+\sqrt{2}\frac{\pi}{2} \; , \quad X\to \overline{X}+\sqrt{2}\frac{\pi}{2} \; , \quad \left( J_3 \, , \; J_{\pm} \right) \to \left( J_3 \, , \; - J_{\pm} \right) \nonumber \\[5pt]
{\mathbb Z}_2^{orb} \; &: \qquad X\to-X \; , \quad X\to -\overline{X} \; , \quad \left( J_3 \, , \; J_{\pm} \right) \to \left( - J_3 \, , \; J_{\mp} \right) \; ,
\end{align}
with a similar action on the anti-holomorphic part. The quotient of ${\mathbb Z}_2^s$, combined with $T$-duality\footnote{Equivalently, one may quotient by a $\mathbb{Z}_2\subset U_w(1)$.}, maps the theory to the KT point ($R=2$). On the other hand, the quotient by ${\mathbb Z}_2^{orb}$ defines the origin of the orbifold branch ($R_{orb}=1$). Since the global symmetry exchanges the two $\mathbb{Z}_2$, the two theories must be equivalent, leading to a multicritical point. In~\cite{Thorngren:2021yso} it is showed that the KT point hosts a non-invertible duality symmetry, that satisfies the structure of a $\mathbb{Z}_4$ Tambara-Tamagami category symmetry. Moreover, the duality symmetry is preserved on the orbifold branch.

Taking the product of two copies of the self-dual theory described above describes the $SU(2)^2$ point $(\tau,\rho)=(i,i)$ on the toridal branch of the $c=2$ conformal manifold. The symmetry algebra of the $SU(2)^2$ point is then simply accounted for by two copies of the above currents, constructed in terms of $(\phi_1,\widetilde\phi_1)$ and $(\phi_2,\widetilde\phi_2)$ respectively.   

\subsection{The $SU(3)$ point at $c=2$}\label{app:enhancedc2}

The point with maximal enhanced symmetry within the $c=2$ toroidal branch corresponds to $(\tau,\rho)=(\omega,\omega
)$, with $\omega=e^{i\frac{2\pi}{3}}$. Given that $\tau$,$\rho$ are in ${\mathbb Q}(\sqrt{-3})$, the corresponding theory is rational and, moreover, diagonal since $\tau=\rho$. The condition $\overline p_i=0$ has eight solutions with $h=1$, which we assemble in the following currents
\begin{align}
    I_{\pm} &= \pm V_{\pm1,\mp1,\pm1,0} = \pm e^{\pm i \sqrt{2}X^1} \; , \quad \overline I_{\pm} = \pm V_{\pm1,0,\mp1,0} = \pm e^{\pm i \sqrt{2} \, \overline X^1} \nonumber \\[5pt]
    K_{\pm} &= \mp V_{\pm1,0,\pm1,\pm1} = \mp e^{\pm i \frac{1}{\sqrt{2}}X^1 \pm i \sqrt{\frac{3}{2}}X^2} \; , \quad \overline K_{\pm} = \mp V_{0,\pm1,\mp1,\mp1} = \mp e^{\pm i \frac{1}{\sqrt{2}} \overline X^1 \pm i \sqrt{\frac{3}{2}} \overline X^2} \; , \nonumber \\[5pt]
    U_{\pm} &= \pm V_{0,\pm1,0,\pm1} = \pm e^{\mp i \frac{1}{\sqrt{2}}X^1 \pm i \sqrt{\frac{3}{2}}X^2} \; , \quad \overline U_{\pm} = \pm V_{\mp1,\pm1,0,\mp1} = \pm e^{\mp i \frac{1}{\sqrt{2}} \overline X^1 \pm i \sqrt{\frac{3}{2}} \overline X^2} \; , \nonumber \\[5pt]
    J^3 &= i \sqrt{2} \partial X^1 \; , \quad  \overline J^3 = i \sqrt{2}\, \overline \partial \, \overline X^1 \nonumber \\[5pt]
    J^8 &= i \sqrt{2} \partial X^2 \; , \quad  \overline J^8 = i \sqrt{2}\, \overline \partial \, \overline X^2 \; ,
\end{align}
which respect the $su(3)$ algebra
\begin{align}
    \left[ J^3 , I_{\pm} \right] &= \pm 2 I_{\pm} \; , \quad \left[ J^3 , U_{\pm} \right] = \mp U_{\pm} \; , \quad \left[ J^3 , K_{\pm} \right] = \pm K_{\pm} \; , \nonumber \\[5pt] 
    \left[ J^8 , I_{\pm} \right] &= 0 \; , \quad \left[ J^8 , U_{\pm} \right] = \pm \sqrt{3} U_{\pm} \; , \quad \left[ J^8 , K_{\pm} \right] = \pm \sqrt{3} K_{\pm} \; , \nonumber \\[5pt] 
    \left[ I_+ , I_- \right] &= J^3 \; , \quad \left[ I_+ , U_+ \right] = K_+ \; \quad \left[ I_+ , K_- \right] = - U_- \; , \nonumber \\[5pt]
    \left[ U_+ , K_- \right] &= I_- \; , \quad \left[ U_+ , U_- \right] = \frac{1}{2}\left( \sqrt{3}J^8 - J^3 \right) \; , \quad \left[ K_+ , K_- \right] = \frac{1}{2}\left( \sqrt{3}J^8 + J^3 \right) \; , 
\end{align}
and similarly for the antiholomorphic currents. The remaining commutators are zero. In terms of the basis 
\begin{align}
    I_{\pm} = \frac{1}{2} \left( J^1 \pm i J^2 \right) \; , \nonumber \\[5pt]
    K_{\pm} = \frac{1}{2} \left( J^4 \pm i J^5 \right) \; , \nonumber \\[5pt]
    U_{\pm} = \frac{1}{2} \left( J^6 \pm i J^7 \right) \; , 
\end{align}
the $su(3)$ chiral algebra reads
\be
J_i(z) J_j(w) = \frac{1}{2}\frac{\delta_{ij}}{(z-w)^2}+\frac{i f_{ijk}J_k(w)}{(z-w)}
\ee 
with the fully antisymmetric structure constants given in the Gell-Mann presentation
\be
f_{123}=1 \quad , \quad f_{147}=-f_{156}=f_{246}=f_{257}=f_{345}=-f_{367}=\frac12 \quad , \quad f_{458}=f_{678}=\frac{\sqrt{3}}{2}
\ee
Note that there are three distinguished $SU(2)$ subalgebras generated by 
\be
\{J^1,J^2,J^3\} \quad , \quad \{J^4, J^5, J^3+\sqrt{3}J^8\} \quad , \quad \{J^6, J^7, +\sqrt{3}J^8-J^3\} \; .
\ee

\section{Partition function for RCFTs at $c=2$}\label{app:PartFunctionc2}

In this section, we briefly review the construction of the partition function in terms of representations of the extended chiral algebra at rational points along the toroidal branch. This construction is well known and we will follow the recent presentation in \cite{Benini:2022hzx}.

At a generic rational point in the toroidal branch, the theory features an extended chiral algebra of the form
\be\label{eq:K chiral alg}
u(1)^2_{K_L}\times u(1)^2_{K_R}
\ee
where $K_L$ and $K_R$ are positive symmetric even integer matrices. In the following, we describe how these matrices are defined and, more importantly, how to construct the representations of \eqref{eq:K chiral alg}. 

We start by writing the scaling dimensions as\footnote{We denote $|v|^2_M=v^iM_{ij}v^j$.} 
\be
h=\frac14\big| \vec{n} + M\vec{w} \big|^2_{G^{-1}} \quad , \quad \bar h=\frac14\big| \vec{n} - M^T\vec{w} \big|^2_{G^{-1}}
\ee
where $M=G+B$, with $G$ and $B$ the metric and $B$-field respectively introduced in section \ref{sec: c=2}. The charge vectors $\vec{n}$, $\vec{w}$ are two dimensional in this case, but one may apply this construction to any $c\in\bZ$. The condition for an operator to be chiral then reads $\vec{n}=M^T\vec{w}$. Notice that for generic $M\in GL(2,{\mathbb Q})$, then $M^T\vec{w}$ is not necessarily in ${\mathbb Z}^2$ for all $\vec{w}$, hence not corresponding to any genuine primary operator. In order to avoid that, we pick a sublattice $\Lambda_L$ such that for any $\vec{w}\in \Lambda_L$ then $M^T\vec{w}\in{\mathbb Z}^2$. We assemble a particular basis for $\Lambda_L$ into the columns of a matrix $P_L$. Of course, the choice of basis is completely arbitrary, hence $P_L$ is itself defined up to conjugation by unimodular integer matrices. The same procedure is applied to the anti-chiral sector, where the condition instead reads $\vec{n}=-M\vec{w}$, hence defining $P_R$ in a similar fashion.

In terms of the objects just introduced, we proceed to define the $K_L$, $K_R$ matrices in \eqref{eq:K chiral alg}, namely
\be
K_L=2P_L^T G P_L \quad , \quad K_R=2P_R^T G P_R \,.
\ee
The (finite) set of representations of \eqref{eq:K chiral alg} are in one-to-one correspondence with the elements of the following lattices 
\be
{\cL}_{L,R}\equiv \left\lbrace {\bf v}\in {\mathbb Z}^2 \, , \, {\bf v}\sim {\bf v}+K_{L,R}{\bf v}' \, , \, {\bf v}' \in {\mathbb Z}^2
 \right\rbrace \,.
\ee
Notice that there is a total of $|\cL_L|=|\cL_R|={\rm det}K_L={\rm det}K_R$ non-equivalent elements in these lattices. A given (anti-)chiral primary of \eqref{eq:K chiral alg} is labeled by a vector ${\bf \lambda}\in {\cL}_L$ (${\bf\lambda}_R \in \overline{\cL}_R$) and the characters of its associated representation then read
\be
\chi_{{\boldsymbol\lambda}_L}(\tau)=\frac{1}{\eta(\tau)^2}\sum_{{\bf l}\in{\mathbb Z}^2} q^{\frac12|{\boldsymbol\lambda}_L+K_L{\bf l}|^2_{K_L^{-1}}}
\quad , \quad 
\bar{\chi}_{{\boldsymbol\lambda}_R}(\bar\tau)=\frac{1}{\eta(\bar\tau)^2}\sum_{{\bf r}\in{\mathbb Z}^2} \bar{q}^{\frac12|{\boldsymbol\lambda}_R+K_R{\bf r}|^2_{K_R^{-1}}}
\ee
Notice that the dimensions associated to the primaries are 
\be
h=\frac12|{\boldsymbol\lambda}_L|^2_{K_L^{-1}} \quad , \quad \bar h=\frac12|{\boldsymbol\lambda}_R|^2_{K_R^{-1}}
\ee
whereas the integer vectors ${\bf l}$ and ${\bf r}$ parametrize their descendants under the action of the higher spin currents of the chiral algebra \eqref{eq:K chiral alg}. 

The modular invariant partition function is determined by a particular pairing between the left- and right-moving sectors. Such a pairing is given by a group isomorphism $\hat\omega: \, {\cL}_L\to{\cL}_R$ defined as follows. First, it is easy to verify that, for a given $M=G+B$, there exist integer $2\times 2$ matrices $N_{1,2}$, such that 
\be
S=\left(\begin{array}{cc}
P_L^T & P_L^T M\\
N_1 & N_2
\end{array}
\right) \in SL(4,{\mathbb Z})
\ee
Its inverse is also a unimodular integer matrix and reads
\be
S^{-1}= S=\left(\begin{array}{cc}
R_0 &  MN_3\\
-S_0 & -N_3
\end{array}
\right)
\ee
for some $R_0$, $S_0$ and $N_3$ (see \cite{Benini:2022hzx} for a proof that these matrices always exist). The left-right pairing $\hat\omega$ is therefore written as
\be
\hat\omega =P_R^TR_0+P_R^TM^TS_0
\ee
and the partition function results
\be
Z=\sum_{\vec{\lambda}_{L}\in{\cal D}_{L}}\chi_{{\boldsymbol\lambda}_L}\overline{\chi}_{\hat\omega{\boldsymbol\lambda}_L}
\ee
Finally, the modular $S$ and $T$ matrices acting on the holomorphic characters are of the form
\be
S_{{\boldsymbol\lambda},{\boldsymbol\mu}}=\frac{1}{|\cL_L|^{1/2}}e^{-2\pi i \boldsymbol\lambda^T K^{-1}_L\boldsymbol\mu} \quad , \quad T_{{\boldsymbol\lambda},{\boldsymbol\mu}}=\delta_{{\boldsymbol\lambda},{\boldsymbol\mu}}e^{-i\frac{\pi}{6}+i\pi|\boldsymbol\lambda|^2_{K^{-1}_L}} \,,
\ee
with analogous counterparts for the antiholomorphic blocks.
A useful identity is the following
\be\label{eq:orthogonal}
\sum_{\vec{\boldsymbol\lambda}\in{\cL}}e^{-i2\pi {\boldsymbol\lambda}^T K^{-1}{\boldsymbol\mu}}=|D|\delta_{{\boldsymbol\mu},{\bf 0}}\,.
\ee
with ${\bf 0}$ denoting the trivial vector in $\cL$.
In all these expressions, the constraints imposed by the delta functions are intended to hold up to lattice identifications. In particular, the above properties ensure the modular invariance of the partition function.

Given the representations introduced above, it is instructive to write the corresponding charges ${\bf n}$, ${\bf w}$ under $U(1)_{\bf n}\times U(1)_{\bf w}$ symmetry in terms of $\boldsymbol\lambda_L$, $\boldsymbol\lambda_R$, ${\bf l}$ and ${\bf r}$. To this purpose, we define the orthogonal matrix $O$ and  such that
\be
K= O^T {\rm diag}(k_1,k_2) O
\ee
for some integers $k_i$. Since $K$ is symmetric, the orthogonal matrix $O$ is guaranteed to exist.
Comparing the expressions for the scaling dimensions in terms of either variables one arrives to the following
\begin{align}\label{eq:gen charge map}
p_i&= 
%\frac{1}{\sqrt{2g_i}}O_{ij}\left(n_j+(Mw)_j\right)= 
\frac{1}{\sqrt{k_i}}{O}_{ij}\left(\lambda_j+(K_L l)_j\right) \\
\overline p_i&= 
%\frac{1}{\sqrt{2g_i}}O_{ij}\left(n_j-(M^Tw)_j\right)= 
\frac{1}{\sqrt{k_i}}{O}_{ij}\left(\bar\lambda_j+(K_R l)_j\right)
\end{align}
where we redefined ${\boldsymbol\lambda} \equiv {\boldsymbol\lambda}_L$ ($\overline{\boldsymbol{\lambda}} \equiv \boldsymbol{\lambda}_R$), and $p_i$ ($\overline p_i$) are the usual left (right) chiral weights. The above equivalence defines a set of four linear equations for $n_i$, $w_i$.

\section{Tambara-Yamagami categories}\label{App:TY} 
In this appendix we review some of the properties of the Tambara Yamagami categories \cite{TAMBARA1998692} which describe categorical symmetries arising from self-dualities under gauging some discrete symmetry. Given a discrete abelian group $\bA$, a symmetric non-degenerate bicharacter 
\be \chi: \bA \times \bA \longrightarrow U(1) \ee
and a class $\epsilon \in H^3(\bZ_2, \, U(1))=\bZ_2$ called Frobenius Shur indicator, the Tambara Yamagami category, also dubbed as  $\text{TY}(\bA)_{\epsilon, \, \chi}$, is a graded fusion category:
\be
\cC = \cC_0 \oplus \cC_1 \, , \ \ \ \cC_0 = \text{Vec}_{\bA} \, , \ \ \ \cC_1 = \cD 
\ee
with fusion rules
\be\label{eq: TY fusions}
a \times b = ab \, , \ \ \ a \times \cD = \cD \times a = \cD \, , \ \ \ \cD \times \cD = \bigoplus_{a \in \bA} a \, 
\ee
and non-trivial F-symbols\footnote{The F-symbols are a base-dependent representation (i.e. after choosing a set of simple objects) of the natural isomorphism 
\be
F_{x,y,z}: (x \otimes y) \otimes z \rightarrow x \otimes (y \otimes z)
\ee
where $x,y,z$ are some objects inside the category. In the physical language of topological lines, it relates to in-equivalent line configurations
}
\be
\Bigl[ F^{a,\, \cD ,\, b}_\cD \Bigr]_{\cD,\, \cD} = \Bigl[ F^{\cD,\, a ,\, \cD}_b \Bigr]_{\cD,\, \cD} = \chi(a,b) \;,\qquad\qquad
\Bigl[F^{\cD,\, \cD,\, \cD}_\cD \Bigr]_{a,\, b} = \frac{\epsilon}{\sqrt{\lvert \bA \rvert}} \, \chi(a, b)^{-1} \;.
\ee
The category $\text{Vec}_{\bA}$, \ie{} the category of $\bA$ graded vector spaces, describes an anomaly free invertible $0$-form symmetry $\bA$ while $\cD$ is the non-invertible defect which grades $\bA$ and with quantum dimension $\dim(\cD) = \sqrt{\vert \bA \vert}$, as evident from the last fusion rule in \eqref{eq: TY fusions}.

 From the structure of $\text{TY}(\bA)_{\epsilon, \, \chi}$ follows that a theory with this categorical symmetry is automatically self-dual under gauging of $\bA$. Indeed the gauging of $\bA$ can be performed by inserting a fine enough mesh of the algebra object $\cA = \bigoplus_{a \in \bA} a $ inside the $2$d space-time. However, because of \eqref{eq: TY fusions} a mesh of $\cA$ can always be reduced to contractible insertions of $\cD$ 
 \begin{equation}
   \begin{tikzpicture}[scale=0.5]
		\draw[line width=1,red] (0,0)node[below, color=red] {$\mathcal{A}$}--(2,2)--(0,4)node[below, color=red] {$\mathcal{A}$} ;
		\draw[line width=1,red] (2,2)--(5,2)node[below, color=red] {$\mathcal{A}$};
	 	\node[below, color=black] at (6.5,2.2) {$=$};
	 
	     \begin{scope}[shift={(8.5, -0.6)}]

		 \draw[line width=0.8, color=black] (4.7, 2.3)node[below, color=black] {$\mathcal{D}$} arc (90: 140: 6);
		 \draw[line width=0.8, color=black] (0.1, 4.7) node[right, color=black] {$\mathcal{D}$}arc (220: 270: 6);
 		 \draw[line width=0.8, color=black] (-0.25, 0.4) node[left, color=black] {$\mathcal{D}$}arc (-55: 55: 2.5);
	 \end{scope}
	 
  \end{tikzpicture}
    \label{eq: resolution of network}
\end{equation}
automatically implying the claimed self-duality. 
 
 Even more interestingly, also the opposite is true: any $2$d theory self-dual under the gauging of an abelian symmetry $\bA$ enjoys an additional topological line $\cD$ which follows the structure described above \cite{Kaidi:2021xfk,Choi:2021kmx}. More concretely, such topological line can be constructed by gauging $\bA$ on half-space and imposing Dirichlet boundary conditions for the gauge field 
 \bea
\begin{tikzpicture}
    \filldraw[color=white, left color=white!50!blue, right color= white] (0,0) -- (2,0) -- (2, 1.5) -- (0, 1.5) -- cycle; 
    \draw[color= black, line width = 1.5] (0,0) -- (0, 1.5);
    \node at (1, 0.75) {$\cT/\bA$}; \node at (-1, 0.75) {$\cT$}; \node[below] at (0,0) {$\cD$};
\end{tikzpicture}
\eea
Because $\cT/\bA \simeq \cT$, the domain-wall $\cD$ actually describes a topological defect within the theory $\cT$.
From this point of view, the choice of the bicharacter $\chi$ has a nice physical interpretation. While on the l.h.s. of $\cD$ we have a theory with global symmetry $\bA$, on the r.h.s. the gauged theory has global symmetry $\bA^\vee \equiv \text{Hom}(\bA,U(1))$ which is \emph{non-canonically} isomorphic to $\bA$. In order to identify the two symmetries, we actually need to choose a group isomorphism $\phi : \bA \rightarrow \bA^\vee$ and the bicharacter is then defined as
\be
\chi(a,b) = \phi(a)b \in U(1)\,.
\ee
Such categorical structure can be generalized by grading $\bA$ with any group $G$ (possibly also non-Abelian). In this case the full category $\cC$ can be decomposed as direct sum of abelian categories 
\be
\cC = \bigoplus_{g \in G} \cC_g
\ee
where $\cC_1 = \bA$ while $\cC_g$ contain different non-invertible duality defects as their set of objects. These kind of symmetries are marginal in our considerations and we refer to \cite{gelaki2009centers,Choi:2023vgk} for more details. We just mention that theories which are self-dual under gauging of $\bA$ with some non-trivial discrete torsion enjoy this kind of categorical symmetries.

Duality symmetries following $\text{TY}(\bA)_{\epsilon, \, \chi}$ can also have anomalies which constrain the possible IR behaviors of symmetry preserving RG flows \cite{Thorngren:2021yso,Antinucci:2023ezl}. Such anomalies can be defined as obstruction to have a trivially gapped phase \emph{or} as obstruction to gauge the categorical symmetry. Even if the two concepts coincides fro invertible symmetries, it was shown in \cite{Choi:2023xjw} that this two notion are generically inequivalent for fusion category symmetries. In the particular case of TY$(\bA)_{\chi,\epsilon}$ with $\bA\equiv \bZ_N$, a symmetric RG flow is compatible with a symmetry preserving gapped phase only if $N = r^2$ for some integer $r$ while the IR theory must either break spontaneously the duality symmetry or must be gapples when $N \not= r^2$. In the former case, the symmetry can be preserved in the IR if and only if $\epsilon$ is trivial. Similar conditions for generic groups $\bA$ can be found in e.g. \cite{Antinucci:2023ezl}.

We conclude this section with a brief overview on how the topological data corresponding to the type of categories described above can be extracted from the modular bootstrap. We refer to \cite{Thorngren:2021yso} for a more detailed treatment. Resorting to modular covariance, applying the modular $T$ transformation ({\it i.e.} $\tau\to\tau+2$) to the twisted Hilbert space, we end up with the configuration depicted on the right of 

\begin{equation}
 \raisebox{-4.8 em}{  \begin{tikzpicture}
		\draw[line width=1] (0,0)--(0,3)--(3,3)--(3,0)--(0,0);
		\draw[line width=1] (1.5,0)node[below]{$\cD$}--(1.5,3);
   \end{tikzpicture}}
   \qquad \xrightarrow{\tau \rightarrow \tau+2}\qquad
   \raisebox{-4.8 em}{  \begin{tikzpicture}
		\draw[line width=1] (0,0)--(0,3)--(3,3)--(3,0)--(0,0);
		\draw[line width=1] (1.5,0)node[below]{$\cD$}--(3,0.75);
		\draw[line width=1] (0,0.75)--(3,3*0.75);
		\draw[line width=1] (0,3*0.75)--(1.5,3);
   \end{tikzpicture}}
    \label{eq: T^2 transf}
\end{equation}
Alternatively, the action $\tau\to \tau+2$ generates a phase proportional to the spin, namely
\be
Z_{({\cal D},1)}(\tau+2)={\rm Tr}_{{\cal H}_{\cal D}}e^{-2\pi\tau_2\hat\Delta}e^{2\pi i(\tau_1+2)\hat s}=\sum_{{\cal H}_{\cal D}}e^{4\pi i s}q^h \overline{q}^{\overline h}
\ee
where $\hat\Delta=L_0+\overline{L}_0$ and $\hat s=L_0-\overline{L}_0$ are the usual dilatation and spin operators. 

Now, the trick amounts to map the configuration on the right of \eqref{eq: T^2 transf} to a trace over the defect Hilbert space ${\cal H}_{\cal D}$. This can be done by combining different F-moves, as shown in \eqref{eq: spin sel rules}.
\begin{equation}
   \raisebox{-4.8 em}{  \begin{tikzpicture}
		\draw[line width=1] (0,0)--(0,3)--(3,3)--(3,0)--(0,0);
		\draw[line width=1] (1.5,0)node[below]{$\cD$}--(3,0.75);
		\draw[dashed, line width=1,red] (1.5,1.5) --(1.5+0.75,0.75/2);
		\node[red] at (1.5+1/2,0.75+0.75/2) {$e$};
		\draw[line width=1] (0,0.75)--(3,3*0.75);
		\draw[line width=1] (0,3*0.75)--(1.5,3);
   \end{tikzpicture}}
    = \frac{\epsilon}{\sqrt{|\bA|}}\sum_{a}
   \raisebox{-4.8 em}{  \begin{tikzpicture}
		\draw[line width=1] (0,0)--(0,3)--(3,3)--(3,0)--(0,0);
		\draw[line width=1] (1.5,0)node[below]{$\cD$}--(0,0.75);
	         \draw[line width=1,red] (1.5+0.75,1.33) --(1.5-0.75,0.75/2);
		 \draw[line width=1] (3, 3*0.75) arc (90: 270: 0.75);
		 \node[red] at (1.5+1/2,1) {$a$};
		\draw[line width=1] (0,3*0.75)--(1.5,3);
   \end{tikzpicture}}
    = \frac{\epsilon}{\sqrt{|\bA|}}\sum_{a}
     \raisebox{-4.8 em}{  \begin{tikzpicture}
		\draw[line width=1] (0,0)--(0,3)--(3,3)--(3,0)--(0,0);
		\draw[line width=1] (1.5,0)node[below]{$\cD$}--(1.5,3);
	        \draw[line width=1,red] (3,1.5)--(1.5,1.5-0.3) ;
	        \draw[line width=1,red] (0,1.5)--(1.5,1.5+0.3);
	         \node[red] at (0.75,1.4) {$a$};
	          \node[red] at (1.5+0.75,1.5) {$a$};
   \end{tikzpicture}}
    \label{eq: spin sel rules}
\end{equation}
We insert an identity line $e$ running from two consecutive pieces of the defect ${\cal D}$. Then we perform an $F$-move to map it to a configuration on which ${\cal D}$ runs parallel to the time direction, subsequently making use of $\chi(e,a)=1$ for any $a$. The resulting diagram computes the trace over the defect Hilbert space with a sum over the elements of $\bA$ inserted. In order to determine the action of the symmetry $\bA$ on the states in ${\cal H}_{\cal D}$, it is required to resolve the four-way junction. The two possibilities are related by an $F$-move, more precisely $\left[F^{a,{\cal D},a}_{{\cal D}}\right]_{{\cal D},{\cal D}}=\chi(a,a)$. Following \cite{Thorngren:2021yso}, we denote by $\widehat{(\eta^a)}_-$ to the resolution in the right hand side of \eqref{eq: spin sel rules}, namely the one in which the left-insertion is placed above the right-insertion.

Finally, modular covariance implies the following constraints over the spin of states in the defect Hilbert space 
\be\label{eq:spin selection rule}
e^{4\pi i \hat{s}} \Big|_{{\cal H}_{\cal D}}= \frac{\epsilon}{\sqrt{|\bA|}}\sum_{a\in\bA} \widehat{(\eta)^a}_- \Big|_{{\cal H}_{\cal D}}
\ee
Note the above equality is written as an operator equation. In practice, one decomposes ${\cal H}_{\cal D}$ in terms of common eigenstates for both operators\footnote{This is always possible as these are commuting operators.} and demands for the eigenvalues to match. Note that the spin selection rule \eqref{eq:spin selection rule} involves both $\chi$ and $\epsilon$ (the bi-character is implicit in the choice of resolution for the junction). Therefore, this piece of topological data can be a priori determined by inspection of the spectrum of spins arising in ${\cal H}_{\cal D}$. We show how it works in several examples in the main body of this article. Let us conclude by mentioning that there may be cases for which \eqref{eq:spin selection rule} is satisfied by more than one inequivalent combination of $(\chi,\epsilon)$. In such a case, a more refined analysis is required to fully fix the categorical data.

\section{Modular functions}\label{app: modular}\label{App:modular functions}

Let us give a brief overview of the modular functions employed in the calculations exposed in the main text. We will mainly follow \cite{DHoker:2022dxx} for the conventions and transformation properties. 

First, the Dedekind eta function is defined as
\be
\eta(\tau)=q^{\tfrac{1}{24}}\prod_{m=1}^{\infty}1-q^m\,,
\ee
with $q$ ($\bar q$) defined as usual $q=e^{2\pi i\tau}$ ($\bar q=e^{2\pi i\bar\tau}$). Under the modular $S$ transformation, the function above transforms as
\be
\eta\left(-\tfrac{1}{\tau}\right)=\sqrt{-i\tau}\eta(\tau) \quad , \quad \eta\left(-\tfrac{1}{\bar\tau}\right)=\sqrt{i\bar\tau}\eta(\bar\tau)\,.
\ee
In addition, we introduce the elliptic Jacobi theta function with characteristics defined as follows\footnote{More generally, these are functions of two complex variables $\tau$ and $z$, where $z$ is usually interpreted as a fugacity. Throughout this work we will always set $z=0$, as it does not play any role in for our purposes. }
\be
\vartheta\left[{\alpha\atop \beta}\right](\tau)=\sum_{n\in\bZ}e^{\pi i \tau(n+\alpha)^2}e^{2\pi i(n+\alpha)\beta}\,,
\ee
where the real parameters $\alpha$, $\beta$ take values on the interval $[0,1)$. These functions have well defined transformations under the action of the modular group, in particular
\be\label{eq:mod transf theta}
\vartheta\left[{\alpha\atop \beta}\right]\left(-\tfrac{1}{\tau}\right)=\sqrt{-i\tau} e^{2\pi i \alpha \beta}\vartheta\left[{\beta\atop -\alpha}\right](\tau)\,.
\ee
A predominant role is usually played by these functions when $\alpha\sim -\alpha$, $\beta\sim-\beta$, that is $\alpha,\beta\in\{0,\tfrac12\}$, hence they are given special names 
\be
\vartheta_1(\tau)\equiv-\vartheta\left[{\tfrac12\atop \tfrac12}\right](\tau) \quad , \vartheta_2(\tau)\equiv \vartheta\left[{\tfrac12\atop 0}\right] \quad , \quad \vartheta_3(\tau)\equiv \vartheta\left[{0\atop 0}\right] \quad , \quad 
\vartheta_4(\tau)\equiv \vartheta\left[{0\atop \tfrac12}\right](\tau)\,.
\ee
From \eqref{eq:mod transf theta}, one can readily obtain their transformation under $\tau\to-1/\tau$, namely
\begin{align}
\vartheta_1\left(-\tfrac{1}{\tau}\right)=-i\sqrt{-i\tau}\vartheta_1(\tau) \quad &, \quad \vartheta_2\left(-\tfrac{1}{\tau}\right)=\sqrt{-i\tau}\vartheta_4(\tau) \nonumber \\
\vartheta_3\left(-\tfrac{1}{\tau}\right)=\sqrt{-i\tau}\vartheta_3(\tau) \quad &, \quad 
\vartheta_4\left(-\tfrac{1}{\tau}\right)=\sqrt{-i\tau}\vartheta_2(\tau) \,.
\end{align}

Finally, there is also a twisted version of the Dedekind eta function which, in the context of the present paper, often appears when the oscillator modes are odd under a particular symmetry operation, as it is the case of $T$-duality at $c=1$. In terms of the elliptic functions just introduced, such a distribution reads
\be
q^{\tfrac{1}{24}}\prod_{m=1}^{\infty}1+q^m =\frac{\eta(\tau)}{\vartheta_4(2\tau)}\,.
\ee

\bibliographystyle{ytphys}
\baselineskip=0.85\baselineskip
\bibliography{myref}

\end{document}